\documentclass[floatfix, preprint, onecolumn, prd,showpacs, showkeys, preprintnumbers, nofootinbib, superscriptaddress]{revtex4-1}
\usepackage{times}
\usepackage{amssymb,amsbsy,amsmath,amsfonts}
\usepackage{graphicx}
\usepackage{float}
\usepackage{color}
\usepackage{morefloats}
\usepackage{rotating}
\usepackage{srcltx}
\usepackage{slashed}
\usepackage{subfigure}
\usepackage{multirow}
\usepackage{verbatim}
\usepackage{hyperref}
\usepackage{tabularx}

\usepackage{color}

\begin{document}
\title{Meson-baryon scattering up to the next-to-next-to-leading order  in covariant baryon chiral perturbation theory}

\author{Jun-Xu Lu}
\affiliation{School of Physics and
Nuclear Energy Engineering \& International Research Center for Nuclei and Particles in the Cosmos \&
Beijing Key Laboratory of Advanced Nuclear Materials and Physics,  Beihang University, Beijing 100191, China}
\affiliation{Groupe de Physique Th\'{e}orique, IPN (UMR8608), Universit\'{e} Paris-Sud 11, Orsay, France}

\author{Li-Sheng Geng}
\email[E-mail: ]{lisheng.geng@buaa.edu.cn} \affiliation{School of
Physics and Nuclear Energy Engineering \& International Research
Center for Nuclei and Particles in the Cosmos \& Beijing Key
Laboratory of Advanced Nuclear Materials and Physics,  Beihang
University, Beijing 100191, China}

\author{Xiu-Lei Ren}
\affiliation{Ruhr-Universit\"{a}t Bochum, Fakult\"{a}t f\"{u}r Physik und Astronomie, Institut f\"{u}r Theoretische Physik II, D-44780 Bochum, Germany}

\author{Meng-Lin Du}
\affiliation{Helmholtz-Institut f\"ur Strahlen- und Kernphysik and Bethe Center for Theoretical Physics, Universit\"at Bonn, D-53115 Bonn, Germany}

\begin{abstract}
We study the scattering of a pseudoscalar meson off one ground state octet baryon in covariant baryon chiral perturbation theory (BChPT) up to the next-to-next-to-leading order.
The inherent power counting breaking terms are removed within extended-on-mass-shell scheme. We perform the first combined study of the pion-nucleon and kaon-nucleon scattering data in covariant BChPT and show that it  can provide a reasonable description of the experimental data.  In addition, we  find
that it is possible to fit the experimental  baryon masses and the pion-nucleon and kaon-nucleon scattering data simultaneously at this order, thus providing a
consistent check on covariant BChPT. We  compare the scattering lengths of all the pertinent channels with available experimental data and those of other approaches. In addition, we have
studied the leading order contributions of the virtual decuplet and found that they can improve the description of the $\pi N$ phase shifts near the $\Delta(1232)$ peak, while
they have negligible effects on the description of the $K N$ phase shifts.

\end{abstract}

\pacs{13.60.Le, 12.39.Mk,13.25.Jx}

\maketitle
\section{Introduction}
Elastic meson-baryon scattering~\footnote{Throughout this work, mesons and baryons refer to the octet of Nambu-Goldstone bosons and the ground-state octet baryons, unless otherwise specified.}  is a fundamental process that not only can test our understanding of the strong interaction but also plays a relevant role in the studies of the
properties of single and multi baryons~\cite{hohler}. For instance, one can derive from pion-nucleon scattering the  nucleon sigma term, which is essential to understand the quark flavor structure of the nucleon in the scalar channel and plays an important role in direct dark matter searches~\cite{Bottino:1999ei,Bottino:2008mf,Ellis:2005mb,Ellis:2008hf,Hill:2011be,Cline:2013gha,Ellis:2018dmb}. In addition, meson-baryon scattering also provides key inputs in the construction of the  chiral baryon-baryon interactions and may affect the equation of state of dense matter at high densities and therefore help to understand
the so-called hyperon puzzle~\cite{Massot:2012pf,Schulze:2011zza,Hu:2013tma,Miyatsu:2013hea} in explaining the existence of two-solar-mass neutron stars~\cite{Demorest:2010bx,Antoniadis:2013pzd}. Furthermore,  meson-baryon scattering appears in the final states of heavier hadron decays and therefore becomes
an integrated part in the test of the standard model~\cite{Oset:2016lyh,Roca:2015tea,Miyahara:2015cja,Ikeno:2015xea,Xie:2017gwc} and in the search of beyond standard model physics~\cite{Doring:2013wka}. Because of these, one has
seen increasing theoretical, such as chiral perturbation theory (ChPT)~\cite{Kaiser:2001hr,Liu:2006xja,Liu:2007ct, Mai:2009ce,Huang:2015ghe, Huang:2017bmx} and  lattice QCD~\cite{Torok:2009dg, Detmold:2015qwf},  as well as experimental interests~\cite{Niiyama:2008rt,Agakishiev:2012xk,Moriya:2013eb,Moriya:2013hwg,Moriya:2014kpv}  in meson-baryon scattering in recent years.

ChPT, as a low-energy effective field theory of QCD, plays an important role in our understanding of the nonperturbative strong interaction physics~\cite{Weinberg:1978kz,Gasser:1983yg,Gasser:1984gg,Gasser:1987rb}.
In particular, it provides a model independent framework to describe the dynamics of the Nambu-Goldstone bosons  interacting
among themselves and with other hadrons containing light ($u$, $d$, and $s$) quarks. For comprehensive reviews, see, e.g., Refs.~\cite{Leutwyler:1994fi,Bernard:1995dp,Pich:1995bw,Ecker:1994gg,Scherer:2002tk,Bernard:2007zu}.

The constraints imposed by  chiral symmetry and its breaking are the most stringent on the self-interactions of the Nambu-Goldstone bosons and therefore ChPT has the largest predictive power in the pure mesonic sector. In the one-baryon sector, its predictive power decreases because a large number of unknown low energy constants (LECs) has to be introduced. As only a finite number of them appears in a particular process, this does not  severely  hamper its applicative power. A further complicating factor is the power counting breaking (PCB) issue. Namely, because of the large nonzero baryon masses $m_0$ in the chiral limit, lower order analytical terms appear in nominal higher order loop calculations, and therefore a consistent power counting is lost~\cite{Gasser:1987rb}. In the past three decades, several solutions have been proposed. The most studied ones are the heavy baryon ChPT~\cite{Jenkins:1990jv,Bernard:1995dp}, the infrared (IR) baryon ChPT~\cite{Becher:1999he}, and the extended-on-mass-shell (EOMS) baryon ChPT~\cite{Gegelia:1999gf,Fuchs:2003qc}. For a short summary and comparison of these different schemes, see, e.g., Ref.~\cite{Geng:2013xn}. In recent years, it has been shown that both formally and empirically, the EOMS BChPT seems to be more appealing because it satisfies all the symmetry and analyticity constraints and
converges relatively faster.

Although the EOMS BChPT has been successfully applied to study pion-nucleon scattering~\cite{ Alarcon:2011zs,Alarcon:2012kn,Chen:2012nx,Yao:2016vbz,Siemens:2016hdi,Siemens:2016jwj,Siemens:2017opr}, it has not been
applied to study kaon-nucleon,  or more generally, meson-baryon scattering. Our present study aims to fill this gap. It is particularly timely given the extensive studies of baryon masses~\cite{Ren:2012aj,Ren:2013dzt,Ren:2013wxa,Ren:2013oaa,Ren:2014vea,Ling:2017jyz} and the recent attempt to construct baryon-baryon interactions using covariant BChPT~\cite{Ren:2016jna,Li:2016mln,Song:2018qqm,Li:2018tbt,Xiao:2018jot}.~\footnote{
In a recent work~\cite{Girlanda:2018xrw}, it was shown that one can achieve a satisfactory description of the polarized p-d scattering data below the deuteron breakup threshold
and solve the long-standing $A_y$ puzzle e with the leading order relativistic 3N interaction (in the power counting of Ref.~\cite{Ren:2016jna}).} As mentioned above,
meson-baryon scattering connects these studies and provides a non-trivial test of the consistency of BChPT.

This article is organized as follows. In Sec. II, we present the theoretical formalism and calculate meson-baryon scattering amplitudes up to the next-to-next-to-leading (NNLO) order.
In Sec. III, we explain in detail the renormalization of the meson-baryon scattering amplitudes. In Sec. IV, we specify how to remove ultraviolet divergences and power-counting breaking terms
in the loop amplitudes.  Fitting results and discussions
are presented in Sec. V, followed by a short summary and outlook in Sec. VI.

\section{Theoretical formalism}
In this section, we explain in detail how to calculate the meson-baryon scattering amplitudes in covariant BChPT with the EOMS scheme. As pion-nucleon scattering
has been studied in this framework previously~\cite{ Alarcon:2011zs,Alarcon:2012kn,Chen:2012nx,Yao:2016vbz}, we will highlight the new ingredients in extending the study
from SU(2) to SU(3). For details similar to the SU(2) case, we refer the reader to Refs.~\cite{ Alarcon:2011zs,Alarcon:2012kn,Chen:2012nx,Yao:2016vbz}.

\begin{figure}
\centering
  \includegraphics[width=0.6\textwidth]{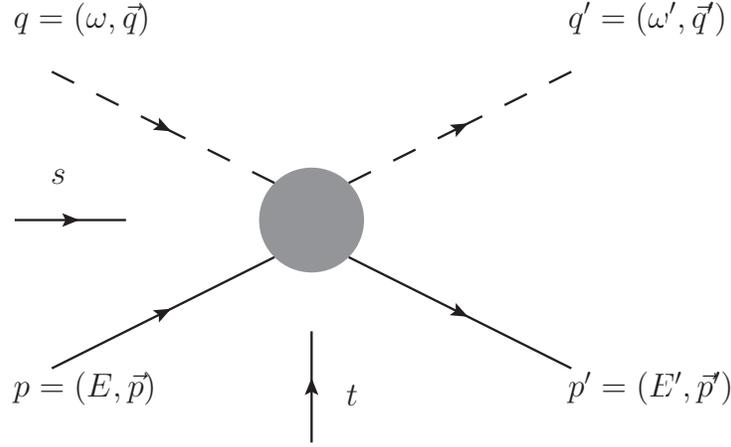}\\
  \caption{Kinematics of meson-baryon scattering, where $p$, $p'$,$ q$, $q'$ are the momenta of
  incoming and outgoing baryons and mesons,  $s$ and $t$ are
 the Mandelstam variables. The solid lines denote baryons, and dashed lines represent mesons .}\label{mb-kinematics}
\end{figure}

\subsection{Scattering amplitudes and partial wave phaseshifts}
In the isospin limit, the standard decomposition of the meson-baryon scattering amplitude reads~\cite{Gasser:1987rb,Becher:2001hv},
\begin{equation}\label{stde1}
  T_{MB}=\overline{u}(p',s')\left[A+\frac{1}{2}(\slashed q+\slashed q')B\right]u(p,s),
\end{equation}
where  $p(p')$ and $q(q')$  are the momentum of the initial (final) baryons and mesons, respectively (see Fig.~\ref{mb-kinematics}).
Introducing the Mandelstam variables $s,t,u$, one can rewrite  Eq.~(\ref{stde1}) in an alternative form~\footnote{This can be easily checked by noting that
$[\gamma^{\mu},\gamma^{\nu}]q_{\mu}q'_{\nu}=2(m_i+m_f)\slashed q -s+u$.}
\begin{equation}\label{stde2}
  T_{MB}=\overline{u}(p',s')\left[D+\frac{i}{m_i+m_f}\sigma^{\mu\nu}q'_{\mu}q_{\nu}B\right]u(p,s),
\end{equation}
where $\sigma^{\mu\nu}=\frac{i}{2}[\gamma^{\mu},\gamma^{\nu}]$ and  $D=A+\frac{s-u}{2(m_i+m_f)}B$.
However, as noted in Ref~\cite{Chen:2012nx}, since the leading part of $A$ and $B$ may cancel each other, one  better use $B$ and $D$ to perform the low energy expansion of the scattering amplitudes when extracting the PCB terms.

The above scattering amplitudes can be projected onto specific partial waves in the following form~\cite{hohler}:
\begin{equation}\label{PWA1}
    \mathcal{T}^{l\pm}_{MB}=\frac{1}{2}(f^l_1+f^{l\pm}_2),
\end{equation}
where $f^l_1$ and $f^l_2$ take the forms~of

\begin{equation}\label{PWAf1}
    f^l_1=\frac{\sqrt{E+m_i}\sqrt{E'+m_f}}{8\pi\sqrt{s}}(A_l +\frac{\omega \omega '}{2}B_l+(\frac{|\vec{q}|}{2(E+m_i)}+\frac{|\vec{q}'|}{2(E'+m_f)})B_l),
\end{equation}
\begin{equation}\label{PWAf2}
    f^l_2=\frac{\sqrt{E+m_i}\sqrt{E'+m_f}|\vec{q}||\vec{q}'|}{8\pi\sqrt{s}}(\frac{B_l}{2(E+m_i)}+\frac{B_l}{2(E'+m_f)} -\frac{A_l-\frac{\omega \omega '}{2}B_l}{(E+m_i)(E'+m_f)}),
\end{equation}
\begin{equation}\label{PWAABl}
    \begin{split}
      A_l(s)= & \int_{-1}^{1} A(s,t)P_l(\cos \theta)d\cos\theta, \\
      B_l(s)= & \int_{-1}^{1} B(s,t)P_l(\cos \theta)d\cos\theta,
    \end{split}
\end{equation}
where $E,E',\omega,\omega'$ are the energy of the incoming and outgoing particles in the center of mass (c.m.) frame, $\vec{q}$  and $\vec{q}'$ are the c.m. momentum of the incoming(outgoing) mesons,
$m_i$ and $m_f$ are the masses of the incoming and outgoing baryons. The $P_l$ above refers to the Legendre polynomials with angular momentum $l$.

From the partial wave amplitudes, one can obtain the corresponding phase shifts~\cite{Fettes:1998ud}
\begin{equation}\label{PWA2}
    \delta_{l\pm}=\arctan \{|\vec{p}|Ref_{l\pm}(s)\}.
\end{equation}
In the present work, we will  rely on the modern partial wave analysis of the George Washington University group~\cite{Arndt:2006bf,Hyslop:1992cs} to fix the relevant LECs.~\footnote{
For pion-nucleon scattering, one may also use the latest analysis based on the Roy-Steiner equation~\cite{Hoferichter:2015hva}. However, as our primary interest is to compare different formulations of BChPT, we choose
to use the same data to fix the relevant LECs as those used by the previous studies~\cite{Chen:2012nx,Huang:2017bmx}.}

\subsection{Power counting}
In ChPT, the relative importance of a certain Feynman diagram contributing  to a particular process is determined by its chiral order, $\nu$, whose size is
of the order of $(p/\Lambda_{\chi})^\nu$, where $p$ denotes a generic small quantity and $\Lambda_{\chi}$ the chiral symmetry breaking scale.  In the one-baryon sector, where only one baryon is
involved in both the initial and the final states, the chiral order for any given Feynman diagram with $L$ loops, $V_n$ $n$-th order vertices, $N_M$ internal meson lines, and $N_B$ internal baryon lines, is
\begin{equation}
\nu=4L+\sum\limits_n n V_n -2 N_M - N_B.
\end{equation}

In the present context, the small quantities or expansion parameters are
\begin{equation}\label{SmallQuantity}
    s-\tilde{m}^2\sim \mathcal{O}(p),\quad t  \sim \mathcal{O}(p^2),\quad m_{\pi}, m_K, m_{\eta}  \sim \mathcal{O}(p), \quad m_{N,\Lambda,\Sigma,\Xi}-\tilde{m}\sim \mathcal{O}(p^2),
\end{equation}
Note that although in principle  $\tilde{m}$ here refers to $m_0$, the chiral limit baryon mass,  in the study of $\pi N$ and $KN$ scattering, we set $\tilde{m}=m_N$.


\subsection{Chiral Lagrangians}
In order to calculate the meson-baryon scattering amplitudes up to the leading one-loop order,  i.e., $\mathcal{O}(p^3)$, we need the following meson-meson and meson-baryon Lagrangians:
\begin{equation}
\mathcal{L}_\mathrm{eff}=\mathcal{L}^{(2)}_{MM}+\mathcal{L}^{(4)}_{MM} + \mathcal{L}^{(1)}_{MB}+\mathcal{L}^{(2)}_{MB}+\mathcal{L}^{(3)}_{MB},
\end{equation}
where the superscripts denote the chiral order. The lowest-order meson-meson Lagrangian is
\begin{equation}\label{MMLag}
  \mathcal{L}_{MM}^{(2)}=\frac{F_{0}^2}{4}\langle D_{\mu}U(D^{\mu}U)^{\dag} \rangle + \frac{F_{0}^2}{4}\langle \chi U^{\dag}+U\chi^{\dag} \rangle,
\end{equation}
where $\chi=\mathrm{diag}(m_{\pi}^2,m_{\pi}^2,2m_K^2-m_{\pi}^2)$,  $U(\phi)=u^2(\phi)=\exp(i\frac{\phi}{F_{0}})$, and $F_0$ is the chiral limit value of the pseudoscalar decay constant. The traceless $3\times3$ matrix contains the pseudoscalar fields:
\begin{equation}\label{phi}
  \phi=\sqrt{2}\left(
                                                   \begin{array}{ccc}
                                                     \frac{1}{\sqrt{2}}\pi^0+\frac{1}{\sqrt{6}}\eta & \pi^+ & K^+ \\
                                                     \pi^- & -\frac{1}{\sqrt{2}}\pi^0+\frac{1}{\sqrt{6}}\eta & K^0 \\
                                                     K^- & \bar{K}^0 & -\frac{2}{\sqrt{6}}\eta \\
                                                   \end{array}
                                                 \right).
\end{equation}
The next-to-leading order meson-meson Lagrangian relevant to our study has the following form
\begin{equation}
\mathcal{L}^{(4)}_{MM}= L_4 \langle D_{\mu}U(D^{\mu}U)^{\dag}\rangle \langle\chi U^{\dag}+U \chi ^{\dag}\rangle + L_5\langle D_{\mu}U(D^{\mu}U)^{\dag}(\chi U^{\dag}+U\chi^{\dag})\rangle.
\end{equation}

The lowest order meson-baryon Lagrangian~\cite{Oller:2007qd,Oller:2006yh} is
\begin{equation}\label{MBLag1}
  \mathcal{L}_{\phi B}^{(1)}=\langle \bar{B}(i\gamma^{\mu}D_{\mu}-m_0)B \rangle +\frac{D/F}{2}\langle \bar{B}\gamma^{\mu}\gamma_5[u_{\mu},B]_{\pm} \rangle,
\end{equation}
where $m_0$ is the chiral limit baryon mass, the covariant derivative $D_{\mu}B=\partial_{\mu}B+[\Gamma_{\mu},B]$, $
  \Gamma_{\mu}=\frac{1}{2}\{u^{\dag}\partial_{\mu} u+u\partial_{\mu}u^{\dag}\}$,  and $u_{\mu}=i\{ u^{\dag}\partial_{\mu}u-u \partial_{\mu} u^{\dag}\}$. The
  $3\times 3$ traceless matrix contains the ground-state octet baryons fields:
  \begin{equation}\label{BB}
  B=\left(
                                                   \begin{array}{ccc}
                                                     \frac{1}{\sqrt{2}}\Sigma^0+\frac{1}{\sqrt{6}}\Lambda & \Sigma^+ & p^+ \\
                                                     \Sigma^- & -\frac{1}{\sqrt{2}}\Sigma^0+\frac{1}{\sqrt{6}}\Lambda & n^0 \\
                                                     \Xi^- & \Xi^0 & -\frac{2}{\sqrt{6}}\Lambda \\
                                                   \end{array}
                                                 \right).
\end{equation}

The meson-baryon Lagrangian at order $\mathcal{O}(p^2)$ relevant to meson-baryon scattering has 14 terms of the following form~\cite{Oller:2007qd,Oller:2006yh,Frink:2006hx}:
\begin{equation}\label{MBLag2}
  \begin{split}
    \mathcal{L}_{\phi B}^{(2)}= & b_D\langle\bar{B}\{\chi_+,B \}\rangle +b_F\langle\bar{B}[\chi_+,B ]\rangle + b_0\langle\bar{B}B\rangle\langle\chi_+\rangle+ \\
      & b_1\langle \bar{B}[u^{\mu},[u^{\mu},B]] \rangle + b_2\langle \bar{B}\{u^{\mu},\{u^{\mu},B\}\} \rangle +\\
      & b_3\langle \bar{B}\{u^{\mu},[u^{\mu},B]\} \rangle + b_4\langle\bar{B}B\rangle\langle u^{\mu}u_{\mu} \rangle+\\
      & ib_5\left( \langle \bar{B}[u^{\mu},[u^{\nu},\gamma_{\mu} D_{\nu}B]] \rangle - \langle \bar{B}\overleftarrow{D}_{\nu}[u^{\nu},[u^{\mu},\gamma_{\mu}B]] \rangle\right)+\\
      & ib_6\left( \langle \bar{B}[u^{\mu},\{u^{\nu},\gamma_{\mu} D_{\nu}B\}] \rangle - \langle \bar{B}\overleftarrow{D}_{\nu}\{u^{\nu},[u^{\mu},\gamma_{\mu}B]\} \rangle\right)+\\
      & ib_{7}\left( \langle \bar{B}\{u^{\mu},\{u^{\nu},\gamma_{\mu} D_{\nu}B\}\} \rangle - \langle \bar{B}\overleftarrow{D}_{\nu}\{u^{\nu},\{u^{\mu},\gamma_{\mu}B\}\} \rangle\right)+\\
      & ib_{8}\left( \langle \bar{B}\gamma_{\mu}D_{\nu}B \rangle - \langle \bar{B}\overleftarrow{D}_{\nu}\gamma_{\mu}B \rangle \right)\langle u^{\mu}u^{\nu} \rangle+\\
      & ic_1\langle \bar{B}\{[u^{\mu},u^{\nu}],\sigma_{\mu\nu}B\} \rangle + ic_2\langle \bar{B}[[u^{\mu},u^{\nu}],\sigma_{\mu\nu}B] \rangle +ic_3\langle \bar{B}u^{\mu} \rangle\langle u^{\nu}\sigma_{\mu\nu}B \rangle.
  \end{split}
\end{equation}

The meson-baryon Lagrangian contributing to $MB\rightarrow MB$ at order $\mathcal{O}(p^3)$ has 13 terms of the following form~\cite{Oller:2007qd,Oller:2006yh,Frink:2006hx}:
\begin{equation}\label{MBLag3}
  \begin{split}
    \mathcal{L}_{MB}^{(3)}= & id_1\left( \langle\bar{B}\gamma_{\mu}D_{\nu\rho}B[u^{\mu},h^{\nu\rho}]\rangle + \langle\bar{B}\overleftarrow{D}_{\nu\rho}\gamma_{\mu}B[u^{\mu},h^{\nu\rho}]\rangle \right) +\\
      & id_2\left( \langle\bar{B}[u^{\mu},h^{\nu\rho}]\gamma_{\mu}D_{\nu\rho}B\rangle + \langle\bar{B}\overleftarrow{D}_{\nu\rho}[u^{\mu},h^{\nu\rho}]\gamma_{\mu}B\rangle \right) +\\
      & id_3\left( \langle\bar{B}u^{\mu}\rangle\langle h^{\nu\rho}\gamma_{\mu}D_{\nu\rho}B\rangle - \langle\bar{B}\overleftarrow{D}_{\nu\rho}h^{\nu\rho}\rangle \langle u^{\mu}\gamma_{\mu}B\rangle \right)+\\
      & id_4\langle \bar{B}[u^{\mu},h^{\mu\nu}]\gamma_{\nu}B \rangle + id_5\langle \bar{B}\gamma_{\nu}B[u^{\mu},h^{\mu\nu}] \rangle +\\
      & id_6\left(  \langle\bar{B}u^{\mu}\rangle\langle h^{\mu\nu}\gamma_{\nu}B\rangle - \langle\bar{B}h^{\mu\nu}\rangle\langle u^{\mu} \gamma_{\nu}B\rangle \right)+\\
      & id_7\left( \langle \bar{B}\sigma_{\mu\nu}D_{\rho}B\{u^{\mu},h^{\nu\rho}\} \rangle - \langle \bar{B}\overleftarrow{D}_{\rho}\sigma_{\mu\nu}B\{u^{\mu},h^{\nu\rho}\} \rangle \right) + \\
      & id_8\left( \langle \bar{B}\{u^{\mu},h^{\nu\rho}\}\sigma_{\mu\nu}D_{\rho}B \rangle - \langle \bar{B}\overleftarrow{D}_{\rho}\{u^{\mu},h^{\nu\rho}\}\sigma_{\mu\nu}B \rangle \right) + \\
      & id_9\left( \langle \bar{B}u^{\mu}\sigma_{\mu\nu}D_{\rho}Bh^{\nu\rho} \rangle - \langle \bar{B}\overleftarrow{D}_{\rho}u^{\mu}\sigma_{\mu\nu}Bh^{\nu\rho} \rangle \right) + \\
      & id_{10}\left( \langle \bar{B}\sigma_{\mu\nu}D_{\rho}B \rangle - \langle \bar{B}\overleftarrow{D}_{\rho}\sigma_{\mu\nu}B \rangle \right)\langle u^{\mu}h^{\nu\rho} \rangle + \\
      & d_{48}\langle \bar{B}\gamma_{\mu}B[\chi_-,u^{\mu}] \rangle +d_{49}\langle \bar{B}[\chi_-,u^{\mu}]\gamma_{\mu}B \rangle +\\
      & d_{50}\left( \langle\bar{B}u^{\mu}\rangle\langle\chi_-\gamma_{\mu}B\rangle - \langle\bar{B}\chi_-\rangle\langle u^{\mu}\gamma_{\mu}B\rangle \right),
  \end{split}
\end{equation}
where $D_{\nu\rho}=D_{\nu}D_{\rho}+D_{\rho}D_{\nu}$ and  $h_{\mu\nu}=D_{\mu}u_{\nu}+D_{\nu}u_{\mu}$.

For Born terms at $\mathcal{O}(p^3)$ and vertex corrections, we also need the following Lagrangian, which contributes to  $B_1\rightarrow M_1 B_2$ and  has 10 terms
\begin{eqnarray}\label{BMBLag3}
      \mathcal{L}_{BMB}^{(3)}&= & d_{38}\langle \bar{B}u^{\mu}\gamma_5\gamma_{\mu}B \chi_+ \rangle +d_{39} \langle \bar{B}\chi_+\gamma_5\gamma_{\mu}B u^{\mu}\rangle  +d_{40} \langle \bar{B}u^{\mu}\gamma_5\gamma_{\mu}B \rangle\langle \chi_+ \rangle +d_{41} \langle \bar{B}\gamma_5\gamma_{\mu}B u^{\mu}\rangle\langle \chi_+ \rangle \nonumber\\
        & +&d_{42} \langle \bar{B}\gamma_5\gamma_{\mu}B \rangle\langle u^{\mu}\chi_+ \rangle +d_{43} \langle \bar{B}\gamma_5\gamma_{\mu}B \{u^{\mu},\chi_+\} \rangle +d_{44} \langle \bar{B}\{u^{\mu},\chi_+\}\gamma_5\gamma_{\mu}B \rangle \nonumber\\
        & +&d_{45} \langle \bar{B}\{\chi_-, \gamma_5 B\} \rangle  +d_{46} \langle \bar{B}[\chi_-, \gamma_5 B] \rangle  +d_{47} \langle \bar{B} \gamma_5 B\rangle\langle \chi_- \rangle.
\end{eqnarray}

It should be noted that not all of the $\mathcal{O}(p^2)$ and $\mathcal{O}(p^3)$ terms contribute to a specific process. Particularly, for pion-nucleon and kaon-nucleon scattering, only 24 out of the total 37 LECs contribute.  They are tabulated in Table ~\ref{final-LEC}.

\begin{table}\label{final-LEC}
\centering
\caption{Independent (combinations of) LECs  contributing to $\pi N$ and $KN$ scattering. For the sake of later reference, we introduce  $\alpha_{1,\cdots,8},\beta_{1,\cdots,8},\gamma_{1,\cdots,8}$ to denote different combinations of LECs. The units of the LECs are given in the last column.}
\begin{tabular}{c|c|c|c}
\hline\hline
$\pi N$  &  $KN_{I=0}$  &  $KN_{I=1}$  & \\
\hline
 $\alpha_1=b_1+b_2+b_3+2b_4$  &  $\beta_1=b_3-b_4$         &  $\gamma_1=b_1+b_2+b_4$    &$[\mathrm{GeV^{-1}}]$\\
 $\alpha_2=b_5+b_6+b_7+b_8$   &  $\beta_2=2b_6-b_8$        &  $\gamma_2=2b_5+2b_7+b_8$  &$[\mathrm{GeV^{-2}}]$\\
 $\alpha_3=c_1+c_2$           &  $\beta_3=4c_1+c_3$        &  $\gamma_3=4c_2+c_3$       &$[\mathrm{GeV^{-1}}]$\\
 $\alpha_4=2b_0+b_D+b_F$      &  $\beta_4=b_0-b_F$         &  $\gamma_4=b_0+b_D$        &$[\mathrm{GeV^{-1}}]$\\
 $\alpha_5=d_2$               &  $\beta_5=d_1+d_2+d_3$     &  $\gamma_5=d_1-d_2-d_3$    &$[\mathrm{GeV^{-4}}]$\\
 $\alpha_6=d_4$               &  $\beta_6=d_4+d_5+d_6$     &  $\gamma_6=d_4-d_5+d_6$    &$[\mathrm{GeV^{-2}}]$\\
 $\alpha_7=d_8+d_{10}$        &  $\beta_7=d_7-d_8+d_{10}$  &  $\gamma_7=d_7+d_8+d_{10}$ &$[\mathrm{GeV^{-3}}]$ \\
 $\alpha_8=d_{49}$            &  $\beta_8=d_{48}+d_{49}+d_{50}$  &  $\gamma_8=d_{48}+d_{49}-d_{50}$  &$[\mathrm{GeV^{-2}}]$\\
\hline
\end{tabular}
\end{table}

For an explicit study of the matching  between SU(3) and SU(2), we refer the reader to Refs.~\cite{Mai:2009ce,Frink:2004ic,Ren:2016aeo}. In doing so, one should note that the Lagrangians in Eqs.~(\ref{MBLag2},\ref{MBLag3},\ref{BMBLag3}) do not share the same Lorentz structures with those used in  SU(2). To obtain the matching relations between the LECs in the SU(2) and SU(3) Lagrangians, the following relation between $D_{\mu}$ and the Dirac matrix $\gamma_{\mu}$ is needed, which reads:
\begin{equation}\label{DmuGammamu}
    \bar{\Psi} A^{\mu} i D_{\mu} \Psi +h.c. \doteq 2m\bar{\Psi}\gamma_{\mu}A^{\mu}\Psi,
\end{equation}
where $A^\mu$ is an external field, and the symbol $\doteq$ means equal up to terms of higher orders. Neglecting the possible higher order corrections, which is beyond our concern here, it is straightforward to reduce the
SU(3) Lagrangians to those of their SU(2) counterparts. We notice that although the application of  Eq.~(\ref{DmuGammamu}) only leads to difference of higher orders, which could be ignored from the point of view of effective field theories, it results in a reorganization of the scattering amplitudes when divided into $A$ and $B$ parts. As a consequence, the explicit expressions of the tree level diagrams will be different.

We would like to  point out  that compared to the 9 free LECs in the $\pi N$ channel in  SU(2)~\cite{Chen:2012nx}, we find that only 8 of them are actually independent. All of the LECs in Eq.~(\ref{BMBLag3}), which correspond to the $d_{16}$ and $d_{18}$ terms of Ref.~\cite{Chen:2012nx}, eventually will not contribute  to the scattering amplitudes. In the $\mathcal{O}(p^3)$ Born diagrams, the contributions from the  $d_{38,\ldots,44}$ terms are canceled by the corrections from vertex renormalization. The remaining part, containing $d_{45},d_{46},d_{47}$, can be absorbed into those of the  $d_{48,\ldots,50}$ terms via
\begin{equation}\label{d45d46d47}
   2m\bar{\Psi}\gamma_{5}\chi_{-}\Psi \doteq -\bar{\Psi}\gamma_5\gamma^{\mu}[iD_{\mu},\chi_{-}]\Psi + \frac{g_A}{2} \bar{\Psi} [\slashed{u},\chi_{-}]\Psi,
\end{equation}
where  $g_A$ refers to the axial-vector current coupling constant. The first term on the right hand side
will be canceled as the  $d_{38,\ldots,44}$ terms do, while the second term is in the form of the $d_{48,\ldots,50}$ terms. Thus in the final scattering amplitudes, only 8 combination of the LECs will survive, which is consistent with the HBChPT study~\cite{Huang:2017bmx}.

In addition, we note that  the $b_5,b_6,b_7$ terms  in the Lagrangians [Eq.~(\ref{MBLag2}) and Eq.~(\ref{MBLag3})] are not symmetric under the exchange of the Lorentz indices $\mu,\nu$, while the $b_8$ term is. As a consequence, these four terms do not share the same expression. The same applies to the $d_1,d_2,d_3$ terms. Considering that the differences are two chiral orders higher, we supplement these terms with the terms with exchanged Lorentz indices  to make these Lagrangians symmetric with respect to the exchange of Lorentz indices. For instance, the modified $b_5$ and $d_3$ terms finally utilized in our calculation read
\begin{equation}\label{LagMod}
    \begin{split}
      \mathcal{L}_{b_5}=& i\left( \langle \bar{B}[u^{\mu},[u^{\nu},\gamma_{\mu} D_{\nu}B]] \rangle - \langle \bar{B}\overleftarrow{D}_{\nu}[u^{\nu},[u^{\mu},\gamma_{\mu}B]] \rangle\right) \\
                       +& i\left( \langle \bar{B}[u^{\nu},[u^{\mu},\gamma_{\mu} D_{\nu}B]] \rangle - \langle \bar{B}\overleftarrow{D}_{\nu}[u^{\mu},[u^{\nu},\gamma_{\mu}B]] \rangle\right), \\
      \mathcal{L}_{d_3}=&i\left( \langle\bar{B}u^{\mu}\rangle\langle h^{\nu\rho}\gamma_{\mu}D_{\nu\rho}B\rangle - \langle\bar{B}\overleftarrow{D}_{\nu\rho}h^{\nu\rho}\rangle \langle u^{\mu}\gamma_{\mu}B\rangle \right) \\
                       -&i\left( \langle\bar{B}h^{\nu\rho}\rangle\langle u^{\mu}\gamma_{\mu}D_{\nu\rho}B\rangle - \langle\bar{B}\overleftarrow{D}_{\nu\rho}u^{\mu}\rangle \langle h^{\nu\rho}\gamma_{\mu}B\rangle \right).
    \end{split}
\end{equation}

\subsection{Feynman diagrams up to $\mathcal{O}(p^3)$}

\subsubsection{Tree level contact terms}
\begin{figure}
\centering
\begin{tabular}{cccc}
$O(p^1)$:
&{\includegraphics[width=0.2\textwidth]{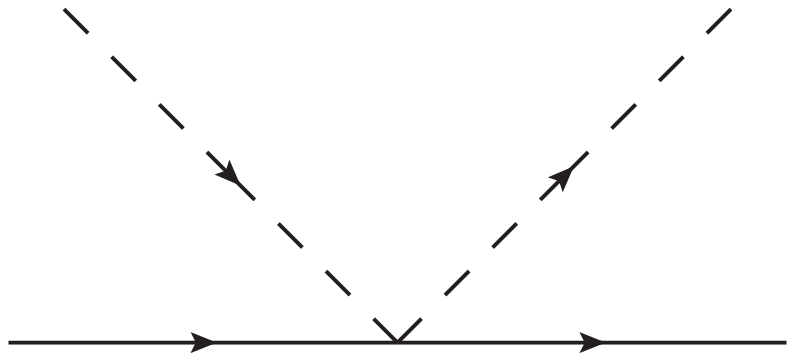}}&{\includegraphics[width=0.2\textwidth]{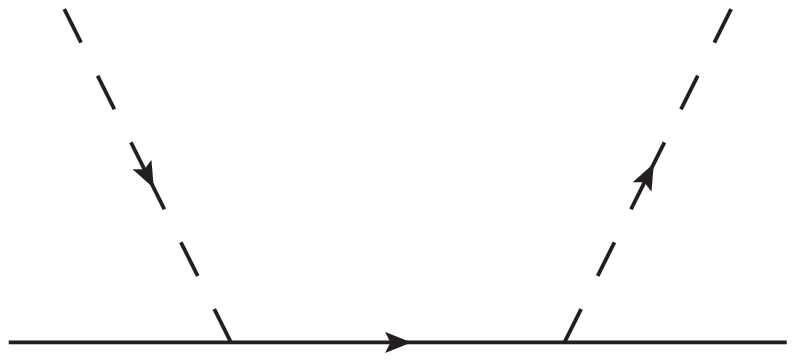}}& \\
& (a) & (b) &\\
$O(p^2)$:
&{\includegraphics[width=0.2\textwidth]{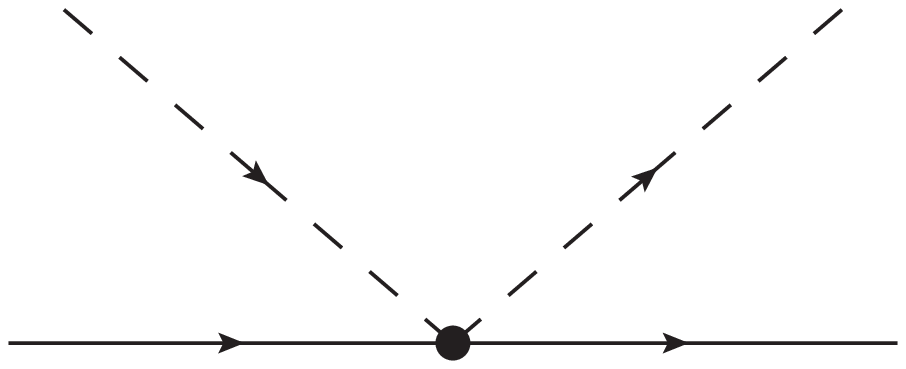}}& & \\
& (c)&  & \\
$O(p^3)$:
&{\includegraphics[width=0.2\textwidth]{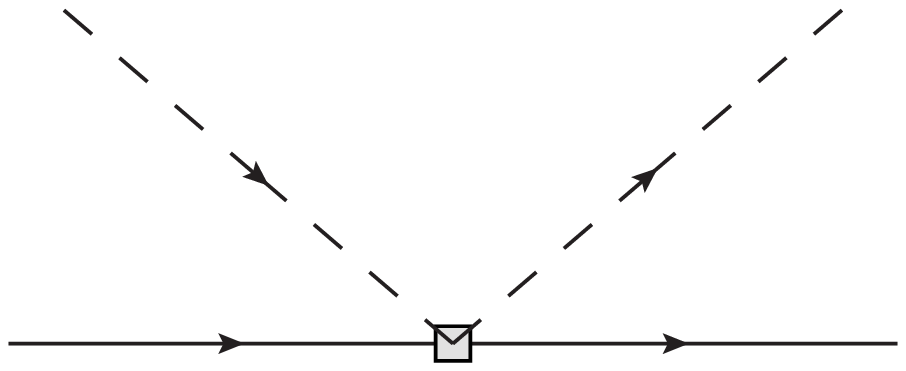}}&{\includegraphics[width=0.2\textwidth]{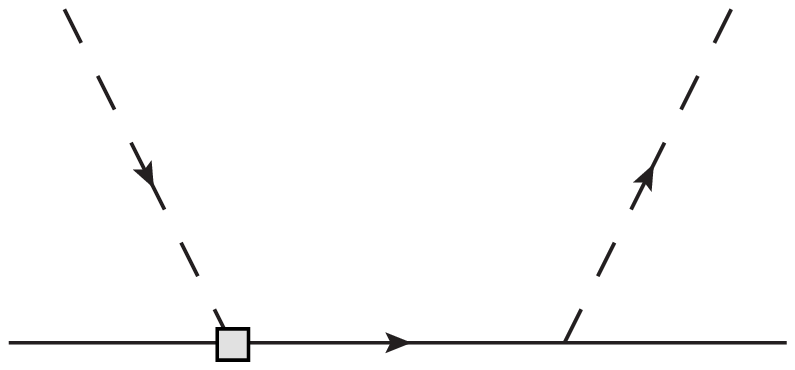}}&{\includegraphics[width=0.2\textwidth]{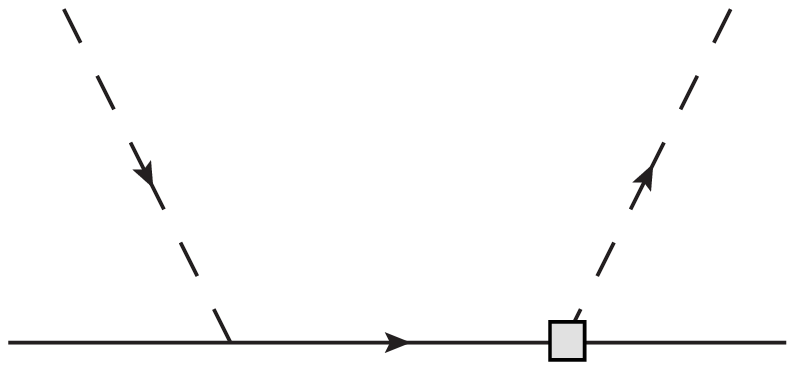}}\\
& (d)  & (e) & (f) \\
\end{tabular}
\caption{Tree level diagrams contributing to meson-baryon scattering up to $O(p^3)$. The solid lines correspond to baryons, and the dashed lines represent mesons. The vertices with filled circles and hollow blocks stem from the $\mathcal{L}^{(2)}_{MB}$ and $\mathcal{L}^{(3)}_{MB}$ Lagrangians, respectively.}
\label{treedia}
\end{figure}

The tree level contributions up to $O(p^3)$ are shown in Fig.~\ref{treedia}. In the present work, we focus on the $\pi N$ and $K N$ sectors. They can be organized into
the following four isospin multiplets: $\pi N^{I=3/2,1/2}$ and $K N^{I=1,0}$. The calculation of the contact terms  is rather straightforward and  the corresponding results are given in Appendix A.

\subsubsection{Tree level Born terms}
In general, the amplitude for a Born diagram could be written as
\begin{equation}\label{bornamplitude}
    \mathcal{A}=\frac{-\bar{u}_f\slashed q_f(\slashed {\mathcal{P}}-m_P)\slashed q_iu_i}{\mathcal{P}^2-m_P^2},
\end{equation}
where $u_i$, $\bar{u}_f$ refer to the spinors of the incoming or outgoing baryons, $q_i$ and  $q_f$ are the momentum of the incoming or outgoing mesons, $m_P$  is the mass of the baryon propagated,  and $\mathcal{P}$ is the total four momentum.

The Born terms at $O(p^3)$ can be categorized into two different groups. The first group contains the LECs $d_{38}\ldots d_{44}$. They share the same expression as that of Eq.~(\ref{bornamplitude}). The second group include the LECs $d_{45},d_{46},d_{47}$  and their form is slightly different:
\begin{equation}\label{bornamplitude2}
    \mathcal{A}=\frac{i\bar{u}_f(-\slashed {\mathcal{P}}+m_P)\slashed q_iu_i}{\mathcal{P}^2-m_P^2}.
\end{equation}

The explicit results for the Born diagrams are given in  Appendix B.

\subsubsection{Mass insertion diagrams}
Mass insertions are induced by the SU(3) breaking corrections to the chiral limit baryon mass $m_0$, which are of order $\mathcal{O}(p^2)$ and have the following explicit form:
\begin{equation}\label{massinsert}
    \begin{split}
      \Delta_N= & 4 m_K{}^2 (b_0+b_D-b_F)+2 m_{\pi }{}^2 (b_0+2 b_F),\\
      \Delta_{\Sigma}= & 2 m_{\pi }{}^2 (b_0+2 b_D)+4 b_0 m_K{}^2,\\
      \Delta_{\Lambda}=& \frac{2}{3} \left(m_K{}^2 (6 b_0+8 b_D)+m_{\pi }{}^2 (3 b_0-2 b_D)\right),\\
      \Delta_{\Xi}=&4 m_K{}^2 (b_0+b_D+b_F)+2 m_{\pi }{}^2 (b_0-2 b_F).
    \end{split}
\end{equation}

One easy way to include these corrections is to supplement the intermediate baryon mass of the Born terms with the $\mathcal{O}(p^2)$ corrections given in Eq.~(\ref{massinsert}).  The contribution from this part can be automatically included if one performs a substitution of $m_0\rightarrow m_2=m_0+\Delta_B$ in the mass renormalization of baryons. Thus we will not  explicitly show the contribution of this part.

\subsubsection{Leading one-loop diagrams}
\begin{figure}
  \centering
  \includegraphics[width=0.85\textwidth]{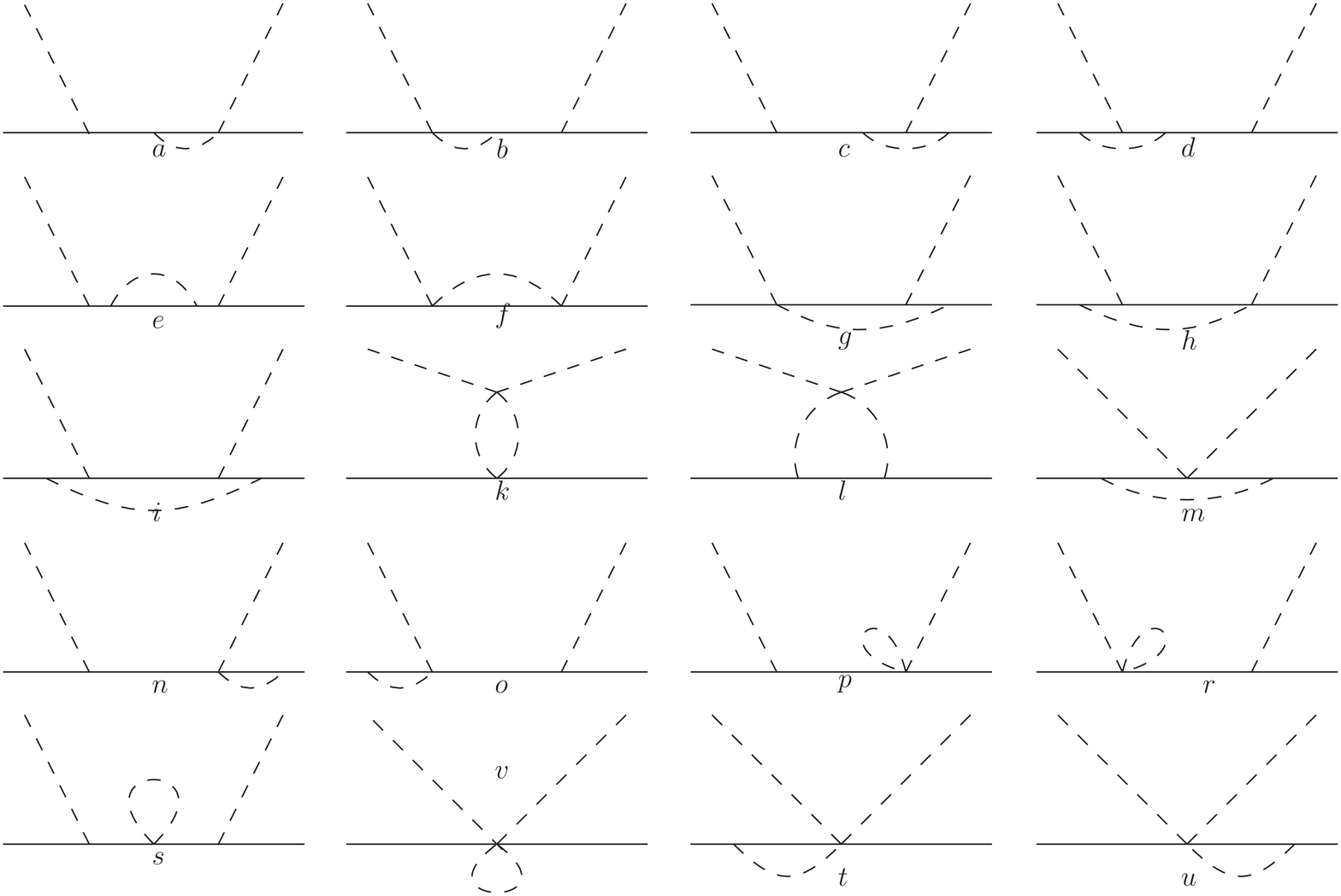}\\
  \caption{Leading one-loop contributions to meson-baryon scattering up to $O(p^3)$. Note that the wave function renormalization and crossed graphs are not shown explicitly.}\label{loopdia}
\end{figure}

The leading one-loop contributions to meson-baryon scattering include the Feynman diagrams shown in Fig.~\ref{loopdia}.

The crossed diagrams, if exist, can be obtained with the same replacement rule as in the case of the crossed Born diagrams:
\begin{equation}\label{crossloop}
    \begin{split}
      B_{Loop}=&B(s) -B(s\leftrightarrow u,M_i\leftrightarrow M_f), \\
      A_{Loop}=&A(s) +A(s\leftrightarrow u,M_i\leftrightarrow M_f),
    \end{split}
\end{equation}
where $M_{i,f}$ refer to the masses of incoming and outgoing mesons. In the numerical  evaluation of all these loop diagrams, we adopt physical values for all the quantities appearing in the amplitudes, including decay constants and masses. Employing their chiral limit values only lead to differences of  higher chiral order.

In numerical calculations, we utilize the package OneLoop~\cite{vanHameren:2009dr,vanHameren:2010cp}. Due to the complexity of the explicit expressions of the one-loop
contributions, they are not explicitly shown in this paper~\footnote{ They can be obtained from the authors upon request. }.

\subsubsection{Wave function renormalization}
\begin{figure}
\centering
\begin{tabular}{ccc}
{\includegraphics[width=0.2\textwidth]{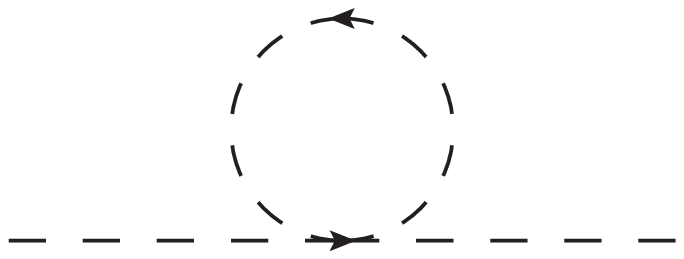}}&{\includegraphics[width=0.2\textwidth]{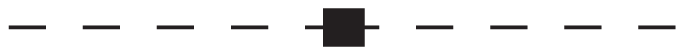}}&{\includegraphics[width=0.2\textwidth]{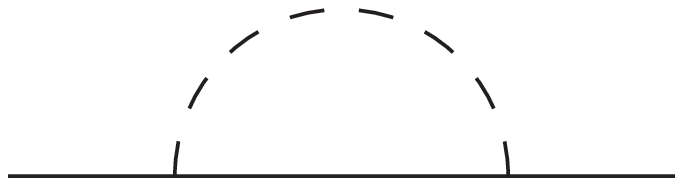}} \\
(a) & (b) & (c)
\end{tabular}
\caption{Wave function renormalization contributions to meson (dashed) and baryon (solid) fields. Counter terms from $\mathcal{L}^{(4)}_{MM}$ are donated by the filled block.}
\label{WFRdia}
\end{figure}

The wave function renormalization of the external mesons and baryons are shown in Fig.~\ref{WFRdia}. From Fig.~\ref{WFRdia}(a)(b),
one obtains the wave function renormalization constants for the Goldstone bosons up to NLO ~\cite{Scherer:2002tk}
\begin{equation}\label{WFRC}
  \begin{split}
    \mathcal{Z}_{\pi} =& 1-\frac{1}{F^2_0}\left[ 8L_4(2m_K^2+m_{\pi}^2)+8L_5m_{\pi}^2+\frac{1}{3}I(m_K^2)+\frac{2}{3}I(m_{\pi}^2) \right],\\
    \mathcal{Z}_{K}   =& 1-\frac{1}{F^2_0}\left[ 8L_4(2m_K^2+m_{\pi}^2)+8L_5m_{\pi}^2+\frac{1}{2}I(m_K^2)+\frac{1}{4}I(m_{\pi}^2)+\frac{1}{4}I(m_{\eta}^2) \right], \\
    \mathcal{Z}_{\eta}=& 1-\frac{1}{F^2_0}\left[ 8L_4(2m_K^2+m_{\pi}^2)+\frac{8}{3}L_5m_{\eta}^2+I(m_K^2) \right],
  \end{split}
\end{equation}
where $I(M^2)=-\frac{M^2}{16 \pi^2}\ln\frac{M^2}{\mu^2}$ is the one point function, and $L_4$ and $L_5$ are the NLO LECs of meson-meson interaction.
For the baryons, as depicted in Fig.~\ref{WFRdia}(c), the wave function renormalization constants up to $\mathcal{O}(p^3)$ are
\begin{equation}\label{WFRB}
  \mathcal{Z}_B=1-(f(m_B^2)+2m_B^2f'(m_B^2)-2m_Bm_Pg'(m_B^2)),
\end{equation}
where  $f(m_B^2)$ and $g(m_B^2)$ come from the baryon self-energy $-i\Sigma_{self}=-\slashed{\mathcal{P}}f(\mathcal{P}^2)+m_P g(\mathcal{P}^2)$ with $\mathcal{P}^2=s=m_B^2$ and can be written as
\begin{equation}\label{massrnf}
  \begin{split}
    f(s)=&-\frac{i}{32 \pi ^2 s} \left(\left(-m_P{}^2 \left(2 s+M_{\phi }{}^2\right)+s \left(s-M_{\phi }{}^2\right)+m_P{}^4\right) B_0\left(s,m_P{}^2,M_{\phi}{}^2\right) \right.\\
    &\left.+\left(m_P{}^2-s\right) A_0\left(M_{\phi }{}^2\right)-\left(s+m_P{}^2\right) A_0\left(m_P{}^2\right)\right),\\
    g(s)=&\frac{i \left(M_{\phi }{}^2 \left(-B_0\left(s,m_P{}^2,M_{\phi }{}^2\right)\right)-A_0\left(m_P{}^2\right)\right)}{16 \pi ^2},
  \end{split}
\end{equation}
where $m_P$ and $M_\phi$ refer to the masses of propagated baryons and mesons, $m_B$ is the mass of incoming or outgoing baryons, and $A_0,B_0$ are the one and two point scalar function in the Passarino-Veltman notation~\cite{Passarino:1978jh}.

The above obtained scattering amplitudes still need some further treatment before being employed to describe meson-baryon scattering. First, since in all the calculations above we used the physical values instead of the corresponding bare ones, the amplitudes must be properly renormalized. Second, the amplitudes involving loop diagrams are ultraviolet divergent. Third, the power counting rule specified previously is violated by the non-zero baryon mass in the chiral limit.  In the following section, we will explain how we renormalize the amplitudes. The $\overline{MS}-1$ dimensional regularization scheme and the EOMS scheme will be employed to treat the ultraviolet divergence and powering counting breaking problem.

\section{renormalization}
The main purpose of renormalization is to compensate the corrections caused by the differences between physical LECs, masses and decay constants and the corresponding
bare ones. In the present work, these corrections will promote the order of the original amplitudes by 2, leading to a contribution at the order of $\mathcal{O}(p^3)$. Thus we only need to study the $\mathcal{O}(p)$ tree level amplitudes.

\subsection{Mass renormalization}

The calculation of the diagram (e) in Fig.~\ref{loopdia} shows a double pole structure in the amplitudes. This unphysical structure can be removed, as it should be, after the masses of the propagating baryons in the Born terms are correctly renormalized. Following the same power counting rule as specified above, the physical baryon masses can be expressed up to the NNLO as
\begin{equation}\label{massrn}
    m_{phys}=m_{0}+\Delta_{N,\Lambda,\Sigma,\Xi}+\Sigma_{\mathcal{O}(p^3)},
\end{equation}
where $m_0$ refers to the chiral limit baryon mass, $\Delta_{N,\Lambda,\Sigma,\Xi}$ and $\Sigma_{\mathcal{O}(p^3)}$, given by Eq.~(\ref{massinsert}) and Eq.~(\ref{massrnf}) respectively, denote the NLO and NNLO contributions. As mentioned before, a replacement of $m_0\rightarrow m_2=m_0+\Delta_{N,\Lambda,\Sigma,\Xi}$ automatically include the contributions from mass insertions.

Specifying the intermediate baryons and mesons in Eq.~(\ref{massrnf}), the $\mathcal{O}(p^3)$ self energy reads
\begin{equation}\label{BaryonMass}
\begin{split}
  \Sigma_{\mathcal{O}(p^3)} & \equiv \Sigma(B,\Phi,P) \\
    & = -i\slashed{\mathcal{P}}f(\mathcal{P}^2)+i m_P g(\mathcal{P}^2)\mid_{\slashed{\mathcal{P}}=m_B,\mathcal{P}^2=m_B^2}  \\
    &-\frac{\tilde{m} \left(2 \tilde{m}^2-2 m_P \tilde{m}+M_{\phi }^2+\log \left(\frac{\mu ^2}{\tilde{m}^2}\right) \left(-2 \tilde{m}^2+3 m_P \tilde{m}+M_{\phi }^2\right)\right)}{8 \pi ^2},
\end{split}
\end{equation}
where $B$ denote the incoming or outgoing baryon, and $\Phi$ and $P$ represent the intermediate meson and baryon. The last term of the above equation is actually the power counting breaking term, which will be absorbed into $m_2$.

Below, we list the expressions for the $\mathcal{O}(p^3)$ baryon masses:
\begin{equation}\label{BayonMass2}
    \begin{split}
      \Sigma_{N}= & \frac{1}{12 f^2}\left(\Sigma (N,\eta,N) (D-3 F)^2+\Sigma (N,K,\Lambda) (D+3 F)^2 \right. \\
                  &\left. +9 \left(\Sigma (N,K,\Sigma) (D-F)^2+\Sigma (N,\pi,N) (D+F)^2\right)\right), \\
      \Sigma_{\Sigma}=  & \frac{1}{6 f^2}\left(3 \Sigma (\Sigma,K,N) (D-F)^2+3 \Sigma (\Sigma,K,\Xi) (D+F)^2\right. \\
                  &\left.+2 \left((\Sigma (\Sigma,\eta,\Sigma)+\Sigma (\Sigma,\pi,\Lambda)) D^2+6 \Sigma (\Sigma,\pi,\Sigma) F^2\right)\right) ,\\
      \Sigma_{\Lambda}= & \frac{1}{6 f^2}\left(2 (\Sigma (\Lambda,\eta,\Lambda)+3 \Sigma (\Lambda,\pi,\Sigma)) D^2+\Sigma (\Lambda,K,\Xi) (D-3 F)^2  \right.  \\
                  &\left.+\Sigma (\Lambda,K,N) (D+3 F)^2 \right) ,\\
      \Sigma_{\Xi}= & \frac{1}{12 f^2}\left(\Sigma (\Xi,K,\Lambda) (D-3 F)^2+\Sigma (\Xi,\eta,\Xi) (D+3 F)^2 \right.\\
                  &\left.+9 \left(\Sigma (\Xi,\pi,\Xi) (D-F)^2+\Sigma (\Xi,K,\Sigma) (D+F)^2\right) \right).
    \end{split}
\end{equation}

One can of course simply replace $m_2$ in Eq.~(\ref{bornamplitude3}) with $m_{phys}-\Sigma_{\mathcal{O}(p^3)}$ to complete the mass renormalization, ignoring the resulting higher order differences. However, the treatment here need to be much more careful. The series of studies on baryon masses show that with the EOMS scheme, one can achieve a pretty good and well-converged description at  the complete one-loop level, which is  N$^3$LO. But when limited to NNLO, the convergence is not as good as expected~\cite{Ren:2013oaa}. Contributions from NNLO and N$^3$LO will largely cancel each other. From another point of view, the $\chi^2/d.o.f.$ from a NNLO fit are much larger than a N$^3$LO fit, which implies a relatively unsatisfying description. As a consequence, although a direct replacement is not WRONG, it is not appropriate since the higher order contribution, N$^3$LO here, to the baryon masses may worsen the description of scattering process at the order of our interest. Thus we expand the amplitudes of the Born terms after the substitution at $s=m_2^2$, in order to cancel the double pole structure strictly and avoid worsening of the convergence.

\subsection{Vertice renormalization}

\begin{figure}
\centering
\begin{tabular}{ccc}
{\includegraphics[width=0.2\textwidth]{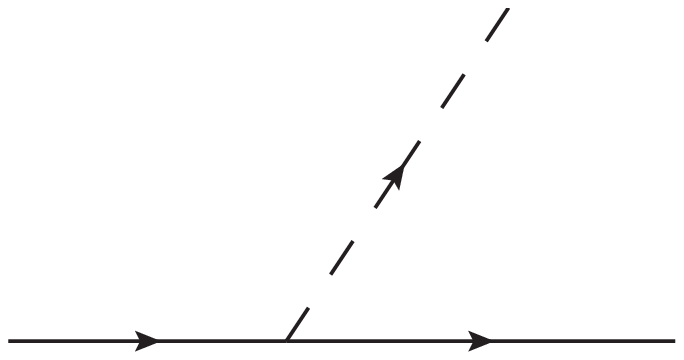}}&{\includegraphics[width=0.2\textwidth]{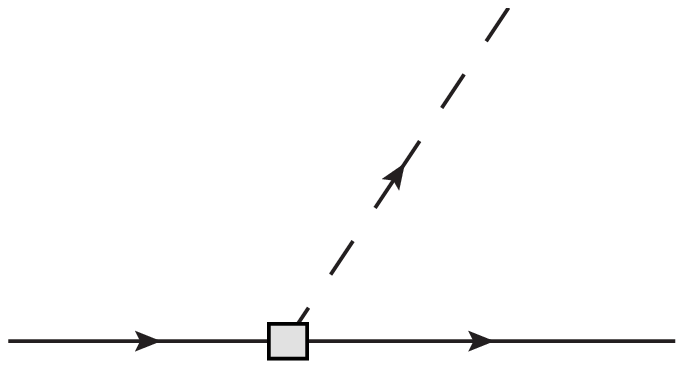}}&{\includegraphics[width=0.2\textwidth]{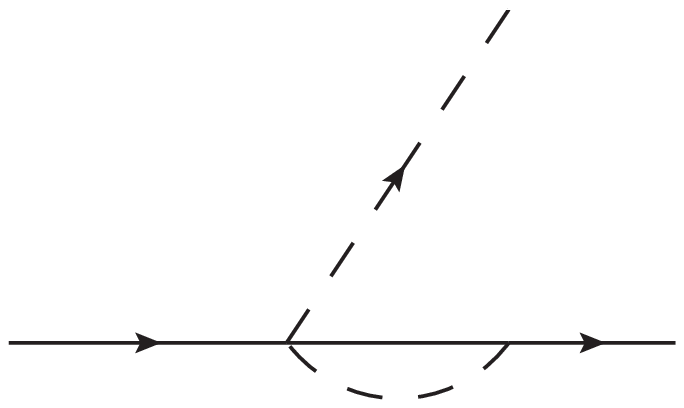}} \\
(a) & (b) & (c) \\
{\includegraphics[width=0.2\textwidth]{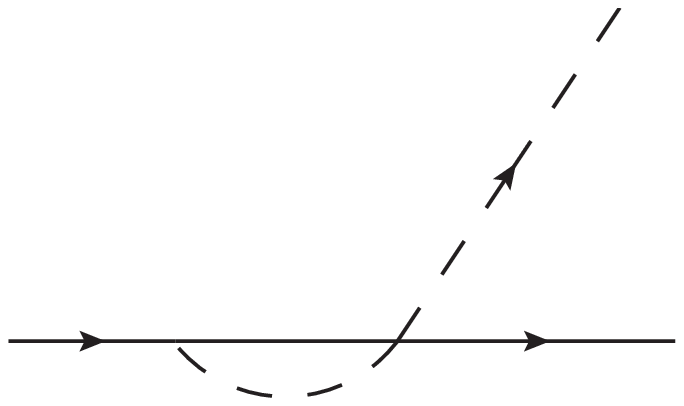}}&{\includegraphics[width=0.2\textwidth]{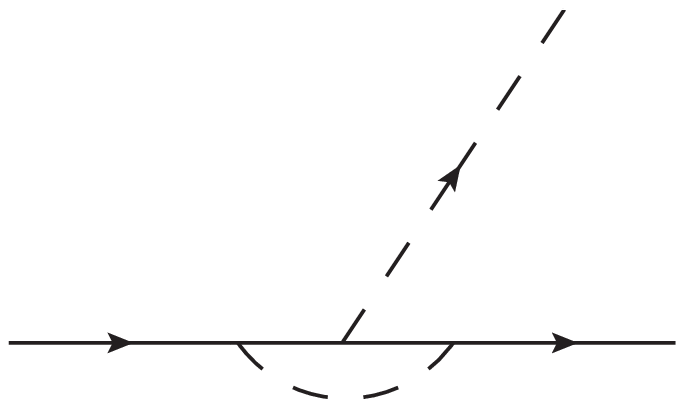}}&{\includegraphics[width=0.2\textwidth]{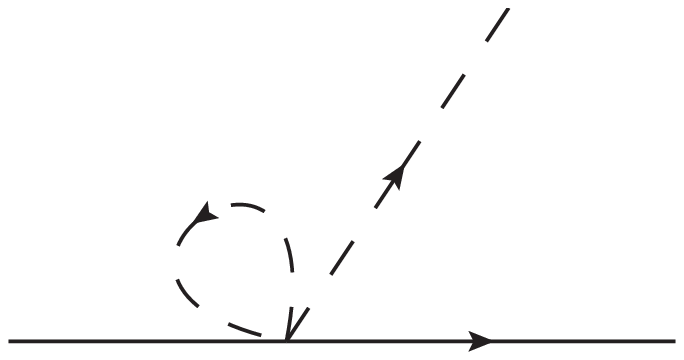}} \\
(d) & (e) & (f) \\
{\includegraphics[width=0.2\textwidth]{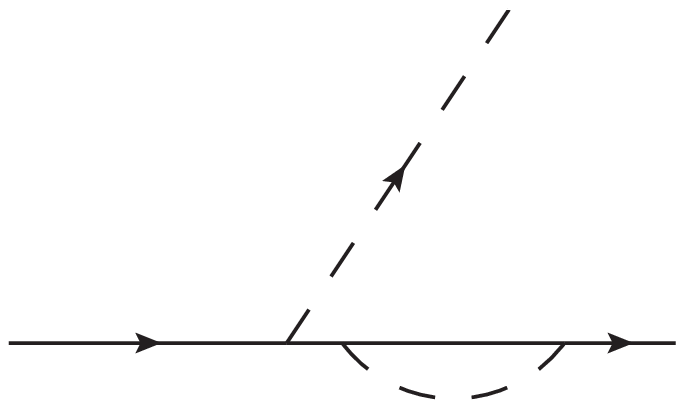}}&{\includegraphics[width=0.2\textwidth]{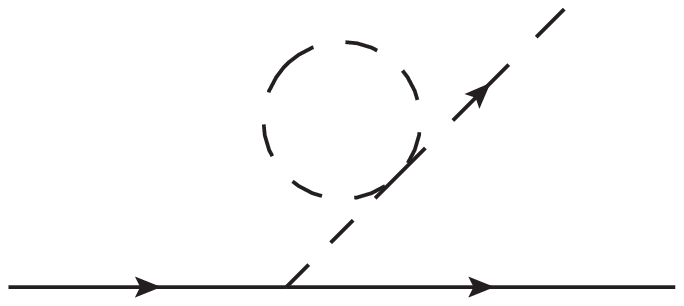}}& \\
(g) & (h) &  \\
\end{tabular}
\caption{Feynman diagrams contributing to vertex renormalizations. The hollow block represents the contributions of the $O(p^3)$ vertices.}
\label{VRdia}
\end{figure}

Several methods are available to renormalize the vertices. Of course, all of the renormalization methods are eventually equivalent. For instance, the authors in Ref.~\cite{Chen:2012nx} chose to renormalize the coupling constants with the axial-vector current. In this work, we choose to achieve the vertex corrections at the one loop level via the two-body decay process, as in Ref.~\cite{Yao:2016vbz}. The two-body decays of a baryon up to $\mathcal{O}(p^3)$ are depicted in Fig.~\ref{VRdia}.

The renormalization of the vertices can be schematically expressed as the following:
\begin{equation}\label{verticern}
    C_{ph}\gamma^5\slashed{q}_f=C_{bare}\gamma^5\slashed{q}_f+C_{ph}\gamma^5\slashed{q}_fZ+C_{ph}\gamma^5\mathcal{A}_{loop}(s)+C_{ph}\gamma^5\slashed{q}_f\Delta_{F}.
\end{equation}
The five terms on the right hand of the above equation come from tree diagrams (a,b), wave function renormalization (g,h), one loop diagrams (c,d,e,f), and the renormalization of the decay constant, respectively, where $Z$ refers to the wave function renormalization constants, $\Delta_F$ is the decay constant at $\mathcal{O}(p^4)$. The last term is indispensable since we only want to renormalize the coupling constant here while the decay process will definitely involve decay constants.
This leads to
\begin{equation}\label{verticern}
\begin{split}
    C_{bare}&=C_{ph}-C_{ph}Z-C_{ph}\Delta_{F}-C_{ph}\frac{\mathcal{A}_{loop}}{m_i+m_f} \\
            &\equiv C_{ph}-C_{re}.
\end{split}
\end{equation}
The explicit form of $\mathcal{A}_{loop}$ can be found in Appendix C.

Substituting Eq.~(\ref{verticern}) into the Born terms leads to
\begin{equation}\label{verticerenormborn}
   \begin{split}
    C^1_{bare}C^2_{bare}\mathcal{A}_{Born}&=(C^1_{ph}-C^1_{re})(C^2_{ph}-C^2_{re})\mathcal{A}_{Born}\\
      &\simeq(C^1_{ph}C^2_{ph}-C^2_{ph}C^1_{re}-C^1_{ph}C^2_{re})\mathcal{A}_{Born},
   \end{split}
\end{equation}
where the last two terms in Eq.(\ref{verticerenormborn}) are the correction parts we need and the $C^1_{re}C^2_{re}\mathcal{A}_{Born}$ has already been omitted since its order is higher.

\subsection{Chiral corrections to the decay constants}
To obtain the full meson-baryon amplitudes, one should also take into account the chiral corrections to the decay constants. In practice, one should use the bare decay constant $F_0$ instead of the corresponding physical ones $F_{\pi}$, $F_K$, and $F_{\eta}$~\cite{Gasser:1984gg}.

Since the chiral  corrections increase the chiral order by $\mathcal{O}(p^2)$, we only need to replace
$F_0$ with $F_\pi$, $F_K$, and $F_\eta$ in the tree level diagrams of order $\mathcal{O}(p^1)$.
\begin{equation}\label{decayconstantre}
    \frac{\mathcal{A}_{\mathcal{O}(p^1)}}{F_0^2}=\frac{\mathcal{A}_{\mathcal{O}(p^1)}}{F_{\pi}^2, F_{K}^2, F_{\eta}^2}(1-2\Delta F_{\pi,K,\eta}),
\end{equation}
where
\begin{equation}\label{decayconstantre2}
    \begin{split}
      \Delta F_{\pi}= & \frac{2 \left(\text{A}_0\left(m_{\pi }{}^2\right)+64 \pi ^2 \left(2 L_4 m_K{}^2+L_4 m_{\pi }{}^2+L_5 m_{\pi}{}^2\right)\right)+\text{A}_0\left(m_K{}^2\right)}{32 \pi ^2 F_0{}^2} ,\\
      \Delta F_{K}=& \frac{3 \text{A}_0\left(m_{\eta }{}^2\right)+6 \text{A}_0\left(m_K{}^2\right)+3 \text{A}_0\left(m_{\pi }{}^2\right)+1024 \pi ^2 L_4 m_K{}^2+512 \pi ^2 L_4 m_{\pi }{}^2+512 \pi ^2 L_5 m_K{}^2}{128 \pi ^2 F_0{}^2},\\
      \Delta F_{\eta}&=\frac{3 \text{A}_0\left(m_K{}^2\right)+128 \pi ^2 \left(L_4 \left(2 m_K{}^2+m_{\pi }{}^2\right)+L_5 m_{\eta }{}^2\right)}{32 \pi ^2 F_0{}^2}.
    \end{split}
\end{equation}

\section{Ultraviolet divergence and power counting breaking terms}
As mentioned above, the one-loop integrals calculated above are ultraviolet divergent. Applying  the $\overline{MS}-1$ dimensional regularization scheme, the ultraviolet divergent part can be absorbed into the LECs at
the corresponding order.

One can separate the LECs in the counter terms into their finite parts and infinite parts as,
\begin{equation}\label{UVab}
  L_i=L_i^r+L_i^dR,
\end{equation}
where $R=\frac{2}{d-4}+\gamma_E-1-\ln(4\pi)$ with $\gamma_E$ being the Euler constant and $d$ the space-time dimension. The $L_i^d$s are listed in  Appendix D.
All other $L_i^d$s not listed are equal to zero. Note that here we approximate all the baryon masses with $\tilde{m}$ just to simplify the expressions.

Since the chiral limit baryon mass does not vanish, the naive  powering counting rule is broken when the $\overline{MS}-1$ scheme is adopted~\cite{Gasser:1987rb}. As we have mentioned above, in the covariant amplitudes, the $A$  and $B$ parts may cancel each other. It is better to use the modified $D$  and $B$ functions when one removes the PCB parts. Here we apply the EOMS scheme~\cite{Fuchs:2003qc} to restore the power counting.

The power counting breaking terms are tightly related to the small quantities in the chiral expansion, which have been listed at the very beginning in Eq.~\ref{SmallQuantity}. In the present work,
since we are working in the $SU(3)$ case, the situation is a bit more complicated because of the mass differences among the octet baryons, which we count as $\mathcal{O}(p^2)$.
To make sure the factor in front of the $B$ part counts as a pure $\mathcal{O}(p^2)$, we rewrite the scattering amplitude in the following way:
\begin{equation}\label{PCB1}
    \begin{split}
      T_{MB}&=\overline{u}(p',s')\left[D+\frac{i}{m_i+m_f}\sigma^{\mu\nu}q'_{\mu}q_{\nu}B\right]u(p,s)\\
        & =\overline{u}(p',s')\left[D+\frac{i}{2\tilde{m}}\sigma^{\mu\nu}q'_{\mu}q_{\nu}\frac{2\tilde{m}}{m_i+m_f}B\right]u(p,s) \\
        & =\overline{u}(p',s')\left[D+\frac{i}{2\tilde{m}}\sigma^{\mu\nu}q'_{\mu}q_{\nu}\tilde{B}\right]u(p,s),
    \end{split}
\end{equation}
where $D=A+\frac{s-u}{2(m_i+m_f)}B$.  Now we can see that we only need to pick up the PCB terms up to $O(p^2)$ for the $D$ functions and those up to $O(p^0)$ for the $\tilde{B}$ functions since the term $\sigma^{\mu\nu}q'_{\mu}q_{\nu}$ is of $O(p^2)$.

Now we are ready to remove all the PCB terms in the $D$ and $B$ functions. The terms that break the power counting  have been shown to originate from the regular part of the loop integrals by Becher and Leutwyler~\cite{Becher:1999he}. This provides a simple way to subtract the PCB terms by working out all the regular parts first. Or alternatively, one can  perform the chiral expansions  of small quantities directly.

Once we get rid of all the PCB terms, we obtain
\begin{equation}\label{PCB2}
    \begin{split}
      T^{'}&=\overline{u}(p',s')\left[D^{'}+\frac{i}{2\tilde{m}}\sigma^{\mu\nu}q'_{\mu}q_{\nu}\tilde{B}^{'}\right]u(p,s),\\
        & =\overline{u}(p',s')\left[D^{'}+\frac{i}{m_i+m_f}\sigma^{\mu\nu}q'_{\mu}q_{\nu}\frac{m_i+m_f}{2\tilde{m}}\tilde{B}^{'}\right]u(p,s).
    \end{split}
\end{equation}
The final $A$ and $B$ functions, where we use $A_f$ and $B_f$ to distinguish them from $A$ and $B$, are then
\begin{equation}\label{PCB3}
   \begin{split}
    B_f&=\frac{m_i+m_f}{2\tilde{m}}\tilde{B}^{'} ,\\
    A_f&=D^{'}-\frac{s-u}{4\tilde{m}}\tilde{B}^{'}.
   \end{split}
\end{equation}

As shown in Ref.~\cite{Fuchs:2003qc}, the PCB terms are all analytical and can be absorbed into the LECs at the corresponding orders.
Assuming that $ \mathrm{LEC}=\mathrm{LEC}_b+\mathrm{LEC}^{PCB}$, we have worked out all the power counting breaking terms, which are explicitly shown
in  Appendix E.

\section{Results and discussion}
The scattering of a pseudoscalar meson off an octet baryon can be
grouped into 11 combinations of isospin and strangeness as tabulated in Table.~\ref{IS}. In the present work we focus on
the $\pi N$ and $K N$ channels, because only for these channels partial wave phase shifts are available.

\begin{table}\label{IS}
\centering
\caption{11 coupled channels of meson-baryon scattering of conserved strangeness ($S$) and isospin ($I$).}
\begin{tabular}{cccccccccccc}
\hline\hline
$(1,1)$&$(1,0)$&$(0,\frac{3}{2})$&(0,$\frac{1}{2})$& $(-1,2)$& $(-1,1)$& $(-1,0)$&($-2,\frac{3}{2}$)& $(-2,\frac{1}{2}$)&$(-3,1)$&$(-3,0)$\\
\hline
$KN$& $KN$&  $K\Sigma$& $K\Sigma$ & $\pi \Sigma$& $\pi \Sigma$ &$\pi \Sigma$  &$\bar{K} \Sigma$&$\bar{K} \Sigma$ &$\bar{K} \Xi$&$\bar{K} \Xi$\\
    &     &  $\pi N$  & $K\Lambda$&             & $\eta \Sigma$&$\eta \Lambda$&$\pi \Xi$       &$\bar{K} \Lambda$&  &  & \\
    &     &           & $\eta N$  &             & $\pi \Lambda$&$\bar{K} N$   &                &$\eta \Xi$       &  &  & \\
    &     &           & $\pi N$   &             & $\bar{K} N$  &$K \Xi$       &                &$\pi \Xi$        &  &  & \\
    &     &           &           &             & $K \Xi$      &              &                &                 &  &  & \\
\hline\hline
\end{tabular}
\end{table}

With the amplitudes properly renormalized, we are now ready to determine the LECs by fitting to the partial wave phase shifts. For $\pi N$, we chose the phase shifts from the analysis of WI08~\cite{Arndt:2006bf} in the $S_{11},S_{31},P_{11},P_{31},P_{13},P_{33}$ partial waves, where in the convention $L_{2I,2J}$   $L$ denotes the total orbit angular momentum, $I$ the total isospin, and $J$ the total angular momentum. Correspondingly, the phase-shift analysis of the SP92 solution~\cite{Hyslop:1992cs} in the $S_{01},P_{01},P_{03},S_{11},P_{11},P_{13}$ partial waves are used for $K N$ where the symbols means $L_{I,2J}$.

For the $\pi N$ channels, we chose the phase shifts with $\sqrt{s}$ between 1082 MeV~\footnote{It was noted in Ref.~\cite{Hoferichter:2009gn} that the renormalized scattering amplitudes in the IR scheme actually diverge at threshold because of the term $t=(p-p')^2$ appearing in the dominators. The same happens  in the EOMS scheme~\cite{Chen:2012nx}. As a result, in the present work the fitting range is
chosen to start from several MeV above the respective thresholds.}, which is slightly above the threshold, and 1130MeV, with an interval of 4 MeV. Thus totally we will have 13 points for each of the 6 partial waves.  For the $KN$ channels we follow the same strategy. Starting from 1435MeV to 1475MeV,  the interval is set to be 2 MeV, with totally 20 points for each partial wave.

Since WI08 does not provide the errors for the data, we follow Refs.~\cite{Meissner:1999vr,Alarcon:2011kh} and take
\begin{equation}\label{dataerr}
    \textit{err}(\delta)=\sqrt{e^2_s+e^2_r\delta^2},
\end{equation}
with the systematic error $e_s=0.1^{\circ}$ and the relative error $e_r=2\% $.

Throughout the numerical study, we use the physical decay constants for the corresponding vertices. The renormalization scale $\mu$ in the loop integrals is chosen to be the average mass of the baryon octet, and the $\tilde{m}$, appearing in the power counting breaking terms via $s-\tilde{m}^2$, is taken to equal to the mass of the nucleon, considering that we focus now on the $\pi N$ and $KN$ channels. The adopted values for $\mu$ and $\tilde{m}$  are somewhat arbitrary. One can of course perform the calculation with the scale $\mu=\tilde{m}=m_N$, similar to the SU(2) case, or $\mu=\tilde{m}=m_0$ where $m_0$ is the baryon mass in the chiral limit. However, the significant difference between $m_N$ and $m_{\Xi}$ reminds us that such a treatment may lead to unusually large PCB terms. The physical values employed in the present work are collected in Table.~\ref{PARAMETER}.

\begin{table*} \label{PARAMETER}
\centering
\caption{Masses and coupling constants (in units of GeV) relevant in the present work. Note the mass of the $K$ meson is taken to be 0.493 GeV to be consistent with the SP92 data, which were originally from $K^+ n$ scattering. }
\begin{tabular}{ccccccccccc}
\hline\hline
$M_{\pi}$ & $M_{K}$ & $M_{\eta}$ & $m_{N}$ & $m_{\Lambda}$ & $m_{\Sigma}$ & $m_{\Xi}$ \\
\hline
0.139 & 0.493 & 0.54765 & 0.939 & 1.1157 & 1.1934 & 1.3183 \\
\hline\hline
$F_{\pi}$ & $F_{K}$ & $F_{\eta}$ & $D$ & $F$ & $\mu$ & $\tilde{m}$\\
\hline
0.0924  &  0.11003 & 0.11088 & 0.8 & 0.467 & 1.16 & $m_N$=0.939 \\
\hline\hline
\end{tabular}
\end{table*}

\subsection{Fitting strategy one:  direct fit to the phase shifts}
We found that to describe the pion-nucleon scattering data, one needs to go to at least $\mathcal{O}(p^3)$. On the other hand, a reasonable reproduction of the kaon-nucleon data
can already be achieved at $\mathcal{O}(p^2)$. We follow the same strategy in the first attempt to provide a simultaneous fit of both the $\pi N$ and $K N$ data.\footnote{As a matter of fact,
different LECs contribute to $\pi N$ and $K N$ scattering independent of each other.}

\begin{figure}
\centering
\begin{tabular}{ccc}
{\includegraphics[width=0.32\textwidth]{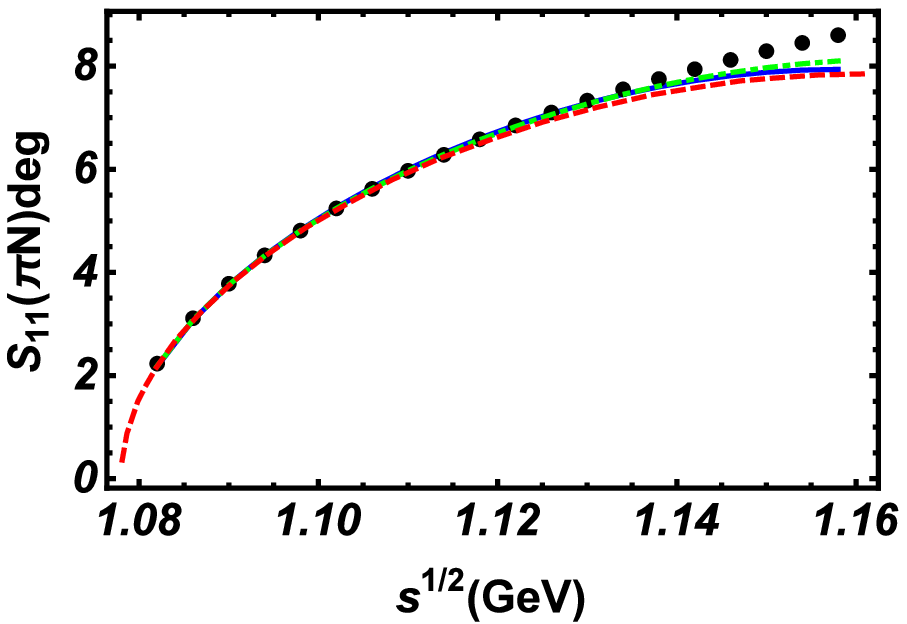}}&{\includegraphics[width=0.32\textwidth]{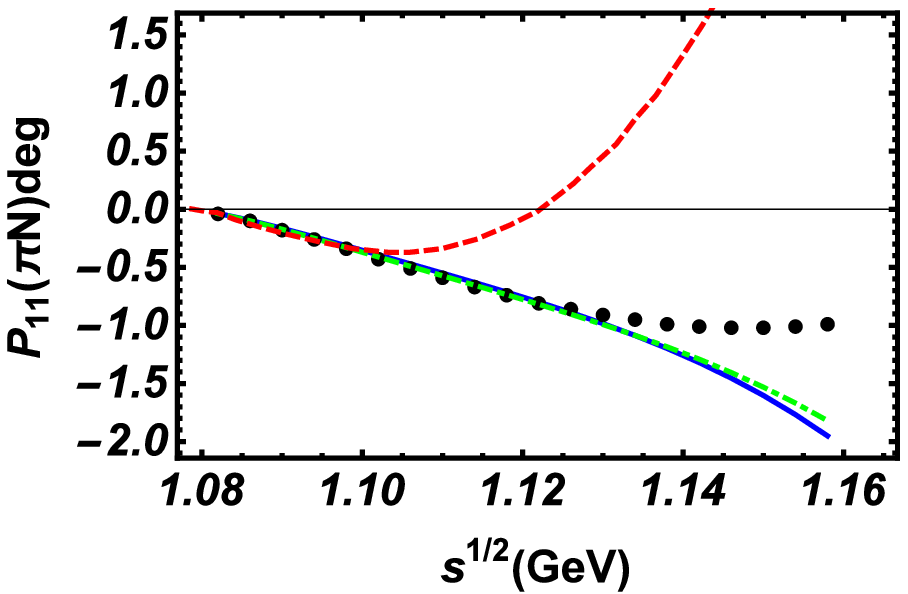}}& {\includegraphics[width=0.32\textwidth]{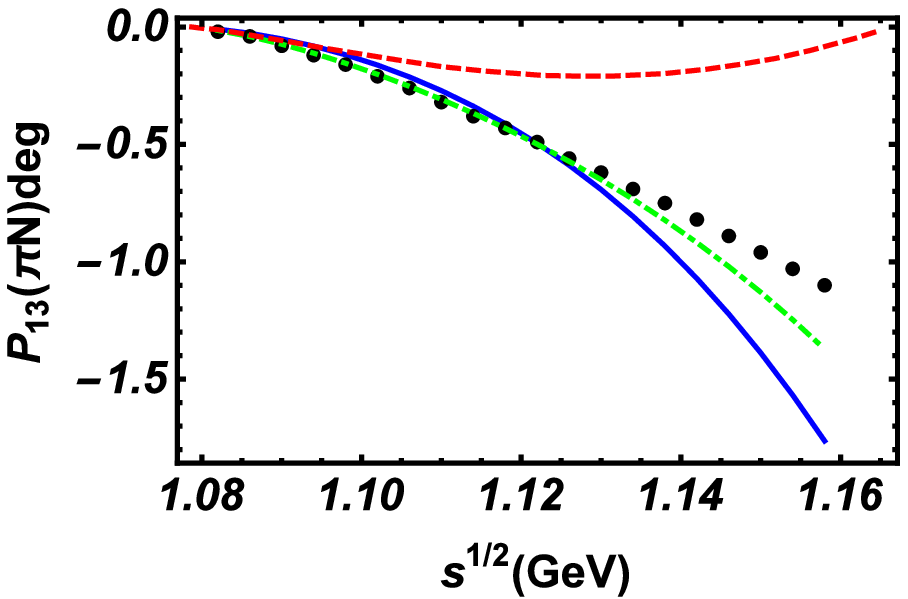}} \\
{\includegraphics[width=0.32\textwidth]{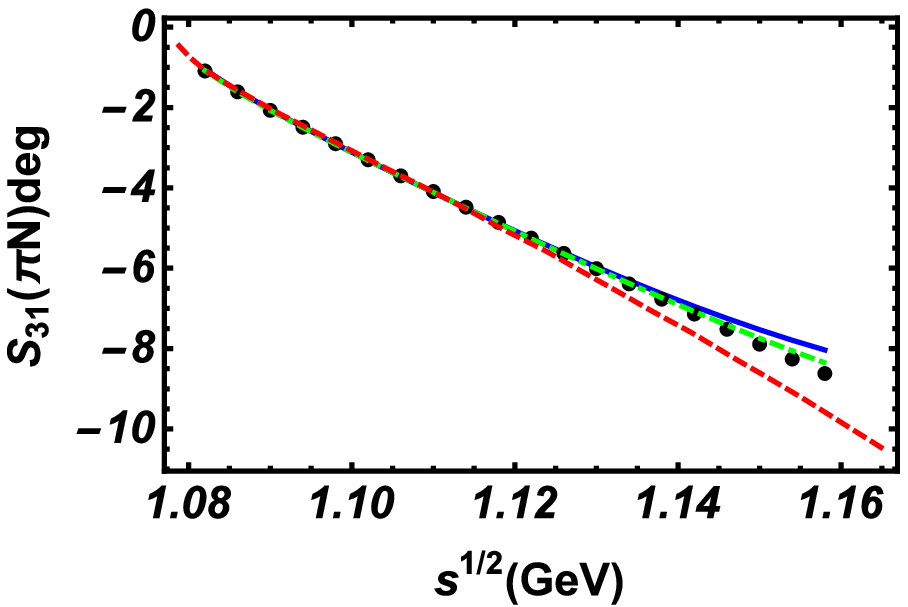}}&{\includegraphics[width=0.32\textwidth]{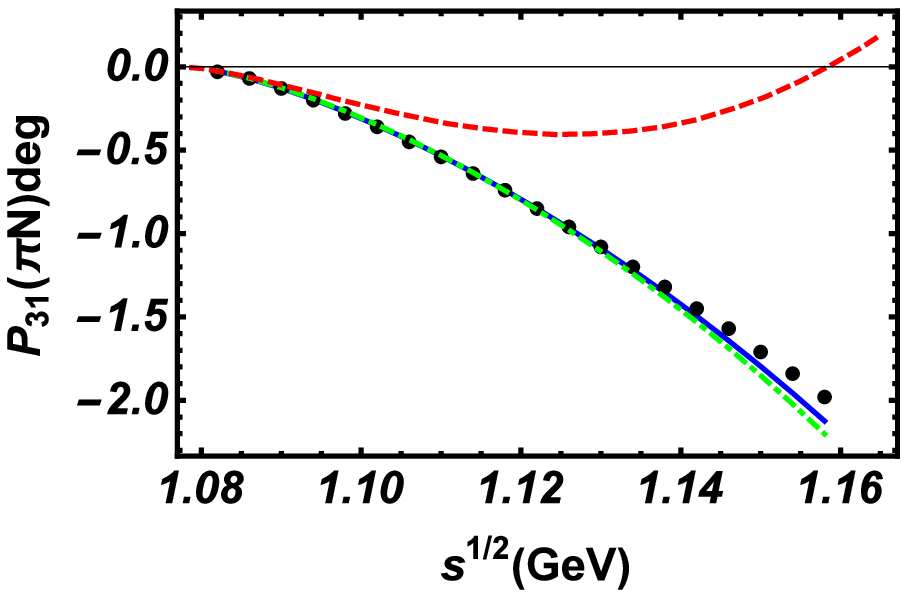}}& {\includegraphics[width=0.32\textwidth]{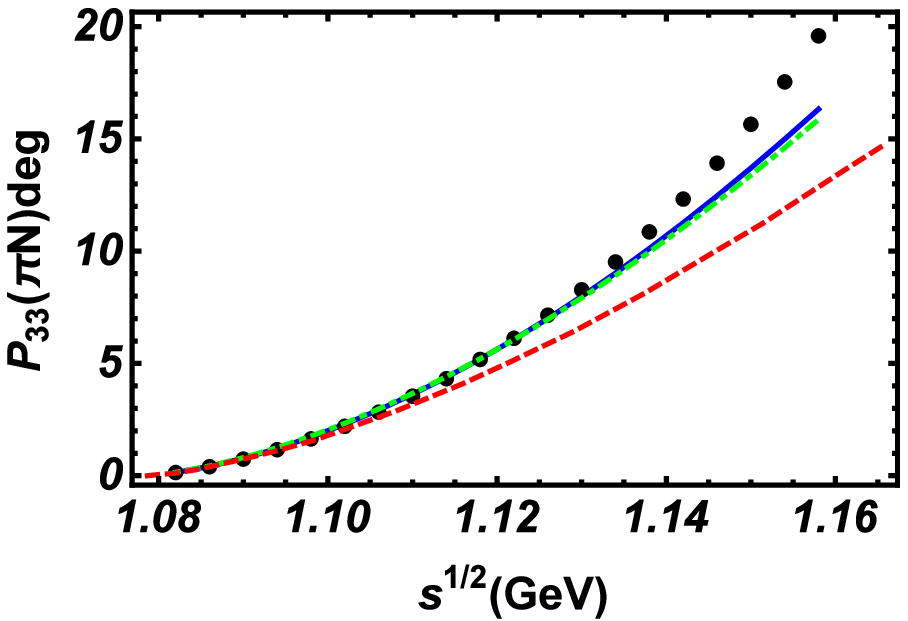}} \\
\end{tabular}
\caption{Pion-nucleon phase shifts. The blue lines denote our results and the black dots represent the WI08 solutions. For the sake of comparison, we show as well
  the EOMS SU(2) results~\cite{Chen:2012nx} (green dot-dashed lines) and the HB SU(3) results~\cite{Huang:2017bmx} (red dashed lines).}\label{fig:pin}
\end{figure}

A least-of-squares fit yielded a $\chi^2/d.o.f.=0.154$ for the 78 data points in the pion-nucleon channel.  The corresponding fit results are compared with the empirical data in Fig.~\ref{fig:pin}.
For the sake of comparison, we show as well the $\mathcal{O}(p^3)$ results of the SU(3) HB ~\cite{Huang:2015ghe,Huang:2017bmx} and the SU(2) EOMS BChPT~\cite{Chen:2012nx}.

Clearly,  the EOMS results can describe the
phase shifts quite well.
Although the data are only fitted up to $\sqrt{s}=1.13$ GeV, the phase shifts are described very well even up to $\sqrt{s}=1.16$ GeV, corresponding to a momentum in the laboratory frame of $|\vec{p}_{lab}|=200$ MeV. In addition, our calculation in SU(3) shows a compatible description compared to that in SU(2), which implies that the inclusion of strangeness has small effects on the fitting results.

We note that in the $P_{11}$ channel, the solution of WI08 tends to increase with energy in the higher energy region while the EOMS results, both in the SU(3) and SU(2) cases, decrease. This disagreement has already been noted in Ref.~\cite{Chen:2012nx}, where the authors point out that including the contribution of the $\Delta(1232)$ may improve the description. Inspired by this, we have checked that in SU(3) the inclusion of the lowest order contribution from the decuplet can have the same positive effect, which is shown in Appendix F. One can achieve a pretty good description even up to $\sqrt{s}=1.2$ GeV, quite close to the region of the $\Delta$ resonance. For a description covering this $\Delta$ resonance region, one needs to include the $\Delta$ explicitly, unitarize the amplitudes, and modify the powering counting rule. For the discussion of these, we refer the reader to Ref.~\cite{Yao:2016vbz}. On the other hand, although the HBChPT can describe the $s$-wave phase shifts,  it fails to describe the $p$-wave phase shifts.

\begin{table}[htpb]
\centering
\caption{LECs in the $\pi N$ channel.}\label{fitpiN}
\begin{tabular}{ccccccccc}
\hline\hline
$\alpha_1$  &  $\alpha_2$  &  $\alpha_3$  &  $\alpha_4$  &  $\alpha_5$  & $\alpha_6$ & $\alpha_7$ & $\alpha_8$  &  $\chi^2/d.o.f.$\\
\hline
$-7.64(6)$  &  $1.42(2)$  &  $1.34(1)$  &  $-1.36(6)$ & $0.61(2)$ & $3.25(6)$ & $1.45(3)$ & $-0.32(12)$  & 0.154\\
\hline
\end{tabular}
\end{table}

For the $KN$ scattering, as noted in the HB study~\cite{Huang:2015ghe,Huang:2017bmx}, a quite good description of the phase shifts can already be achieved at NLO.
 In the present work, we will present two studies of the $KN$ scattering. One is performed up to $\mathcal{O}(p^2)$ and
 the other is performed up to $\mathcal{O}(p^3)$ but only the loop contributions are included, because the phase shifts data are not enough to fix the relevant
 $\mathcal{O}(p^3)$ LECs.  Other inputs in addition to the $KN$ phase shifts are needed. The second study will be denoted by $\mathcal{O}(p^3)^*$.

\begin{figure}
\centering
\begin{tabular}{ccc}
{\includegraphics[width=0.32\textwidth]{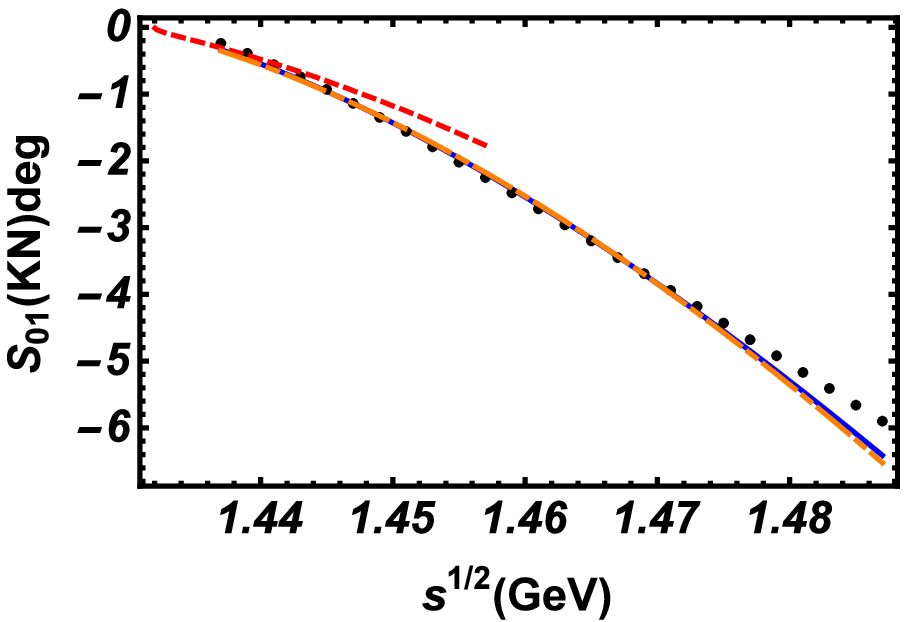}}&{\includegraphics[width=0.32\textwidth]{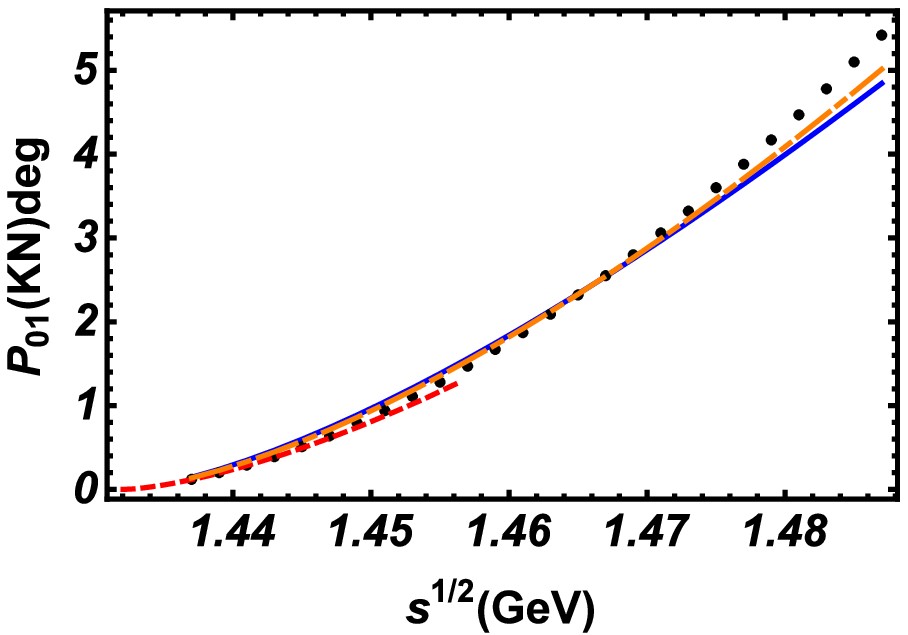}}& {\includegraphics[width=0.32\textwidth]{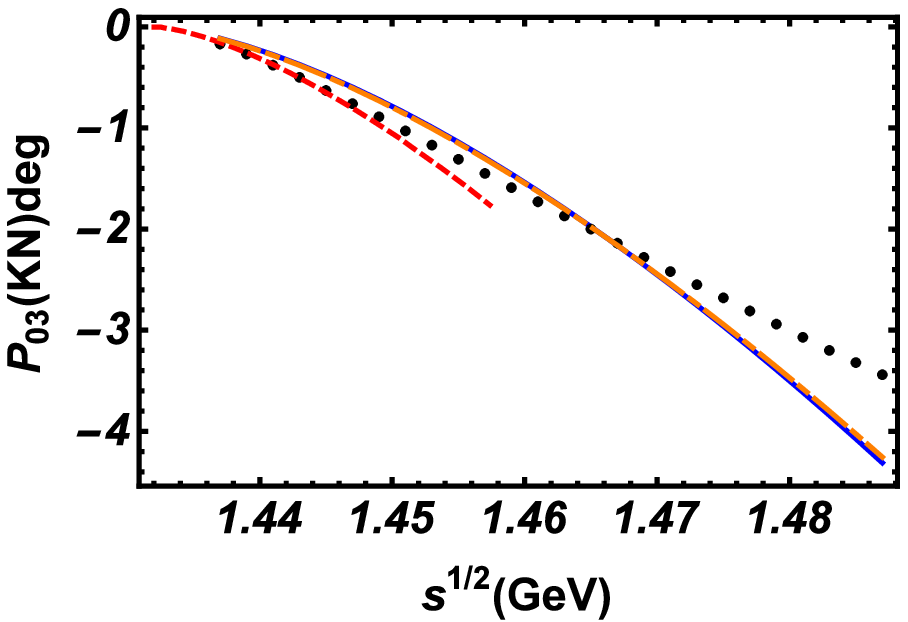}} \\
{\includegraphics[width=0.32\textwidth]{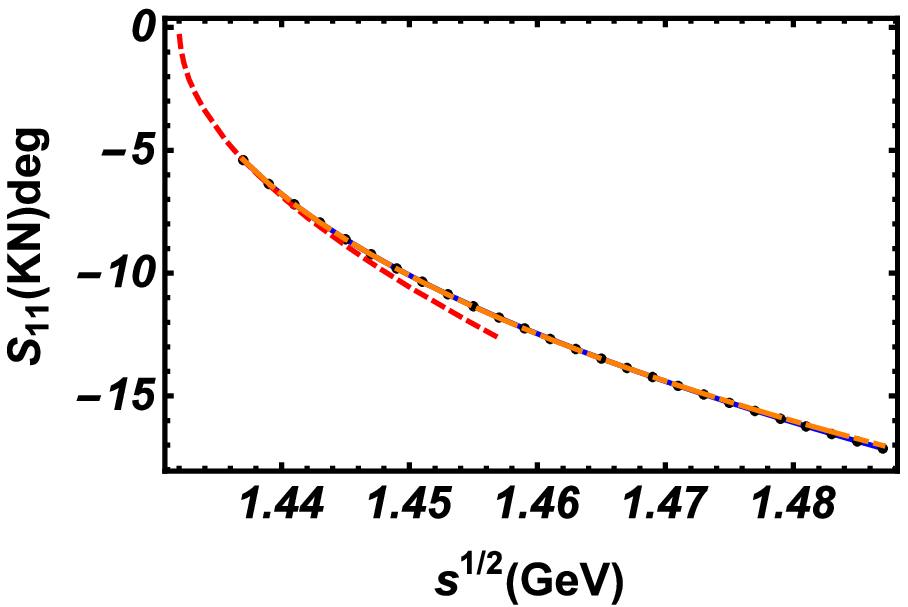}}&{\includegraphics[width=0.32\textwidth]{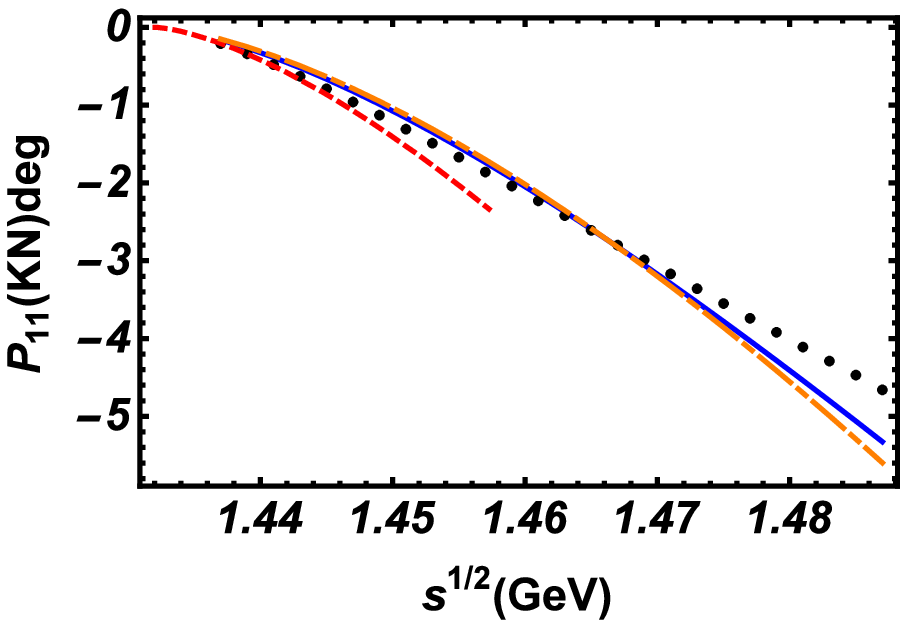}}& {\includegraphics[width=0.32\textwidth]{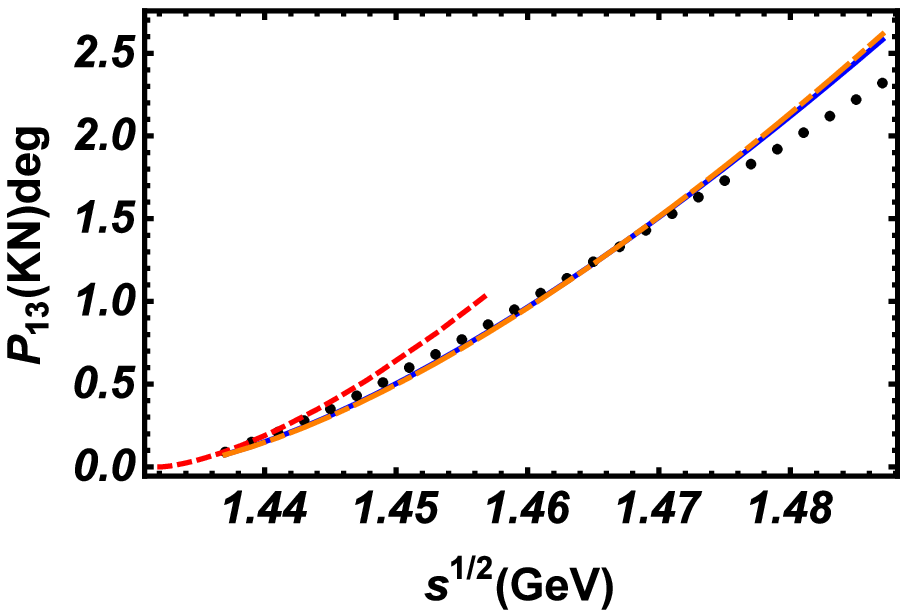}} \\
\end{tabular}
\caption{$I=0$ (upper panel) and $I=1$ (lower panel) $KN$ phase shifts. The orange long-short dashed lines and blue solid lines  represent our $\mathcal{O}(p^2)$ and $\mathcal{O}(p^3)^*$ results while the red dashed lines denote those of the  HB ChPT~\cite{Huang:2017bmx}. }\label{fig:kn}
\end{figure}

 In Fig.~\ref{fig:kn} we show our fitted results together with the experimental data. For the sake of comparison, we show as well the HB results of Refs.~\cite{Huang:2015ghe,Huang:2017bmx}. It is clear that the EOMS descriptions
are slightly better that the HB results when extended to higher energies.

From the above discussions, it is clear that the EOMS provides a satisfactory description of both the pion-nucleon and kaon-nucleon scattering data up to $\mathcal{O}(p^3)$, while the SU(3) HB ChPT fails.

\begin{figure}
\centering
\begin{tabular}{ccc}
{\includegraphics[width=0.32\textwidth]{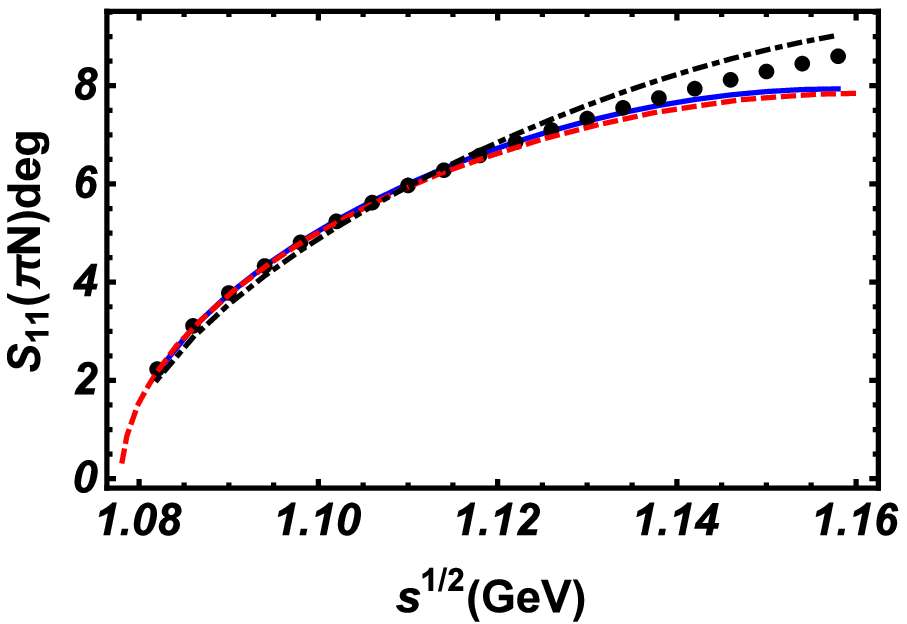}}&{\includegraphics[width=0.32\textwidth]{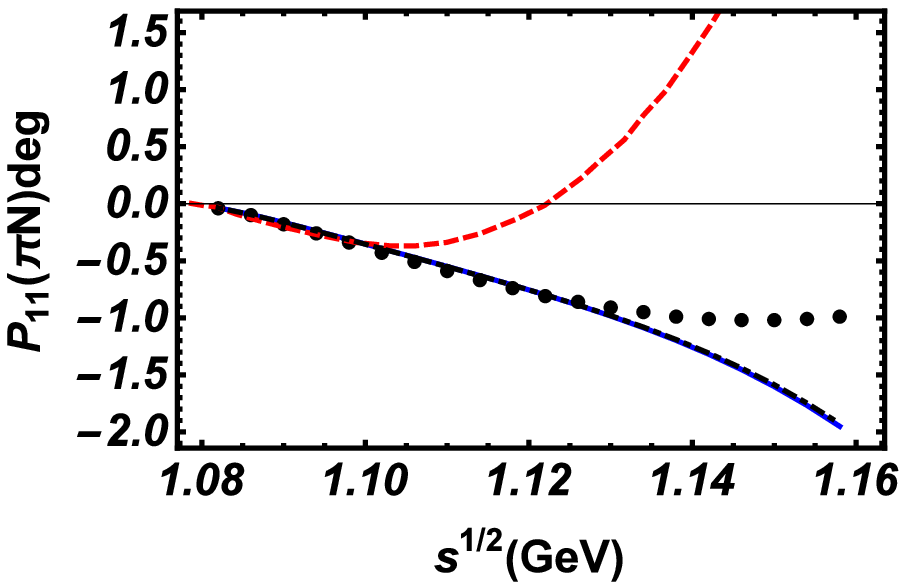}}& {\includegraphics[width=0.32\textwidth]{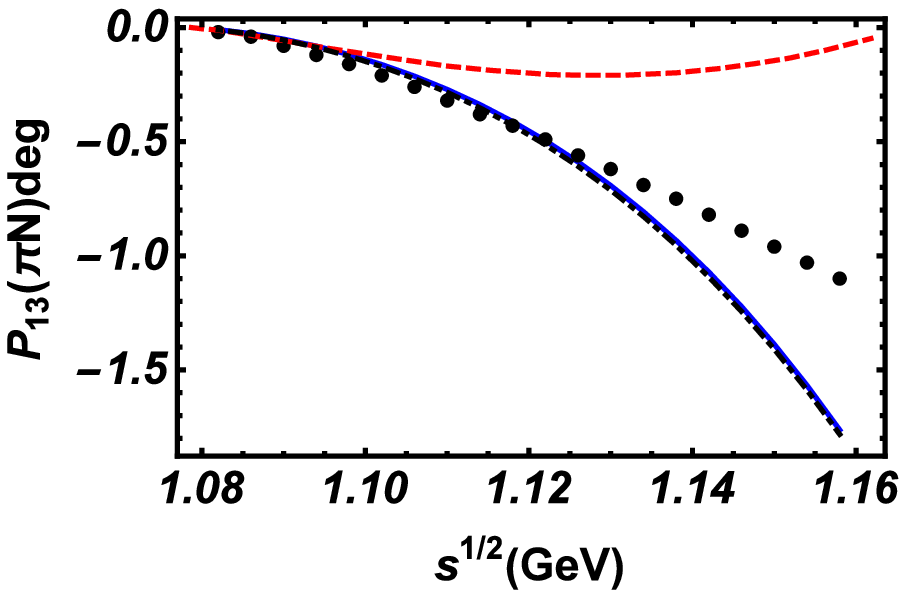}} \\
{\includegraphics[width=0.32\textwidth]{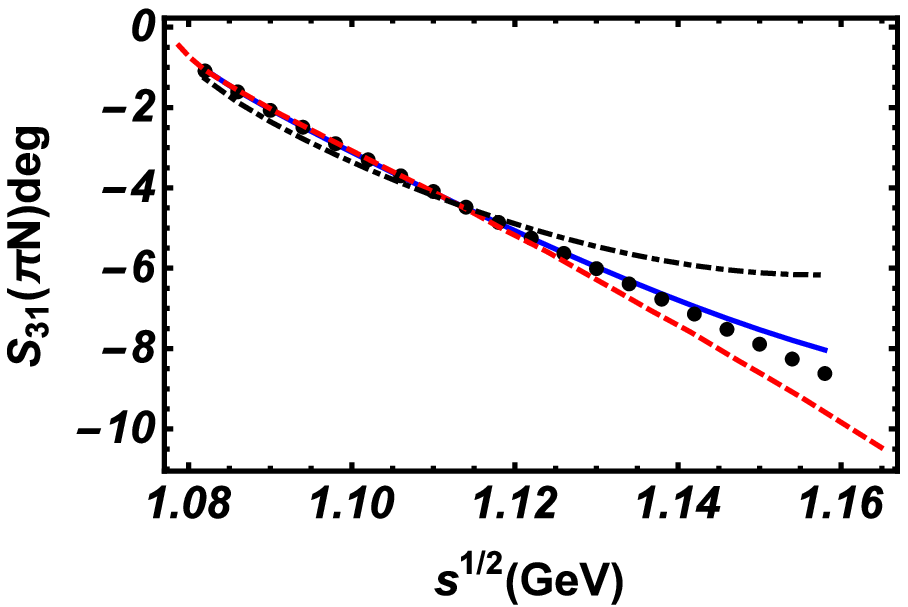}}&{\includegraphics[width=0.32\textwidth]{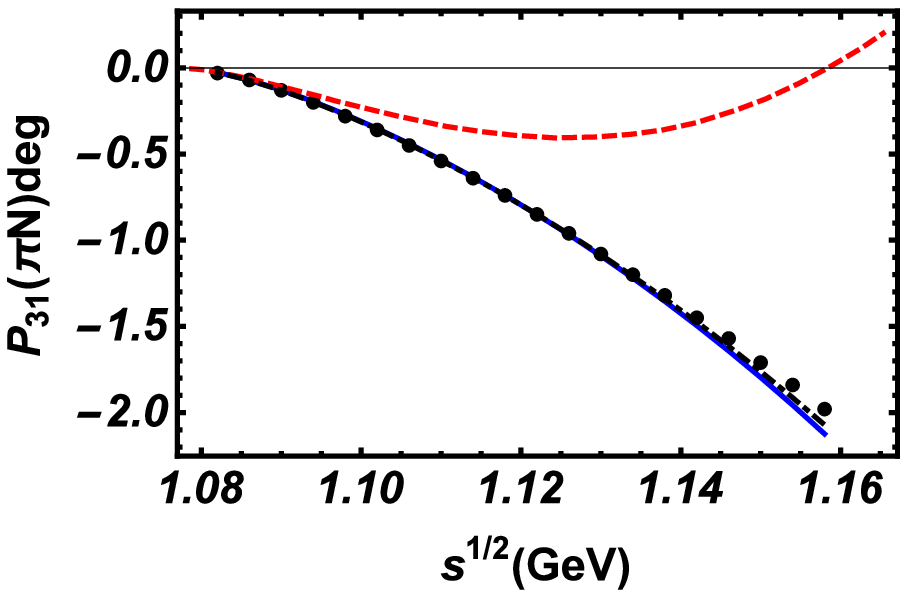}}& {\includegraphics[width=0.32\textwidth]{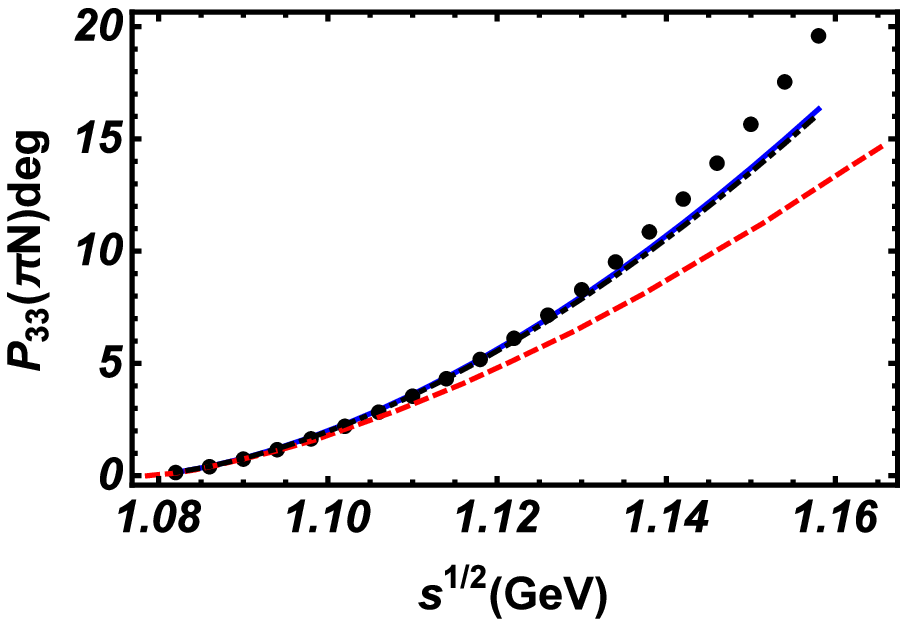}} \\
\end{tabular}
\caption{Same as Fig.~\ref{fig:pin}, but the black dot-dashed  lines denote the EOMS results with $b_0$, $b_D$, and $b_F$ fixed by fitting to the physical (isospin averaged) octet baryon masses at NNLO. }\label{fig:pin2}
\end{figure}

\begin{figure}
\centering
\begin{tabular}{ccc}
{\includegraphics[width=0.32\textwidth]{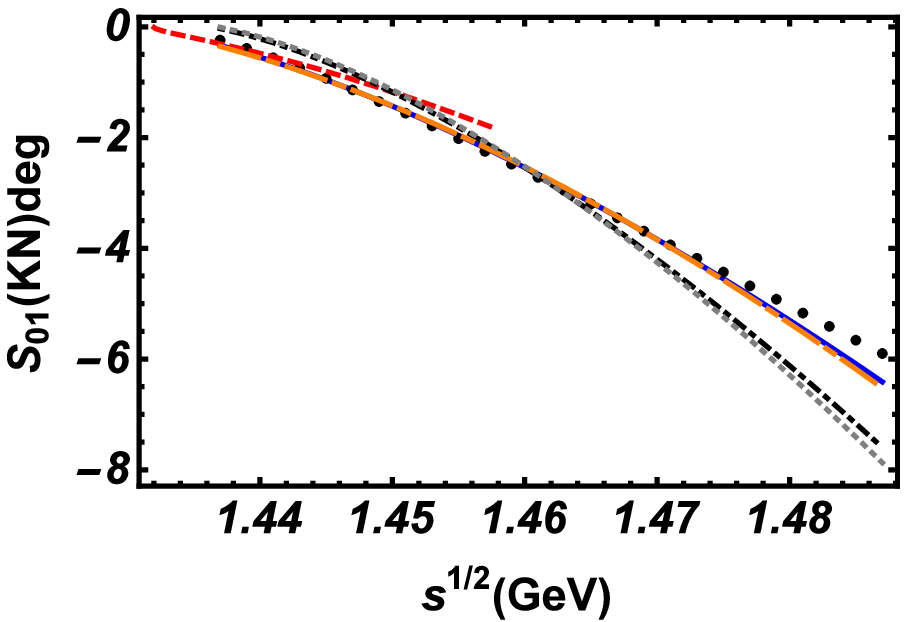}}&{\includegraphics[width=0.32\textwidth]{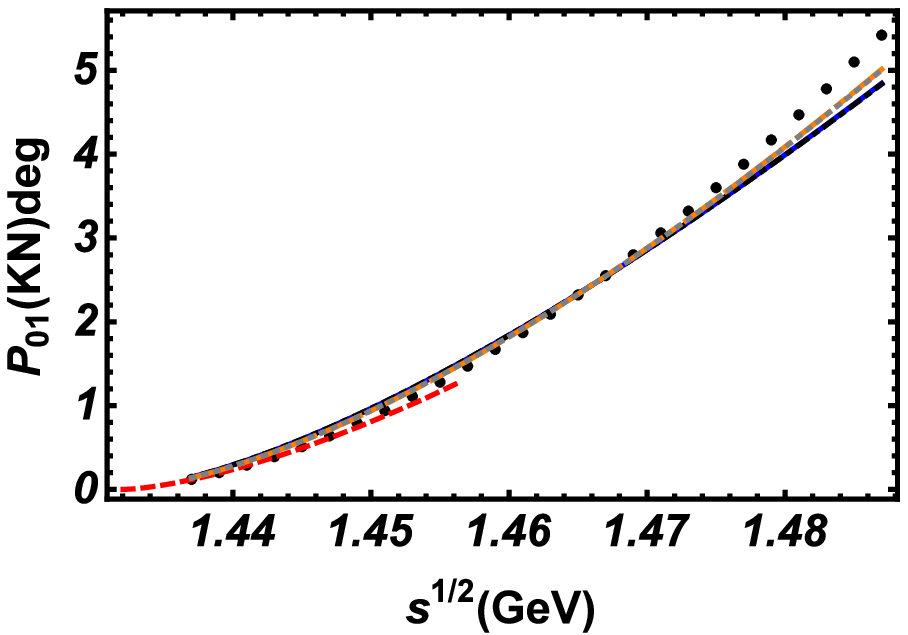}}& {\includegraphics[width=0.32\textwidth]{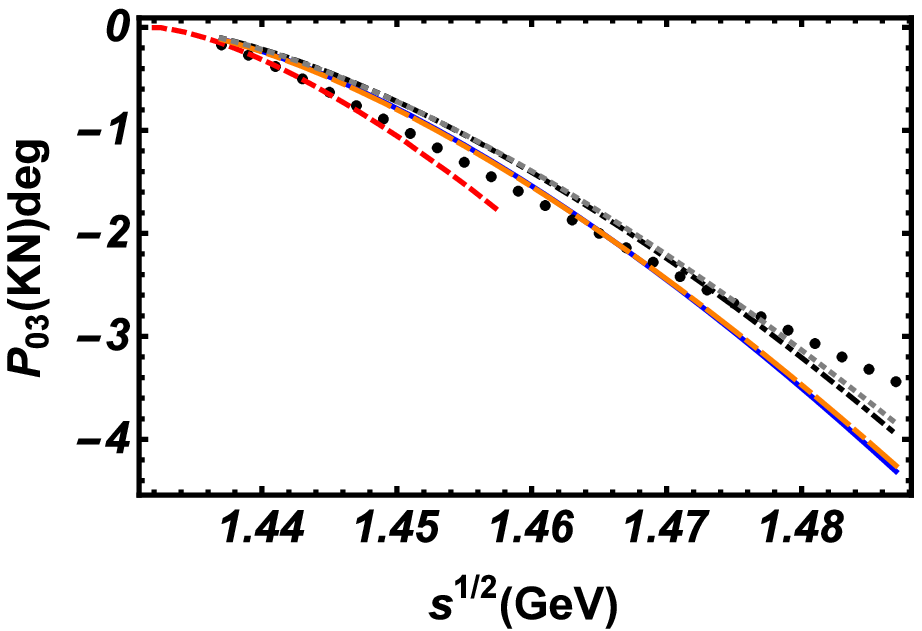}} \\
{\includegraphics[width=0.32\textwidth]{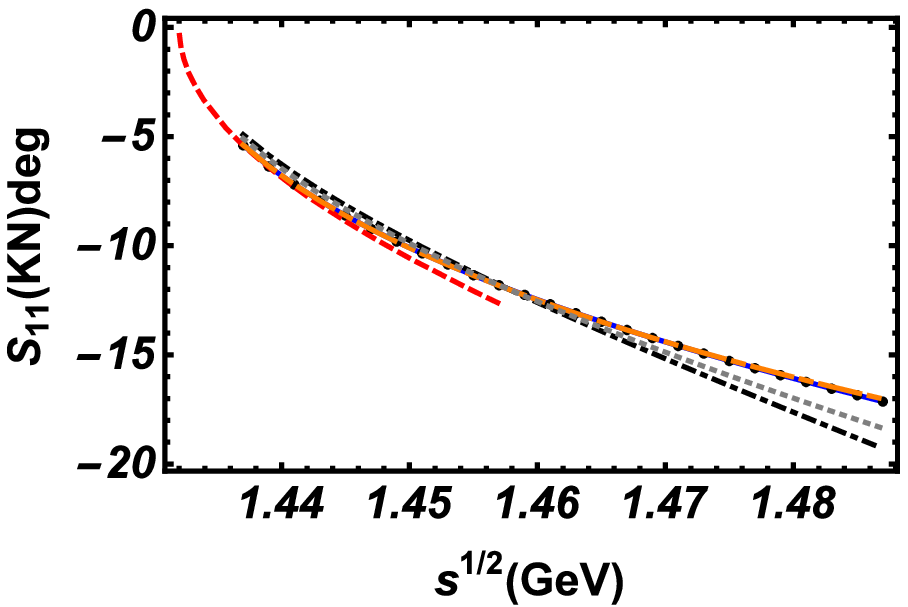}}&{\includegraphics[width=0.32\textwidth]{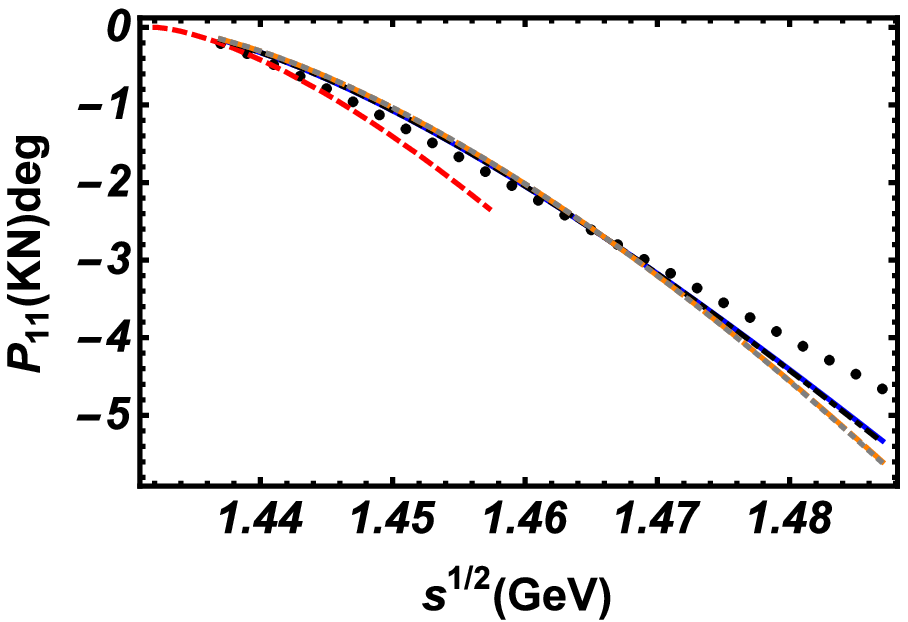}}& {\includegraphics[width=0.32\textwidth]{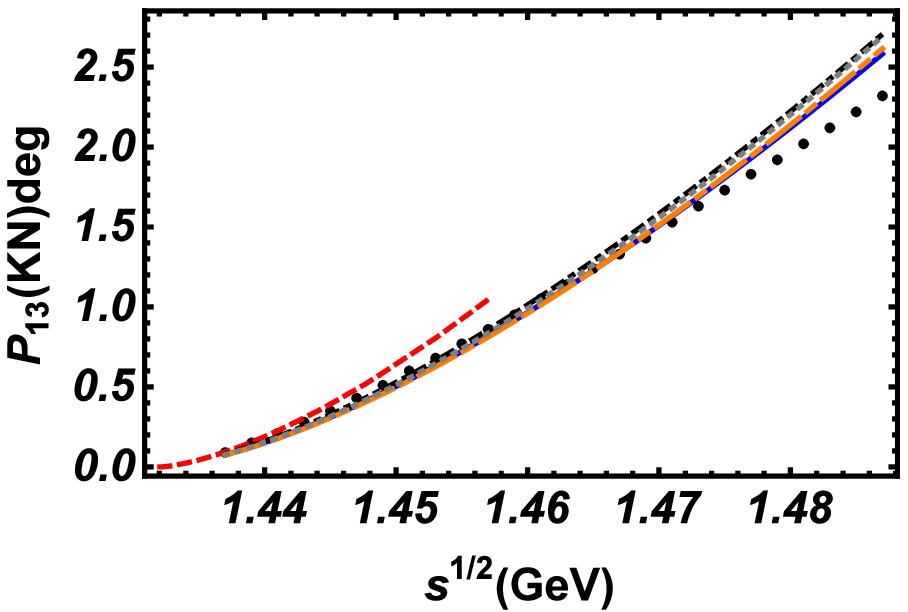}} \\
\end{tabular}
\caption{Same as Fig.~\ref{fig:kn}, but the gray dotted lines and black dot-dashed lines denote the $\mathcal{O}(p^2)$ and $\mathcal{O}(p^3)^*$ results in the EOMS scheme with $b_0$, $b_D$, and $b_F$ fixed by fitting to the physical (isospin averaged) octet baryon masses at NNLO. }\label{fig:kn2}
\end{figure}

\begin{table}
\centering
\caption{LECs contributing to the $I=0$ $KN$ scattering. }\label{fitKN0}
\begin{tabular}{ccccccccc}
\hline\hline
&$\beta_1$  &  $\beta_2$  &  $\beta_3$  &  $\beta_4$  & $\chi^2/d.o.f.$\\
\hline
$\mathcal{O}(p^2)$ &$-0.495(1)$  &  $0.113(1)$  &  $0.447(2)$  &  $0.136(1)$ & 0.829 \\
$\mathcal{O}(p^3)^*$    &$-0.767(1)$  &  $0.126(1)$  &  $0.604(3)$  &  $0.093(1)$ & 0.971 \\

\hline
\end{tabular}
\end{table}

\begin{table}
\centering
\caption{LECs contributing to $I=1$ $KN$ scattering. }\label{fitKN1}
\begin{tabular}{ccccccccc}
\hline\hline
&$\gamma_1$  &  $\gamma_2$  &  $\gamma_3$  &  $\gamma_4$ & $\chi^2/d.o.f.$\\
\hline
$\mathcal{O}(p^2)$&$-0.122(3)$  &  $0.0084(1)$  &  $0.264(1)$  &  $-0.270(1)$  &0.765\\
$\mathcal{O}(p^3)^*$   &$-0.419(2)$  &  $0.429(2)$  &  $0.616(1)$  &  $-0.090(3)$  &0.471\\
\hline
\end{tabular}
\end{table}

\subsection{Fitting strategy two:  combined study of the baryon masses and meson-baryon scattering}

One merit of ChPT (or any other EFT) is that it connects different observables with the same set of LECs.
Thus it is interesting to explore how one observable imposes restrictions on others. In this covariant baryon ChPT framework, baryon masses and scattering process are such a pair of observables which
are described by the same Lagrangians.
Most of the LECs appear  in both the meson-baryon scattering and  the baryon masses, such as, $b_0,b_D, b_F$ and $b_{1,\cdots,8}$. A naive idea for a combined study of these two observables can be performed in two
ways. First, calculating baryon masses at $\mathcal{O} (p^3)$ and using the experimental data as well as the pion-nucleon sigma term to constrain $b_0, b_D, b_F$, and then with these LECs fixed, study the pion-nucleon and kaon-nucleon scattering. Or conversely one can study the baryon masses with some LECs determined via meson-baryon scattering and, furthermore, make predictions on sigma terms~\footnote{One can of course calculate the sigma terms directly from scattering amplitudes via the corresponding subthreshold parameters using the Cheng-Dashen theorem~\cite{Cheng:1970mx}.}.

However, we note that the LECs actually contribute at different chiral orders to these two observables.
In meson baryon scattering, all of these LECs appear at $\mathcal{O}(p^2)$, the order of the chiral Lagrangians. On the other hand, $b_0, b_D, b_F$ contribute
to the baryon masses both at $\mathcal{O}(p^2)$ and $\mathcal{O}(p^4)$ via tree level as well as mass insertions, while $b_{1,\cdots,8}$ only contribute to the
baryon masses via tadpole diagrams at $\mathcal{O}(p^4)$. This complicates things a lot. In principle, from the point of view of effective field theories, to achieve a fully self-consistent and combined study of  baryon masses and meson-baryon scattering, one needs to renormalize the LECs in the same framework.
In other words, the calculation for baryon masses and meson-baryon scattering ought to be performed up to the same order. Otherwise the LECs in these two sectors are mismatched.
Thus if one tries to determine  $b_{1,\cdots, 8}$ through baryon masses, a calculation up to $\mathcal{O}(p^4)$ will be needed, which should be matched with scattering amplitudes also at $\mathcal{O}(p^4)$.
As a consequence, the number of LECs will be too large compared with the number of data available both for baryon masses and meson-baryon scattering from experiments and lattice QCD simulations.
On the other hand, if one is not so ambitious and only calculates the scattering amplitudes and baryon masses up to $\mathcal{O}(p^3)$, new problems show up. In this case, only 3 parameters($b_0,b_D,b_F$)
in addition to  $m_0$ appear in the baryon masses. Although the physical baryon masses can be accurately reproduced, the study in Ref.~\cite{Ren:2012aj} showed that
it is not possible to provide a satisfactory description of the LQCD baryon masses up to this order. In addition, the constraints from baryon masses to meson-baryon
 scattering will be very weak because there are 24 LECs in meson-baryon scattering up to $\mathcal{O}(p^3)$.

Taking all these into account, we calculate the baryon masses up to $\mathcal{O}(p^3)$ in the present work. Using the chiral limit baryon mass determined in Ref.~\cite{Ren:2012aj}, $m_0=0.880$ GeV, we determine $b_0$, $b_D$, $b_F$ by
fitting to the experimental octet baryon masses, with the pseudoscalar decay constants fixed as explained above. The resulting LECs and the fitted octet baryon masses are given in
Table \ref{bmasses}.

\begin{table}
\centering
\caption{LECs determined by fitting to the experimental baryon masses up to NLO in the EOMS BChPT and the corresponding fitted results, in comparison with the experimental data. All of the masses are  in units of GeV.}\label{bmasses}
\begin{tabular}{ccccc}
\hline\hline
&$m_0$ &$b_0$ & $b_D$ & $b_F$ \\
\hline
Fit &$0.88$(FIX)& $-0.6232(9)$ & $0.0570(7)$ &$-0.4022(71)$   \\
\hline
&   $m_N$  &  $m_\Lambda$  &  $m_\Sigma$  &  $m_\Xi$  \\
\hline
Fit&$0.9392$& $1.1157$     &  $1.1862$    & $1.3272$  \\
Exp.&0.938925(645)&$1.115683(6)$&$1.19315(430)$&$1.31828(343)$ \\
\hline
\end{tabular}
\end{table}

\begin{table}
\centering
\caption{LECs in the $\pi N$ channel with $\alpha_4=b_0+\frac{b_D}{2}+\frac{b_F}{2}$ fixed by fitting to the baryon masses.}\label{fitMApiN}
\begin{tabular}{ccccccccc}
\hline\hline
$\alpha_1$  &  $\alpha_2$  &  $\alpha_3$  &  $\alpha_4$ & $\alpha_5$ & $\alpha_6$ & $\alpha_7$ & $\alpha_8$  & $\chi^2/d.o.f.$\\
\hline
$-7.41(7)$  &  $1.56(2)$  &  $1.33(1)$  &  $-0.80$ & $0.63(2)$ & $3.18(6)$ & $1.45(3)$ & $-0.096(120)$ & 1.26 \\
\hline
\end{tabular}
\end{table}

\begin{table}
\centering
\caption{LECs in the $I=0$ $KN$ channel with $\beta_4=b_0-b_F$ fixed by fitting to the baryon masses.}\label{fitMAKN0}
\begin{tabular}{ccccccc}
\hline\hline
&$\beta_1$  &  $\beta_2$  &  $\beta_3$  &  $\beta_4$ & $\chi^2/d.o.f.$ \\
\hline

$\mathcal{O}(p^2)$ &$-0.284(1)$  &  $0.144(2)$  &  $0.443(3)$  &  $-0.221$  & 4.66\\
$\mathcal{O}(p^3)^*$     &$-0.582(11)$  &  $0.153(3)$  &  $0.601(5)$  &  $-0.221$  & 3.93\\
\hline
\end{tabular}
\end{table}

\begin{table}
\centering
\caption{LECs in the $I=1$ $KN$ channel with $\gamma_4=b_0+b_D$ fixed by fitting to the baryon masses.}\label{fitMAKN1}
\begin{tabular}{ccccccc}
\hline\hline
&$\gamma_1$  &  $\gamma_2$  &  $\gamma_3$  &  $\gamma_4$      & $\chi^2/d.o.f.$  \\
\hline
$\mathcal{O}(p^2)$&$-0.236(11)$  &  $-0.033(3)$  &  $0.246(5)$  &  $-0.0566$ & 1.45 \\
$\mathcal{O}(p^3)^*$    &$-0.604(14)$  &  $0.364(3)$  &  $0.588(5)$  &  $-0.0566$ & 2.24 \\
\hline
\end{tabular}
\end{table}

Compared to the  fit up to $\mathcal{O}(p^3)$ to the scattering phase shifts, a combined fit of the baryon masses and scattering amplitudes yields a slightly worse  description of
the scattering phase shifts to some extent. Particularly, the fitting results are worse in the $KN$ channel where the $\chi^2/d.o.f.$ increases by a factor of about 4. This is understandable
as the number of free LECs decreases. Despite of this, the negative effects do not spoil the description. For the $p$-wave,
the descriptions of the phase shifts are of very similar quality, whether one fixes $b_0,b_D,b_F$ and treats them as free LECs.
For the $s$-wave, the differences are rather moderate, particularly in the low energy region. This study indicates that the EOMS BChPT is
able to describe the baryon masses and meson-baryon scattering simultaneously, as it should be.  Nevertheless, as mentioned at the beginning of this sector, to draw a firm conclusion,
more systematic studies are needed.

As for the sigma terms, we find that meson-baryon scattering up to $\mathcal{O}(p^3)$ is not very useful at this moment because  the tree level contributions at $\mathcal{O}(p^3)$ in the $KN$ channels are neglected, leading to unusually large $b_D,b_F$ compared to an independent study of the baryon masses in, e.g., Ref.~\cite{Ren:2012aj}. Thus we will refrain from performing such a study here.

\subsection{Scattering lengths}
Scattering lengths, also known as $s$-wave threshold parameters, can be predicted with the LECs determined above. The general form of the effective range expansion reads
\begin{equation}\label{EREform}
    |\mathbf{p}|^{2l+1}\cot \delta_{l\pm}^{I}=\frac{1}{a_{l\pm}^{I}}+\frac{1}{2}r_{l\pm}^I |\mathbf{p}|^2 + \sum_{n=2}^{\infty}v_{n,l\pm}^I|\mathbf{p}|^{2n},
\end{equation}
where $|\mathbf{p}|$ refers to the three-momentum of the baryon in the c.m. frame, $\ell$ is the angular momentum, $a$ is the threshold parameter, $r$ is the effective range and $v_n$ are the shape parameters. We can easily obtain the expression of threshold parameters from Eq.~(\ref{EREform}) by taking the limit of $|\mathbf{p}|\rightarrow 0$ as
\begin{equation}\label{SCform}
    a_{l\pm}^{I}=\lim_{|\mathbf{p}|\rightarrow 0}\frac{\tan \delta_{l\pm}^{I}}{|\mathbf{p}|^{2l+1}}=\lim_{|\mathbf{p}|\rightarrow 0}\frac{\text{Re}f_{l\pm}^{I}}{|\mathbf{p}|^{2l}}.
\end{equation}

With the phase shifts obtained in ChPT, one can easily compute the $\ell=0$ scattering lengths. {As mentioned earlier, $\text{Re}f_{l\pm}^I$ cannot be calculated at exactly the threshold because the term $t=(p-p')^2$ appearing in the dominator diverges at that point}.
Thus we follow the strategy of  Ref.~\cite{Yao:2016vbz}. We firstly calculate the scattering lengths for energies very close to the threshold and then extrapolate them to the threshold. The scattering lengths for these channels are collected in Table.~\ref{SC}.

\begin{table}
\centering
\caption{$\pi N$ and $KN$ scattering lengths in units of fm. Note that we did not
associate any uncertainties to the $\mathcal{O}(p^3)^*$  contributions of the $KN$ channel because we have not included the tree level contributions at this order for the $KN$ channels. }\label{SC}
\begin{tabular}{cccccccc}
\hline\hline
Channel &$\mathcal{O}(p^1)$  &  $\mathcal{O}(p^2)$  &  $\mathcal{O}(p^3)$  &  Total  & Huang(HB)~\cite{Huang:2017bmx} & Mai(IR)~\cite{Mai:2009ce} & EXP\\
\hline
$a_{\pi N}^{3/2}$& $-0.126$  &  $0.026(11)$  &  $-0.011(8)$  &  $-0.111(16)$ & $-0.110(2)$  & $-0.04(7)$ & $-0.125(3)$~\cite{Schroder:2001rc}  \\
$a_{\pi N}^{1/2}$& $0.212$  &  $0.025(10)$  &  $0.003(16)$  &  $0.240(22)$ & $0.240(2)$  & $0.07(3)$ & $0.250^{+0.006}_{-0.004}$~\cite{Schroder:2001rc}  \\
$a_{KN}^{1}(\mathcal{O}(p^2))$& $-0.476$  &  $0.149(1)$  &  $-/-$  &  $-0.327(1)$ & $-0.330(5)$  & $-0.33(32)$ & $-0.33$~\cite{Arndt:2006bf}  \\
$a_{KN}^{0}(\mathcal{O}(p^2))$& $0.043$  &  $-0.057(2)$  &  $-/-$  &  $-0.014(2)$ & $0.000(4)$  & $0.02(64)$ & $0.02$~\cite{Arndt:2006bf}  \\
$a_{KN}^{1}(\mathcal{O}(p^3)^*)$& $-0.476$  &  $1.067(5)$  &  $-0.919$  &  $-0.328(5)$ & $-/-$  & $-/-$ & $-/-$  \\
$a_{KN}^{0}(\mathcal{O}(p^3)^*)$ & $0.043$  &  $0.164(2)$  &  $-0.219$  &  $-0.012(2)$ & $-/-$  & $-/-$ & $-/-$  \\
\hline
\end{tabular}
\end{table}

It is clear that our results based on the EOMS scheme are in very good agreement with the experimental data and the  HB results, while the  IR results~\cite{Mai:2009ce} seem to be  compatible with data only in the $KN$ channels.

\subsection{Convergence of BChPT}
The  convergence of SU(3) BChPT has remained an issue of heated debate for many years. See, e.g., Ref~\cite{Bernard:2003rp} for early discussions, and Refs.~\cite{Ren:2013dzt, Xiao:2018rvd} for
more recent studies of baryon magnetic moments and masses. From the latter studies, it seems that the EOMS scheme can speed up the convergence of BChPT, particularly, in the SU(3) sector.

\begin{figure}
\centering
\begin{tabular}{ccc}
{\includegraphics[width=0.32\textwidth]{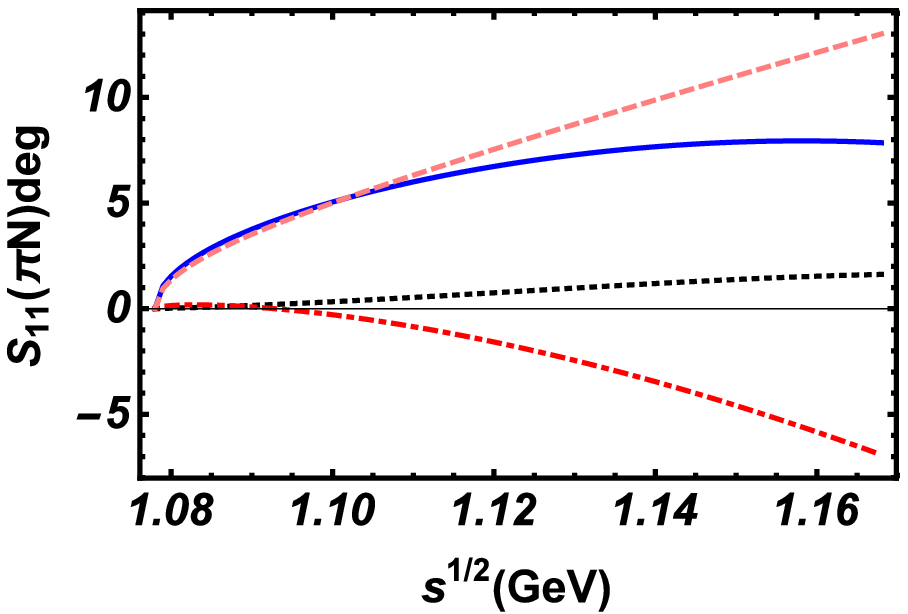}}&{\includegraphics[width=0.32\textwidth]{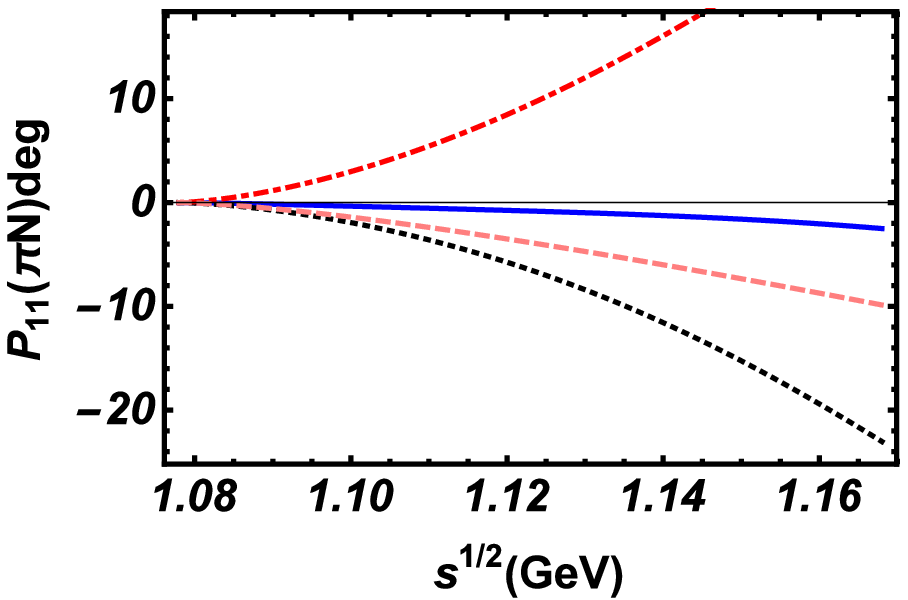}}& {\includegraphics[width=0.32\textwidth]{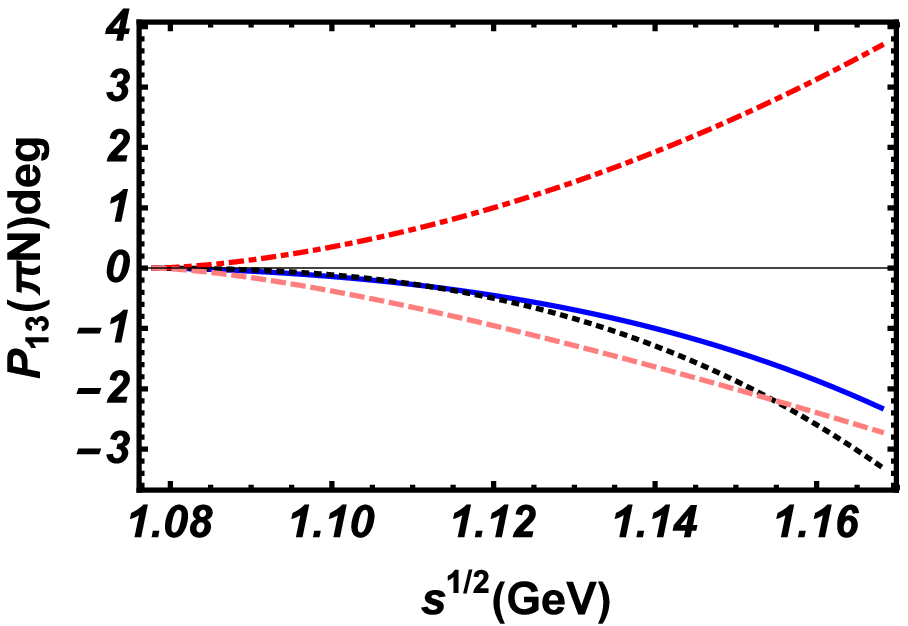}} \\
{\includegraphics[width=0.32\textwidth]{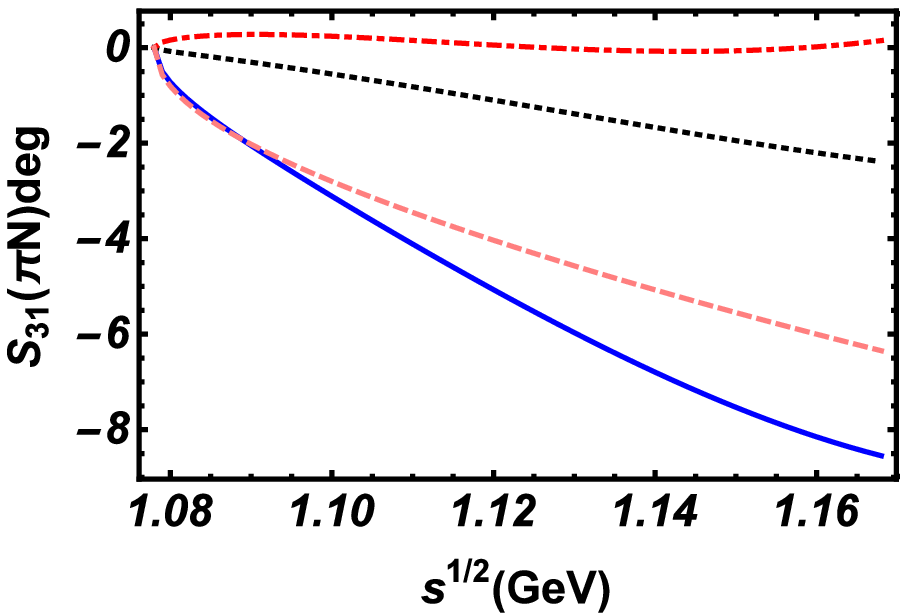}}&{\includegraphics[width=0.32\textwidth]{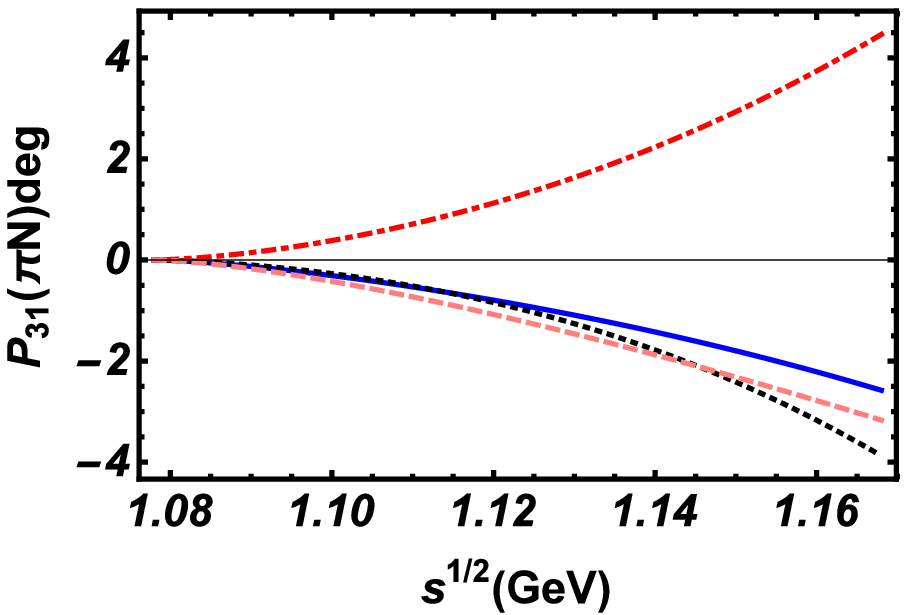}}& {\includegraphics[width=0.32\textwidth]{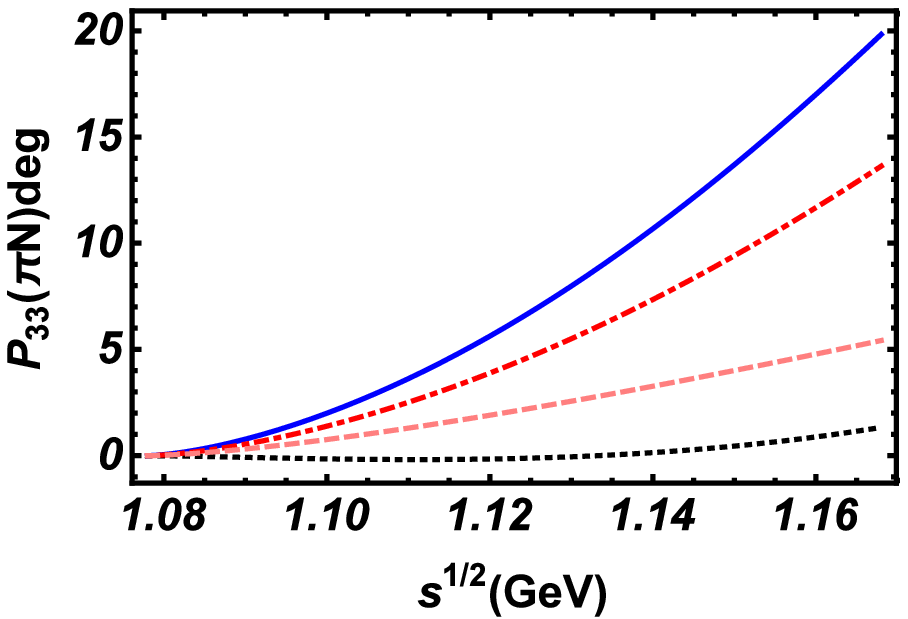}} \\
\end{tabular}
\caption{Order by order decomposition of the $\pi N$ phase shifts. The blue lines donate the total results, while
those of the $\mathcal{O}(p)$, $\mathcal{O}(p^2)$, and $\mathcal{O}(p^3)$ are represented by the pink-dashed, read-dot-dashed, and black-dotted lines, respectively.}
\label{con:pin}
\end{figure}

\begin{figure}
\centering
\begin{tabular}{ccc}
{\includegraphics[width=0.32\textwidth]{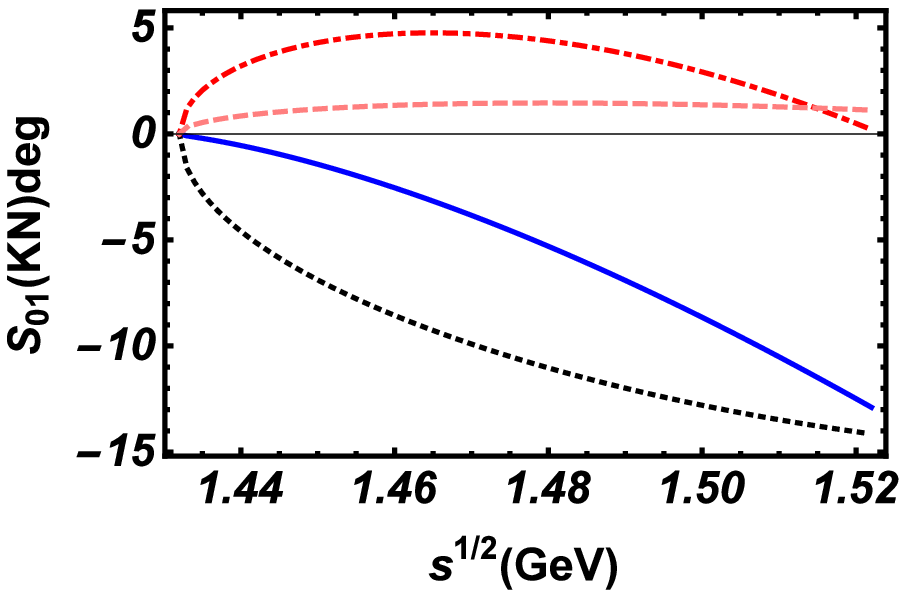}}&{\includegraphics[width=0.32\textwidth]{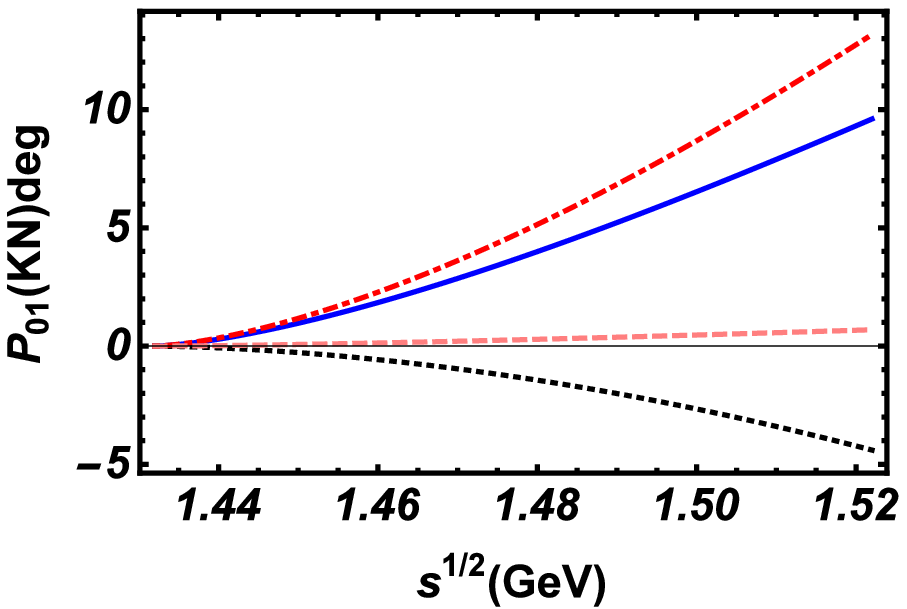}}& {\includegraphics[width=0.32\textwidth]{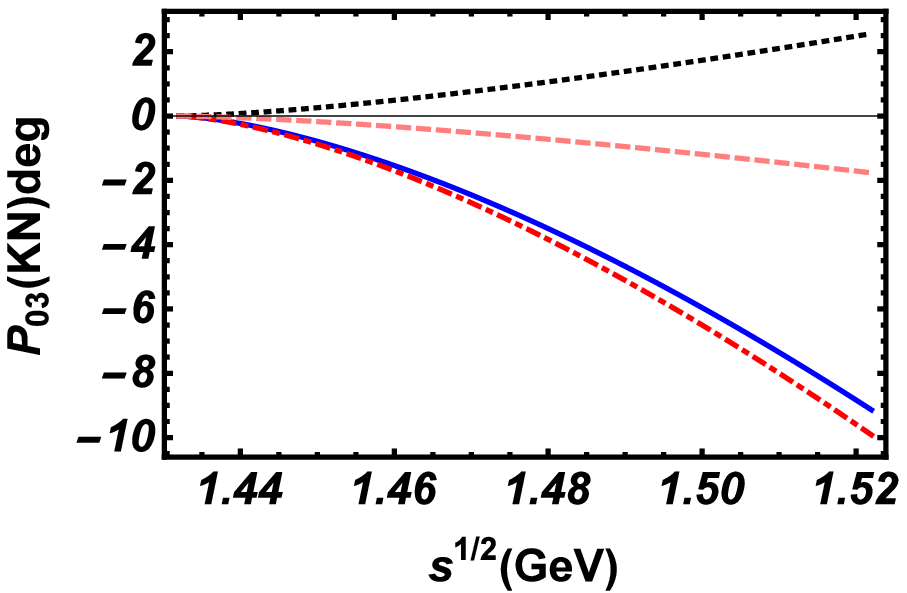}} \\
{\includegraphics[width=0.32\textwidth]{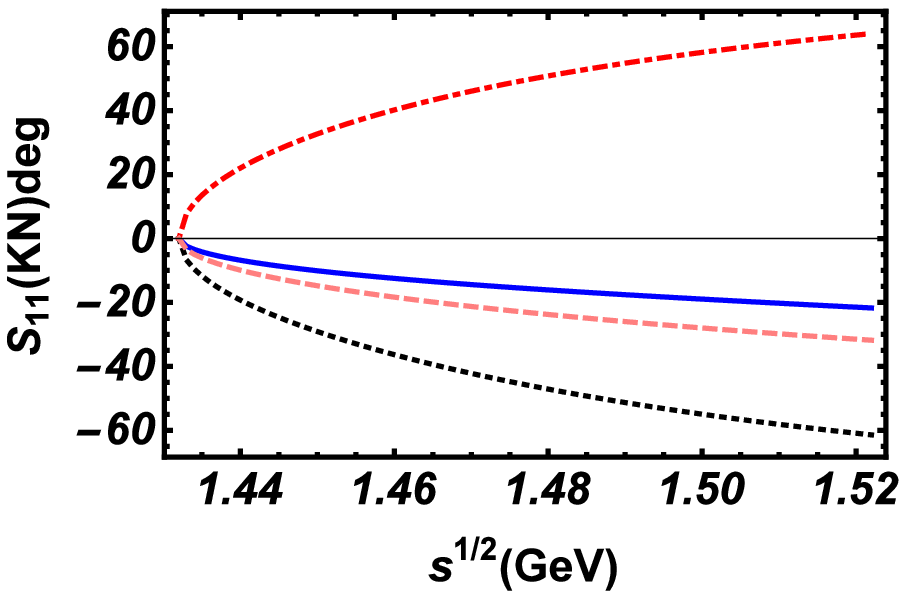}}&{\includegraphics[width=0.32\textwidth]{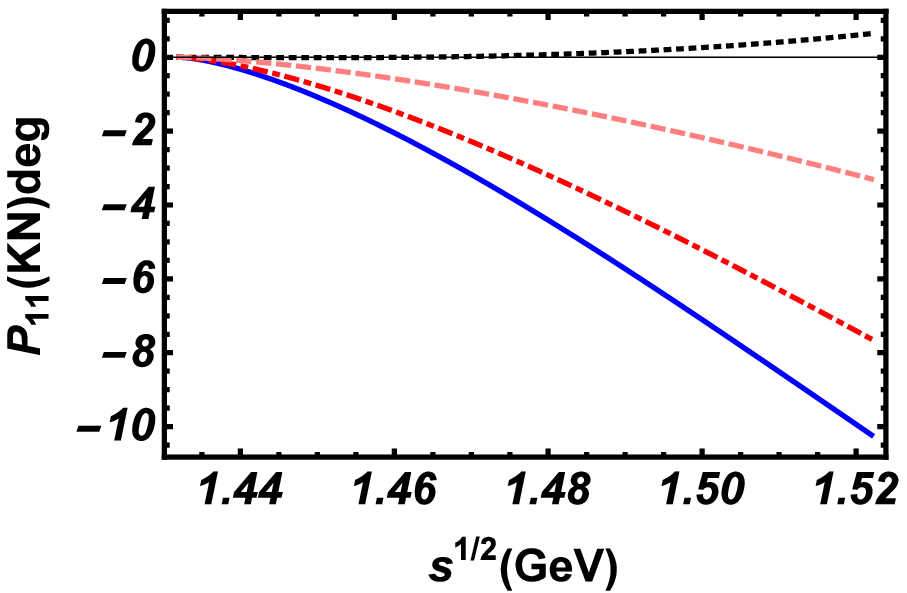}}& {\includegraphics[width=0.32\textwidth]{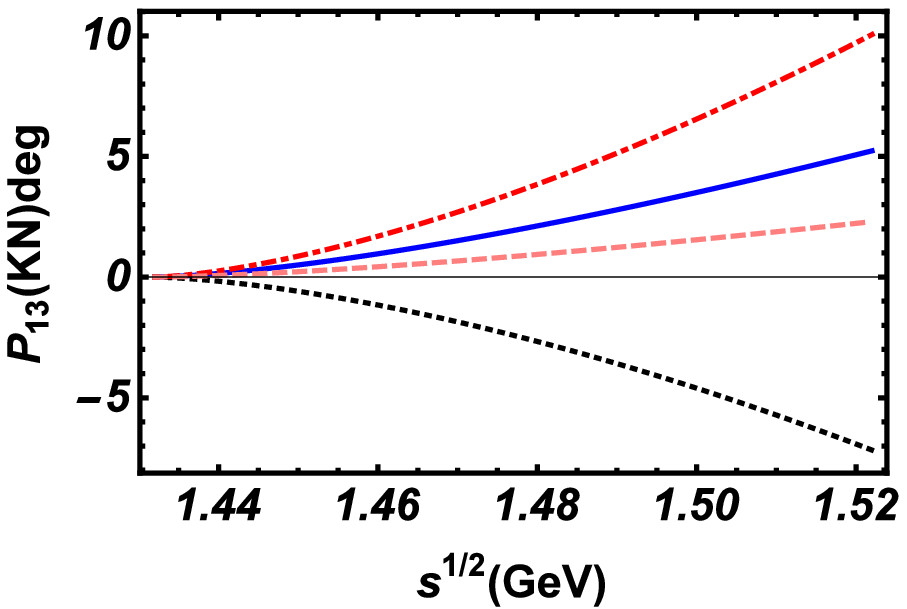}} \\
\end{tabular}
\caption{Order by order decomposition of the $KN$ phase shifts. The blue lines donate the total results, while
those of the $\mathcal{O}(p)$, $\mathcal{O}(p^2)$, and $\mathcal{O}(p^3)^*$ are represented by the pink-dashed, red-dot-dashed, and black-dotted lines, respectively.}
\label{con:kn}
\end{figure}

Nonetheless, even in the EOMS scheme, the convergence turns out to be relatively slow. The origin of this slow convergence in the SU(3) sector is the large expansion parameter $\frac{M_K}{\Lambda_{\chi PT}}$, which is approximately $1/2$ in the physical world. For a LQCD simulation, the situation can become even worse. In the present work, we show the phase shifts of each order, collected in Fig.~\ref{con:pin} for the $\pi N$ channels and Fig.~\ref{con:kn} for the $KN$ channels. Both figures show relatively large contributions from $\mathcal{O}(p^2)$ and $\mathcal{O}(p^3)$, and they tend to cancel each other, which indicates that the convergence is not as good as one would like. Actually this problem already showed up in the SU(2) study. In Ref.~\cite{Alarcon:2012kn}, the authors performed a detailed study of  $\pi N$ scattering up to $\mathcal{O}(p^3)$ in the EOMS scheme. They concluded that the convergence is indeed not  very satisfactory. Thus a even slower convergence is expected in
the SU(3) case. However, the convergence seems much better once the scattering amplitudes are calculated up to $\mathcal{O}(p^4)$ as shown in Ref.~\cite{Chen:2012nx}.

\section{Summary and outlook}
In this work, we performed a SU(3) study of the meson-baryon elastic scattering up to $\mathcal{O}(p^3)$ in covariant baryon chiral perturbation theory. Due to  lack of experimental data, we focus only on the $\pi N^{I=3/2,1/2}$ and $KN^{I=0,1}$ channels. We applied the extended-on-mass-shell (EOMS) scheme to restore the power counting and determined the corresponding low energy constants by fitting to the experimental phase shifts. We achieved a pretty good description in these channels simultaneously up to 1.16GeV for $\pi N$ and 1.52GeV for $KN$. For $\pi N$ channels, our study in SU(3) shows a compatible description as that in SU(2) and much better compared to the HB SU(3) results. For the $KN$ channels, we found that with only phase shifts one can not uniquely determine all the LECs.
Nevertheless, neglecting $\mathcal{O}(p^3)$ tree level contributions, we obtained a description in good agreement with the experimental data.

We attempted a combined study of the baryon masses and meson-baryon scattering up to $\mathcal{O}(p^3)$. We first determined $b_0,b_F,b_D$ using the baryon masses and then kept them fixed
in the fitting of the partial wave phase shifts. Our study showed indeed that the EOMS BChPT can
describe simultaneously the baryon masses and meson-baryon scattering, but a firm conclusion needs more systematic studies at higher orders.

The predicted scattering lengths for the $\pi N$ and $KN$ channels are in good agreement with the HB results and the experimental data. In addition, we explored the convergence of BChPT in meson-baryon scattering. The large cancelation between the NLO and NNLO contributions implies an unsatisfying convergence rate, similar to  that of the SU(2) sector up to $\mathcal{O}(p^3)$. On the other hand,
since in the one baryon sector, both $\mathcal{O}(p^3)$ and $\mathcal{O}(p^4)$ contribute at the one-loop level, it might well be the case that one will see cancelations 
between the $\mathcal{O}(p^3)$ and $\mathcal{O}(p^4)$ contributions, as already noted
in the study of the masses of the decuplet baryons. The convergence pattern in the SU(3) case  should be further examined by going to $\mathcal{O}(p^4)$, as already done for pion-nucleon scattering.

The predicted phase shifts and scattering lengths for other channels listed in Table.~\ref{IS} for the case of $\mathcal{O}(p^3)^*$ should be taken with caution  since the  $\mathcal{O}(p^3)$  LECs are not fully determined. Thus additional data, such as the cross sections in the $\bar{K}N$ channel, ought to be taken into account. As
the interaction in this channel is by nature non-perturbative, tiled to the existence of a shallow bound state of $\bar{K}N$, the $\Lambda(1405)$, we leave such a study to a future work.

\section{Acknowledgements}

We thank Ulf-G. Mei{\ss}ner for a careful reading of the manuscript and for many useful comments. This work is partly supported by the National Natural Science Foundation of China under Grants No.11522539, No. 11735003,
and No. 11775099,
and the fundamental Research Funds for the Central Universities.
XLR and MLD acknowledge supports form DFG and NSFC through funds provided to the Sino-German CRC 110 ``Symmetries and the Emergence of Structure in QCD'' (Grant No. TRR110).

\section{Appendix}

\subsection{Tree level contact terms}
In this subsection, we list the contributions of the tree-level contact terms.
To simplify the expressions, we first define
\begin{equation}\label{nupi}
\begin{split}
    \nu_{\pi}&=\left(-s+m_N{}^2+m_{\pi }{}^2\right){}^2+\left(m_N{}^2+m_{\pi }{}^2-u\right){}^2, \\
    \nu_{K}&=\left(-s+m_N{}^2+m_{K }{}^2\right){}^2+\left(m_N{}^2+m_{K }{}^2-u\right){}^2.
\end{split}
\end{equation}
The contributions in the respective channels are
\begin{itemize}
  \item $\pi N^{I=3/2}$

  \begin{equation}\label{Npi32B}
    \begin{split}
      B_{\pi N}^{I=3/2}= & -\frac{1}{2 f^2}+\frac{2 (s-u) (b_5+b_6+b_7+b_8)}{f^2} -\frac{8 m_N (c_1+c_2)}{f^2}\\
      &\frac{4 \left(d_2 \nu _{\pi }+d_4 \left(t-2 m_{\pi }{}^2\right)-2 d_{49} m_{\pi }{}^2\right)}{f^2}-\frac{8 m_N (s-u) (d_{10}+d_8)}{f^2},
    \end{split}
  \end{equation}
  \begin{equation}\label{Npi32A}
    \begin{split}
      A_{\pi N}^{I=3/2}= & \frac{2 m_{\pi }{}^2 (-2 b_0+b_1+b_2+b_3+2 b_4-b_D-b_F)-t (b_1+b_2+b_3+2 b_4)}{f^2} \\
      &+\frac{2 (s-u) (c_1+c_2)}{f^2}+\frac{2 (s-u)^2 (d_{10}+d_8)}{f^2}.
    \end{split}
  \end{equation}

  \item $\pi N^{I=1/2}$

  \begin{equation}\label{Npi12B}
    \begin{split}
      B_{\pi N}^{I=1/2}= &\frac{1}{f^2} +\frac{2 (s-u) (b_5+b_6+b_7+b_8)}{f^2}+\frac{16 m_N (c_1+c_2)}{f^2} \\
      &\frac{8 \left(-d_2 \nu _{\pi }+d_4 \left(2 m_{\pi }{}^2-t\right)+2 d_{49} m_{\pi }{}^2\right)}{f^2}-\frac{8 m_N (s-u)  (d_{10}+d_8)}{f^2},
    \end{split}
  \end{equation}
  \begin{equation}\label{Npi12A}
    \begin{split}
      A_{\pi N}^{I=1/2}= & \frac{2 m_{\pi }{}^2 (-2 b_0+b_1+b_2+b_3+2 b_4-b_D-b_F)-t (b_1+b_2+b_3+2 b_4)}{f^2} \\
      &-\frac{4 (s-u) (c_1+c_2)}{f^2} +\frac{2 (s-u)^2 (d_{10}+d_8)}{f^2}.
    \end{split}
  \end{equation}

  \item $K N^{I=1}$

  \begin{equation}\label{K N1B}
    \begin{split}
      B_{K N}^{I=1}= & -\frac{1}{f^2}+\frac{2  (2 b_5+2 b_7+b_8)(s-u)}{f^2} -\frac{4 m_N (4 c_2+c_3)}{f^2}-\frac{8 m_N (s-u) (d_{10}+d_7+d_8)}{f^2}\\
      &+\frac{4 \left(\nu _K (-d_1+d_2+d_3)+2 m_K{}^2 (-d_4+d_{48}-d_{49}+d_5+d_{50}-d_6)+t (d_4-d_5+d_6)\right)}{f^2},
    \end{split}
  \end{equation}
  \begin{equation}\label{K N1A}
    \begin{split}
      A_{K N}^{I=1}= & \frac{4 m_K{}^2 (-b_0+b_1+b_2+b_4-b_D)-2 t (b_1+b_2+b_4)}{f^2} +\frac{(s-u) (4 c_2+c_3)}{f^2}\\
      &+\frac{2 (s-u)^2 (d_{10}+d_7+d_8)}{f^2}.
    \end{split}
  \end{equation}

  \item $K N^{I=0}$

  \begin{equation}\label{K N0B}
    \begin{split}
      B_{K N}^{I=0}= & \frac{2 (b_8-2 b_6)(s-u)}{f^2} + \frac{4 m_N (4 c_1+c_3)}{f^2}-\frac{8 m_N (s-u) (d_{10}+d_7-d_8)}{f^2}\\
      &+\frac{4 \left(-\nu _K (d_1+d_2+d_3)+2 m_K{}^2 (d_4+d_{48}+d_{49}+d_5-d_{50}+d_6)-t (d_4+d_5+d_6)\right)}{f^2},
    \end{split}
  \end{equation}
  \begin{equation}\label{K N0A}
    \begin{split}
      A_{K N}^{I=0}= & \frac{4 m_K{}^2 (-b_0-b_3+b_4+b_F)+2 t (b_3-b_4)}{f^2} -\frac{(s-u) (4 c_1+c_3)}{f^2}\\
      &+\frac{2 (s-u)^2 (d_{10}+d_7-d_8)}{f^2}.
    \end{split}
  \end{equation}
\end{itemize}

\subsection{Tree level Born diagrams}

Once simplified with the on-shell condition, the amplitude for the Born diagram could be rewritten as
\begin{equation}\label{bornamplitude3}
\begin{split}
      B_{Born}(s,B_i,B_f,P) =& -\frac{s+m_P (m_f+m_i)+m_f m_i}{s-m_P{}^2}, \\
      A_{Born}(s,B_i,B_f,P) =& -\frac{m_P \left(-2 s+m_f{}^2+m_i{}^2\right)+(m_f+m_i) (m_f m_i-s)}{2 \left(s-m_P{}^2\right)},
\end{split}
\end{equation}
where $s$ is the invariant mass squared,  $m_i$, and $m_f$ are the masses of the initial and final baryons, $B_i,B_f,P$ are the incoming, outgoing, propagating baryons respectively.
For a crossed Born diagram, one can obtain the amplitude from the corresponding direct one with the following replacement $s\rightarrow u$.

For the $d_{45},d_{46},d_{47}$ terms, the expressions are slightly different
\begin{equation}\label{BornAmp}
    \begin{split}
      B^2_{Born}(s,B_i,B_f,P) =& \frac{m_i+m_P}{s-m_P{}^2}, \\
      A^2_{Born}(s,B_i,B_f,P) =& \frac{-2 s+m_P (m_f-m_i)+m_f m_i+m_i{}^2}{2 \left(s-m_P{}^2\right)}.
    \end{split}
\end{equation}

For the $A$ parts of the Born terms, one would need to perform two replacements
\begin{equation}\label{BoenApart}
    B(s)\leftrightarrow A(s), \quad B(u)\leftrightarrow -A(u).
\end{equation}

The contributions of the Born diagrams are
\begin{itemize}
  \item $\pi N_{Born}^{I=3/2}$

\begin{equation}\label{Npi32BB}
    \begin{split}
      B_{\pi N}^{I=3/2} & =-\frac{B(u,N,N,N) (D+F)^2}{2 f^2} \\
        & +\frac{4 B(u,N,N,N) (D+F) \left(2 d_{38} m_K{}^2-d_{38} m_{\pi }{}^2+2 d_{40} m_K{}^2+d_{40} m_{\pi }{}^2+2 d_{44} m_{\pi }{}^2\right)}{f^2} \\
        & -\frac{4 m_{\pi }{}^2 B^2(u,N,N,N) (D+F) (d_{45}+d_{46})}{f^2}.
    \end{split}
\end{equation}

  \item $\pi N_{Born}^{I=1/2}$

    \begin{equation}\label{Npi12BB}
    \begin{split}
      B_{\pi N}^{I=1/2}= & \frac{(D+F)^2 (3 B(s,N,N,N)+B(u,N,N,N))}{4 f^2} \\
                         & -\frac{2 (D+F) (3 B(s,N,N,N)+B(u,N,N,N)) }{f^2}\left(2 d_{38} m_K{}^2-d_{38} m_{\pi }{}^2 \right. \\
                         &\left.+2 d_{40} m_K{}^2+d_{40} m_{\pi }{}^2+2 d_{44} m_{\pi }{}^2\right) \\
                         &\frac{2 m_{\pi }{}^2 (D+F) (d_{45}+d_{46}) (3 B^2(s,N,N,N)+B^2(u,N,N,N))}{f^2}.
    \end{split}
  \end{equation}

  \item $K N^{I=1}_{Born}$

  \begin{equation}\label{K N1BB}
    \begin{split}
      B_{K N}^{I=1}= & -\frac{B(u,N,N,\Lambda) (D+3 F)^2+3 B(u,N,N,\Sigma) (D-F)^2}{12 f^2} \\
                     & -\frac{2}{3 f^2} \left(m_{\pi }{}^2 \left(B(u,N,N,\Lambda) (D+3 F) (2 d_{38}+d_{39}-2 d_{40}+d_{41}) \right.\right.\\
                     &\left.\left. +3 B(u,N,N,\Sigma) (F-D)(d_{39}+d_{41})\right) \right.\\
                     &\left.-2 m_K{}^2 \left(B(u,N,N,\Lambda) (D+3 F) (2 d_{38}+2 d_{40}-d_{41}-d_{43}+2 d_{44}) \right.\right.\\
                     &\left.\left. -3 B(u,N,N,\Sigma) (F-D)( d_{41}+d_{43})\right)\right) \\
                     & -\frac{2 m_K{}^2 (B^2(u,N,N,\Lambda) (D+3 F) (d_{45}+3 d_{46})-3 B^2(u,N,N,\Sigma) (F-D) (d_{45}-d_{46}))}{3 f^2}.
    \end{split}
  \end{equation}

  \item $K N^{I=0}_{Born}$
  \begin{equation}\label{K N0BB}
    \begin{split}
      B_{K N}^{I=0}= & \frac{B(u,N,N,\Lambda) (D+3 F)^2-9 B(u,N,N,\Sigma) (D-F)^2}{12 f^2} \\
                     & -\frac{2}{3 f^2} \left(2 m_K{}^2 \left(B(u,N,N,\Lambda) (D+3 F) (2 d_{38}+2 d_{40}-d_{41}-d_{43}+2 d_{44}) \right.\right. \\
                     &\left.\left.+9 B(u,N,N,\Sigma) (F-D) (d_{41}+d_{43})\right) \right.\\
                     &\left.-m_{\pi }{}^2 \left(B(u,N,N,\Lambda) (D+3 F) (2 d_{38}+d_{39}-2 d_{40}+d_{41}) \right.\right.\\
                     &\left.\left.-9 B(u,N,N,\Sigma) (F-D)
   (d_{39}+d_{41})\right)\right)  \\
                     & \frac{2 m_K{}^2 (B^2(u,N,N,\Lambda) (D+3 F) (d_{45}+3 d_{46})+9 B^2(u,N,N,\Sigma) (F-D) (d_{45}-d_{46}))}{3 f^2}.
    \end{split}
  \end{equation}
\end{itemize}

\subsection{Vertex renormalization}
To simplify the final expression, we provide the amplitudes without integrating over the intermediate momentum. The integral can be easily performed with the help of the
OneLoop package~\cite{vanHameren:2009dr,vanHameren:2010cp}. The contributions are

\begin{equation}\label{vreAab}
    \begin{split}
      Re_{ab}(B_i,B_f,\Phi,P)= & \frac{-i(\slashed k+\slashed q_f)(\slashed P-\slashed k +m_P)\gamma^5\slashed k}{(k^2-m_{\Phi}^2)((P-k)^2-m_P^2)},
    \end{split}
\end{equation}

\begin{equation}\label{vreAcd}
    \begin{split}
      Re_{cd}(B_i,B_f,\Phi,\Phi_f,P_1,P_2)= & \frac{i\gamma^5\slashed k(\slashed p_f-\slashed k+m_2)\gamma^5\slashed q_f(\slashed P-\slashed k+m_1)\gamma^5\slashed k }{(k^2-m_{\Phi}^2)((p_f-k)^2-M_2^2)((P-k)^2-M_1^2)},
    \end{split}
\end{equation}

\begin{equation}\label{vreAno}
    \begin{split}
      Re_{no}(B_i,B_f,\Phi,)= & \frac{-i\gamma^5\slashed k(\slashed p_f+\slashed k+m_P)(\slashed q_f+\slashed k)}{(k^2-m_{\Phi}^2)((p_f+k)^2-m_P^2)},
    \end{split}
\end{equation}

\begin{equation}\label{vreApr}
    \begin{split}
      Re_{pr}(\Phi)= & \frac{i\gamma^5\slashed q_f}{k^2-m_{\Phi}^2},
    \end{split}
\end{equation}
where $B_i,B_f,\Phi_f$ refer to initial and final state baryons and mesons, $\Phi$ is the propagated meson, and $P,P_1,P_2$ are the propagated baryons.

In numerical calculations, when limited to $\pi N$ and $KN$ channels, only vertices listed below are needed,
\begin{equation}\label{VRE115}
\begin{split}
    V^{Re}_{p\rightarrow \pi^0 p}=&-\frac{\Delta F_{\pi} (D+F)}{2 f} -\frac{(D+F)}{24 f^3} \left(Z_B(N,\eta,N) (D-3 F)^2+Z_B(N,K,\Lambda) (D+3 F)^2 \right. \\
                                  &\left.+9 \left(Z_B(N,K,\Sigma)(D-F)^2+Z_B(N,\pi,N) (D+F)^2\right)\right) -\frac{(D+F) Z_{\Phi}(\pi)}{8 f} \\
                                  &-\frac{Re_{ab}(N,N,K,\Lambda) (D+3 F)}{16 f^3}+\frac{Re_{ab}(N,N,K,\Sigma) (F-D)}{16 f^3}-\frac{Re_{ab}(N,N,\pi,N) (D+F)}{4 f^3} \\
                                  &-\frac{Re_{cd}(N,N,\eta,\pi,N,N) (D-3 F)^2 (D+F)}{24 f^3}-\frac{D Re_{cd}(N,N,K,\pi,\Lambda,\Sigma) (F-D) (D+3 F)}{12 f^3} \\
                                  &-\frac{D Re_{cd}(N,N,K,\pi,\Sigma,\Lambda) (F-D) (D+3 F)}{12 f^3}-\frac{F Re_{cd}(N,N,K,\pi,\Sigma,\Sigma) (F-D)^2}{2 f^3} \\
                                  &+\frac{Re_{cd}(N,N,\pi,\pi,N,N) (D+F)^3}{8 f^3} +\frac{Re_{no}(N,N,K,\Lambda) (D+3 F)}{16 f^3} \\
                                  &-\frac{Re_{no}(N,N,K,\Sigma) (F-D)}{16 f^3}+\frac{Re_{no}(N,N,\pi,N) (D+F)}{4 f^3}\\
                                  &+\frac{(D+F) (Re_{pr}(K)+2 Re_{pr}(\pi))}{6 f^3},
\end{split}
\end{equation}

\begin{equation}\label{VRE116}
    V^{Re}_{p\rightarrow \pi^+ n}=\sqrt{2}V^{Re}_{p\rightarrow \pi^0 p},\quad V^{Re}_{n\rightarrow \pi^- p}=\sqrt{2}V^{Re}_{p\rightarrow \pi^0 p}, \quad V^{Re}_{n\rightarrow \pi^0 n}=-V^{Re}_{p\rightarrow \pi^0 p},
\end{equation}

\begin{equation}\label{VRE111}
\begin{split}
    V^{Re}_{p\rightarrow K^+ \Sigma^0}=&\frac{(F-D) \Delta F_K}{2 f} -\frac{D-F}{48 f^3} \left(Z_B(N,\eta,N) (D-3 F)^2+Z_B(N,K,\Lambda) (D+3 F)^2 \right.\\
    &\left.+9 Z_B(N,K,\Sigma) (D-F)^2+9 Z_B(N,\pi,N) (D+F)^2+4 Z_B(\Sigma,\eta,\Sigma) D^2 \right.\\
    &\left.+6 Z_B(\Sigma,K,N) (D-F)^2+6 Z_B(\Sigma,K,\Xi) (D+F)^2+4 Z_B(\Sigma,\pi,\Lambda) D^2+24 Z_B(\Sigma,\pi,\Sigma) F^2\right) \\
    &-\frac{(D-F) Z_{\Phi}(K)}{8 f} -\frac{Re_{ab}(m,\Sigma,\eta,N) (D-3 F)}{16 f^3}+\frac{Re_{ab}(N,\Sigma,K,\Sigma) (F-D)}{4 f^3} \\
    &-\frac{Re_{ab}(N,\Sigma,\pi,N) (D+F)}{16 f^3} -\frac{D Re_{cd}(N,\Sigma,\eta,K,N,\Sigma) (D-3 F) (F-D)}{12 f^3} \\
    &-\frac{Re_{cd}(N,\Sigma,K,K,\Lambda,\Xi) (D-3 F) (D+3 F) (D+F)}{24 f^3}-\frac{Re_{cd}(N,\Sigma,K,K,\Sigma,\Xi) (F-D) (D+F)^2}{8 f^3} \\
    &+\frac{D Re_{cd}(N,\Sigma,\pi,K,N,\Lambda) (D+3 F) (D+F)}{12f^3}+\frac{F Re_{cd}(N,\Sigma,\pi,K,N,\Sigma) (F-D) (D+F)}{2 f^3} \\
    & -\frac{Re_{no}(N,\Sigma,K,N) (D-F)}{8 f^3}-\frac{Re_{no}(N,\Sigma,K,N) (F-D)}{4 f^3} \\
    &+\frac{D Re_{no}(N,\Sigma,\eta,\Sigma)}{8
    f^3}+\frac{D Re_{no}(N,\Sigma,\pi,\Lambda)}{8 f^3}-\frac{F Re_{no}(N,\Sigma,\pi,\Sigma)}{4 f^3} \\
    &\frac{(D-F) (Re_{pr}(\eta)+2 Re_{pr}(K)+Re_{pr}(\pi))}{8 f^3},
\end{split}
\end{equation}

\begin{equation}\label{VRE112}
    V^{Re}_{p\rightarrow K^0 \Sigma^+}=\sqrt{2}V^{Re}_{p\rightarrow K^+ \Sigma^0},\quad V^{Re}_{n\rightarrow K^0 \Sigma^0}=-V^{Re}_{p\rightarrow K^+ \Sigma^0},\quad V^{Re}_{n\rightarrow K^+ \Sigma^-}=\sqrt{2}V^{Re}_{p\rightarrow K^+ \Sigma^0},
\end{equation}

\begin{equation}\label{VRE113}
\begin{split}
    V^{Re}_{p\rightarrow K^+ \Lambda}=&\frac{(D+3 F) \Delta F_K}{2 \sqrt{3} f}+ \frac{D+3 F}{4 \sqrt{3} f}\left( \frac{Z_B(N,\eta,N) (D-3 F)^2}{12 f^2}+\frac{Z_B(N,K,\Lambda) (D+3 F)^2}{12 f^2} \right.\\
                   &\left.+\frac{3 Z_B(N,K,\Sigma) (F-D)^2}{4 f^2}+\frac{3 Z_B(N,\pi,N) (D+F)^2}{4 f^2} + \frac{Z_B(\Lambda,\eta,\Lambda) D^2}{3 f^2} \right.\\
                   &\left.+\frac{Z_B(\Lambda,K,N) (D+3 F)^2}{6 f^2}+\frac{Z_B(\Lambda,K,\Xi) (D-3 F)^2}{6f^2}+\frac{Z_B(\Lambda,\pi,\Sigma) D^2}{f^2} \right) \\
                   &+\frac{(D+3 F)Z_{\Phi}(K)}{8 \sqrt{3} f} +\frac{3 \sqrt{3}Re_{ab}(N,\Lambda,\pi,mN) (D+F)}{16 f^3} \\
                   &-\frac{\sqrt{3} Re_{ab}(N,\Lambda,\eta,mN) (D-3 F)}{16 f^3}+ \frac{D Re_{cd}(N,\Lambda,\eta,K,N,\Lambda) (D+3 F) (D-3 F)}{12 \sqrt{3} f^3} \\
                   &+\frac{Re_{cd}(N,\Lambda,K,K,\Lambda,\Xi) (D+3 F) (D-3 F)^2}{24 \sqrt{3} f^3}-\frac{\sqrt{3} Re_{cd}(N,\Lambda,K,K,\Sigma,\Xi) (F-D) (D+F) (D-3 F)}{8 f^3} \\
                   &+\frac{\sqrt{3} D Re_{cd}(N,\Lambda,\pi,K,N,\Sigma) (F-D) (D+F)}{4 f^3} -\frac{\sqrt{3} Re_{no}(N,\Lambda,K,N) (D+3 F)}{8 f^3} \\
                   &-\frac{\sqrt{3} D Re_{no}(N,\Lambda,\eta,\Lambda)}{8 f^3}+\frac{\sqrt{3} D Re_{no}(N,\Lambda,\pi,\Sigma)}{8 f^3} \\
                   &-\frac{(D+3 F) (Re_{pr}(\eta)+2 Re_{pr}(K)+Re_{pr}(\pi))}{8 \sqrt{3} f^3},
\end{split}
\end{equation}

\begin{equation}\label{VRE513}
    V^{Re}_{n\rightarrow K^0 \Lambda^0}=V^{Re}_{p\rightarrow K^+ \Sigma^0}.
\end{equation}

\subsection{Divergent parts of the LECs}
As mentioned in the main text, the LECs can be divided into finite parts and infinite parts as shown in Eq.~(\ref{UVab}), The divergent parts absorb all the ultraviolet divergence from the
loop diagrams. Their explicit expressions are
\begin{equation}\label{UVlec2}
  \begin{split}
    c_{1d} =& -\frac{-27 m D F^3-31 m D^3 F+9 m D F}{192 f^2 \pi ^2},\\
    c_{2d} =& -\frac{-27 m F^4-106 m D^2 F^2+18 m F^2-19 m D^4-6 m D^2+9 m}{768 f^2 \pi ^2},\\
    c_{3d} =& -\frac{11 m D^4+9 m F^2 D^2+9 m D^2}{72 f^2 \pi ^2},\\
    b_{1d} =& -\frac{207 m F^4+738 m D^2 F^2+18 m F^2-73 m D^4-6 m D^2-45 m}{2304 f^2 \pi ^2},\\
    b_{2d} =& -\frac{-45 m F^4+90 m D^2 F^2+90 m F^2-5 m D^4-30 m D^2-9 m}{256 f^2 \pi ^2},\\
    b_{3d} =& -\frac{3 \left(3 m D F^3-m D^3 F+m D F\right)}{32 f^2 \pi ^2},\\
    b_{0d} =& -\frac{m \left(9 F^2+13 D^2\right)}{144 f^2 \pi^2},\\
    b_{4d} =& -\frac{m D^4-81 m F^2 D^2+27 m D^2}{144 f^2 \pi ^2},\\
    b_{Dd} =& -\frac{3 m F^2-m D^2}{32 f^2 \pi ^2},\\
    b_{Fd} =& -\frac{5 m F D}{48 f^2 \pi ^2},\\
    b_{5d} =& -\frac{-15 F^4+30 D^2 F^2+30 F^2+9 D^4-10 D^2-15}{768 f^2 \pi ^2},\\
    b_{6d} =& -\frac{-9 D F^3-13 D^3 F+9 D F}{96 f^2 \pi ^2},\\
    b_{7d} =& -\frac{-9D^4+18 D^2 F^2+18 F^2-D^4-6 D^2-9}{256 f^2 \pi ^2},\\
    b_{8d} =& -\frac{-5 D^4-27 F^2 D^2+9 D^2}{144 f^2 \pi ^2},
  \end{split}
\end{equation}
\begin{equation}\label{UVlec3}
  \begin{split}
    d_{5d}=&-d_{48d}-\frac{9 F^4-36 D F^3+46 D^2 F^2-18 F^2-20 D^3 F+36 D F+17 D^4-26 D^2-9}{1536 f^2 \pi ^2},\\
    d_{6d}=&d_{50 d}-\frac{D^4+3 F^2 D^2-3 D^2}{96 f^2 \pi ^2},\\
    d_{4d}=&-d_{49 d}-\frac{-9 F^4-36 D F^3-46 D^2 F^2+18 F^2-20 D^3 F+36 D F-17 D^4+26 D^2+9}{1536 f^2 \pi ^2},\\
    d_{10d}=& \frac{b_{8 d}}{4 m}-\frac{5 D^4+27 F^2 D^2-9 D^2}{576 f^2 m \pi ^2}, \\
    d_{7d} =& \frac{b_{5 d}}{8 m}-\frac{b_{6 d}}{8 m}-\frac{21 F^4-36 D F^3-42 D^2
   F^2-42 F^2-52 D^3 F+36 D F-3 D^4+14 D^2+21}{3072 f^2 m \pi ^2},\\
    d_{9d} =& -\frac{b_{5 d}}{2 m}+\frac{b_{7 d}}{2 m}-\frac{3 F^4-6 D^2 F^2-6
   F^2+3 D^4+2 D^2+3}{384 f^2 m \pi ^2},\\
    d_{8d} =& \frac{b_{5 d}}{8 m}+\frac{b_{6 d}}{8 m}+\frac{b_{7 d}}{8 m}-\frac{21 F^4+36 D F^3-42 D^2 F^2-42D_r^2+52 D^3 F-36 D F-3 D^4+14 D^2+21}{3072 f^2 m \pi ^2}.
  \end{split}
\end{equation}

\subsection{Power counting breaking terms of the one-loop diagrams}
In this subsection, we list the power counting breaking terms in the $\pi N$ and $KN$ channels.

\begin{itemize}
  \item $\pi N_{PCB}^{I=3/2}$

  \begin{equation}\label{NpiPCBD32}
    \begin{split}
       D_{PCB}=&-\frac{1}{1152 \pi ^2 f^4 \tilde{m}} \left(2 \tilde{m}{}^2 \left(2 m_{\pi }{}^2 \left(369 D^4+108 D^3 F+18 D^2 \left(43 F^2-4\right) \right.\right.\right.\\
       &\left.\left.\left.+12 D F \left(25 F^2+2\right)+5 F^2 \left(29F^2-8\right)\right)-t \left(369 D^4+108 D^3 F+9 D^2 \left(86 F^2-1\right) \right.\right.\right.\\
       &\left.\left.\left.+6 D F \left(50 F^2+9\right)+F^2 \left(145 F^2+3\right)\right)\right)-\sigma^2 \left(171 D^4+108 D^3 F+138 D^2 F^2 \right.\right.\\
       &\left.\left.+60 D F^3+35 F^4+63\right)\right),
    \end{split}
  \end{equation}
  \begin{equation}\label{NpiPCBB32}
    B_{PCB}=-\frac{ \tilde{m}{}^2 \left(9 D^4+9 D^3 F+2 D^2 F^2-3 D F^3+9 D F+3 F^4-2 F^2\right)}{12 \pi ^2 f^4}.
  \end{equation}

  \item $\pi N_{PCB}^{I=1/2}$

  \begin{equation}\label{NpiPCBD12}
    \begin{split}
       D_{PCB}=&-\frac{1}{1152 \pi ^2 f^4 \tilde{m}} \left(2 \tilde{m}{}^2 \left(2 m_{\pi }{}^2 \left(369 D^4+108 D^3 F+18 D^2 \left(43 F^2-4\right) \right.\right.\right.\\
       &\left.\left.\left.+12 D F \left(25 F^2+2\right)+5 F^2 \left(29F^2-8\right)\right)-t \left(369 D^4+108 D^3 F+9 D^2 \left(86 F^2-1\right) \right.\right.\right.\\
       &\left.\left.\left.+6 D F \left(50 F^2+9\right)+F^2 \left(145 F^2+3\right)\right)\right)-\sigma^2 \left(171 D^4+108 D^3 F+138 D^2 F^2 \right.\right.\\
       &\left.\left.+60 D F^3+35 F^4+63\right)\right),
    \end{split}
  \end{equation}
  \begin{equation}\label{NpiPCBB12}
    B_{PCB}=\frac{ \tilde{m}{}^2 \left(9 D^4+9 D^3 F+2 D^2 F^2-3 D F^3+9 D F+3 F^4-2 F^2\right)}{6 \pi ^2 f^4}.
  \end{equation}

  \item $K N_{PCB}^{I=1}$

  \begin{equation}\label{K NPCBD1}
    \begin{split}
       D_{PCB}=&-\frac{1}{576 \pi ^2 f^4 \tilde{m}} \left(2 \tilde{m}{}^2 \left(2 m_K{}^2 \left(369 D^4+18 D^2 \left(F^2-3\right)+F^2 \left(85 F^2-14\right)\right) \right.\right. \\
       &\left.\left.-t \left(369 D^4+9 D^2 \left(2F^2-1\right)+F^2 \left(85 F^2+3\right)\right)\right)-\sigma ^2 \left(171 D^4-6 D^2 F^2+19 F^4+63\right)\right),
    \end{split}
  \end{equation}
  \begin{equation}\label{K NPCBB1}
    B_{PCB}=-\frac{i \tilde{m}{}^2 \left(27 D^4-3 D^2 F^2+8 F^4+3 F^2\right)}{18 \pi ^2 f^4}.
  \end{equation}

  \item $K N_{PCB}^{I=0}$

  \begin{equation}\label{K NPCBD0}
    \begin{split}
       D_{PCB}=&\frac{1}{144 \pi ^2 f^4 \tilde{m}} \left(F \sigma ^2 \left(-27 D^3+36 D^2 F-15 D F^2+4 F^3\right) \right.\\
       &\left.+\tilde{m}{}^2 \left(2 m_K{}^2 \left(54 D^3 F+D^2 \left(9-378 F^2\right)+6 D \left(25F^2+2\right) F-30 F^4+13 F^2\right)\right.\right. \\
       &\left.\left.+3 F t \left(-18 D^3+126 D^2 F-D \left(50 F^2+9\right)+10 F^3\right)\right)\right),
    \end{split}
  \end{equation}
  \begin{equation}\label{K NPCBB0}
    B_{PCB}=-\frac{ F \tilde{m}{}^2 \left(-27 D^3+9 D^2 F+9 D \left(F^2-3\right)+F \left(F^2-9\right)\right)}{18 \pi ^2 f^4}.
  \end{equation}

\end{itemize}
In the above equations, $\sigma=s-\tilde{m}^2$. We have already set the scale $\mu$ in the $\overline{MS}$ scheme to be equal to the chiral limit baryon mass $\tilde{m}$.
All these power counting breaking terms are absorbed into the corresponding LECs in the EOMS scheme, i.e.,
\begin{equation}\label{PCBabsorb}
    \begin{split}
      b^{PCB}_0 & =b^{PCB}_4+\frac{\tilde{m} \left(9 D^2 \left(42 F^2-1\right)+F^2 \left(30 F^2-13\right)\right)}{288 \pi ^2 f^2}, \\
      b^{PCB}_D & =\frac{\tilde{m} \left(F^2-3 D^2\right)}{64 \pi ^2 f^2}, \\
      b^{PCB}_F & =-\frac{5 D F \tilde{m}}{96 \pi ^2 f^2}, \\
      b^{PCB}_1 & =-\frac{\tilde{m} \left(333 D^4-9 D^2 \left(74 F^2-7\right)+F^2 \left(5F^2-21\right)\right)}{1152 \pi ^2 f^2}, \\
      b^{PCB}_2 & =-\frac{\tilde{m} \left(45 D^4-9 D^2 \left(10 F^2+1\right)+F^2 \left(5 F^2+3\right)\right)}{128 \pi ^2 f^2}, \\
      b^{PCB}_3 & =-\frac{D F \tilde{m} \left(18 D^2+50 F^2+9\right)}{96 \pi ^2 f^2}, \\
      b^{PCB}_5 & =-\frac{-87 D^4+46 D^2 F^2+F^4-15}{768 \pi ^2 f^2}, \\
      b^{PCB}_6 & =\frac{9 D^3 F+5 D F^3}{96 \pi ^2 f^2}, \\
      b^{PCB}_7 & =\frac{9 D^4-18 D^2 F^2+F^4+9}{256 \pi ^2 f^2}, \\
      b^{PCB}_8 & =\frac{144 \pi ^2 b^{PCB}_4 f^2 \tilde{m} t+2 F^2 \sigma ^2 \left(9 D^2+F^2\right)+3 F^2 \tilde{m}{}^2 t \left(63 D^2+5 F^2\right)}{144 \pi ^2 f^2 \sigma ^2}, \\
      c^{PCB}_1 & =\frac{D F \tilde{m} \left(3 D^2-F^2+3\right)}{32 \pi ^2 f^2}, \\
      c^{PCB}_2 & =\frac{\tilde{m} \left(9 D^4+2 D^2 F^2+3 F^4-2 F^2\right)}{96 \pi ^2 f^2}, \\
      c^{PCB}_3 & =-\frac{F^2 \tilde{m} \left(9 D^2+F^2-9\right)}{72 \pi ^2 f^2}.
    \end{split}
\end{equation}

\subsection{Decuplet contributions}

In this section, we evaluate the contributions of the virtual decuplet  to meson-baryon scattering by including the lowest order exchange diagrams. For the construction of HBChPT  and covariant BChPT with $\Delta$ or decuplet fields in general, we refer
the reader to Refs.~\cite{Jenkins:1990jv,Butler:1992ci}, and Ref.~\cite{Pascalutsa:2006up}, respectively.

In the following we  show the lowest order contributions of   the decuplet adopting the so-called $\delta$-expansion~\cite{Pascalutsa:2002pi}. That is, the decuplet contributions
are counted differently for energies well below the resonances or around the resonance peaks:

1. low-energy: $m_{\phi}\sim p$, $m_{\Delta}-m_N\sim p^{1/2}$

2. resonance peak: $m_{\phi}\sim p^2$, $m_{\Delta}-m_N\sim p$

\begin{figure}
\centering
\includegraphics[width=0.5\textwidth]{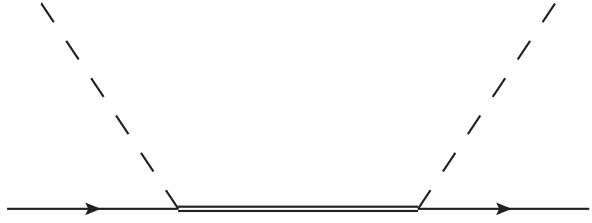}
\caption{Lowest order contribution from the intermediate decuplet. The double line refers to the spin-$\frac{3}{2}$ propagator.}
\label{Deexchange}
\end{figure}

At the lowest order, the decuplet exchange diagram is shown  in Fig.\ref{Deexchange}.
The relevant effective Lagrangian with decuplet as explicit degrees of freedom is
\begin{equation}\label{LphiBD}
    \mathcal{L}_{\Phi BD}=\frac{i \mathcal{C}}{m_{D0} F_{\Phi}}\epsilon^{abc}(\partial_{\alpha}\bar{T}_{\mu}^{ade}) \gamma^{\alpha\mu\nu}B_{c}^{e}\partial_{\nu}\Phi_b^d + H.c.
\end{equation}
where we have adopted the so-called ``consistent" coupling scheme for the meson-octet-decuplet vertices~\cite{Pascalutsa:1998pw,Pascalutsa:1999zz}. $T$ is the tensor collecting the decuplet baryons, $B$ is the baryon octet, and $\Phi$ is
 the Gold-stone boson octet. $m_{D0}$ here refers to the chiral limit decuplet mass. The total antisymmetric gamma matrix products are defined as
\begin{equation}\label{gamma3}
    \gamma^{\mu\nu}=\frac{1}{2}[\gamma^{\mu},\gamma^{\nu}], \quad \gamma^{\mu\nu\rho}=\frac{1}{2}\{\gamma^{\mu\nu},\gamma^{\rho}\}.
\end{equation}

Similar to the Born terms, the contribution of the decuplet exchange diagram reads
\begin{equation}\label{decupletamp}
\begin{split}
      B^D_{Born}(s,B_i,\Phi_i,B_f,\Phi_f,D) =& -\frac{1}{6 \left(s-m_D^2\right)}\left(-m_D m_i \left(m_f^2-M_f^2+s\right)-m_D m_f \left(m_i^2-M_i^2+s\right) \right.\\
                                             &\left.+\left(m_f^2-M_f^2+s\right) \left(m_i^2-M_i^2+s\right)-3 s
   \left(m_f^2+m_i^2-t\right)-2 s m_f m_i \right), \\
      A^D_{Born}(s,B_i,\Phi_i,B_f,\Phi_f,D) =& -\frac{1}{12 \left(s-m_D^2\right)}\left(s \left(m_D \left(4 m_f m_i-6 m_f^2-6 m_i^2+6 t\right) \right.\right.\\
                                             &\left.\left.+m_f \left(-3 m_i^2-2 M_i^2+2 s+3 t\right)+m_i \left(-2 M_f^2-3 m_i^2+2 s+3 t\right) \right.\right.\\
                                             &\left.\left.-3 m_f^2 m_i-3 m_f^3\right)-m_D m_f \left(m_f+m_i\right) \left(m_i^2-M_i^2+s\right) \right.\\
                                             &\left.+\left(m_f^2-M_f^2+s\right) \left(\left(4 m_D+m_f+m_i\right) \left(m_i^2-M_i^2+s\right)-m_D m_i \left(m_f+m_i\right)\right) \right),
\end{split}
\end{equation}
where the symbols $B_i,B_f,D,\Phi_i,\Phi_f$ refer to incoming/outgoing baryons, intermediate decuplet, and incoming/outgoing mesons. Note that in this notation, the coupling constant $\mathcal{C}$ of the meson-octet-decuplet vertices
has  a factor of $2\sqrt{2}$ compared to the $h_A$ in, for instance, Ref.~\cite{Chen:2012nx}. Taking this into account, we set $m_{D0}=m_{\Delta}$ in the fitting process. The $B$ parts of the
scattering amplitudes are
\begin{equation}\label{decupletamppiN}
\begin{split}
    B_{\pi N}^{I=3/2}=& \frac{2 C^2 (B^D(u,N,\pi,N,\pi,\Delta)-3 B^D(s,t,N,\pi,N,\pi,\Delta))}{3 f^2 m_{D0}^2}, \\
    B_{\pi N}^{I=1/2}=& \frac{8 C^2 B^D(u,N,\pi,N,\pi,\Delta)}{3 f^2 m_{D0}^2},
\end{split}
\end{equation}
\begin{equation}\label{decupletampKN}
\begin{split}
    B_{K N}^{I=1}=& \frac{C^2 B^D(u,N,K,N,K,\Sigma^{*})}{3 f^2 m_{D0}^2}, \\
    B_{K N}^{I=0}=& \frac{C^2 B^D(u,N,K,N,K,\Sigma^{*})}{f^2 m_{D0}^2}.
\end{split}
\end{equation}
The $A$ parts can be easily obtained with the same replacement as that in the Born terms.

We first follow exactly the same fitting strategy as in the decuplet-less case except enlarging the fitting range up to 1.2GeV in the $\pi N$ channel. A direct fit of these 9 parameters($\alpha_{1,\ldots,8}$ , $\mathcal{C}$) yields a $h_A=2.900(44)$, which is almost exactly equal to the value t determined by fitting to the Breit-Wigner width of the $\Delta$, i.e., $\Gamma_{\Delta}=118$MeV~\cite{Alarcon:2012kn,Chen:2012nx}. Thus we fix $h_A$ to be exactly $2.9$ and fit(refit) the LECs in the $KN$($\pi N$) channel. The fitting results are collected in Tables~\ref{fitDepiN},~\ref{fitDeKN1},~\ref{fitDeKN0}.

\begin{figure}
\centering
\begin{tabular}{ccc}
{\includegraphics[width=0.32\textwidth]{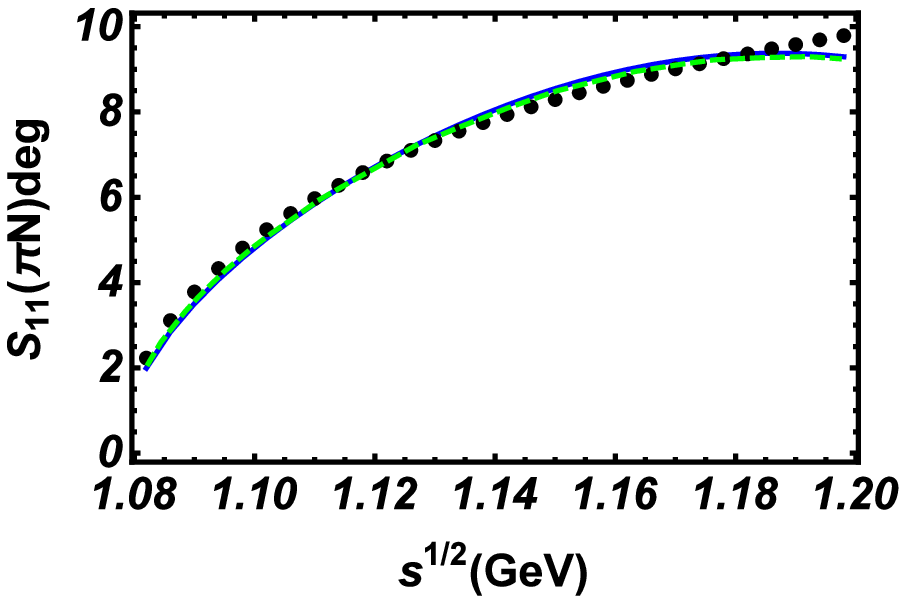}}&{\includegraphics[width=0.32\textwidth]{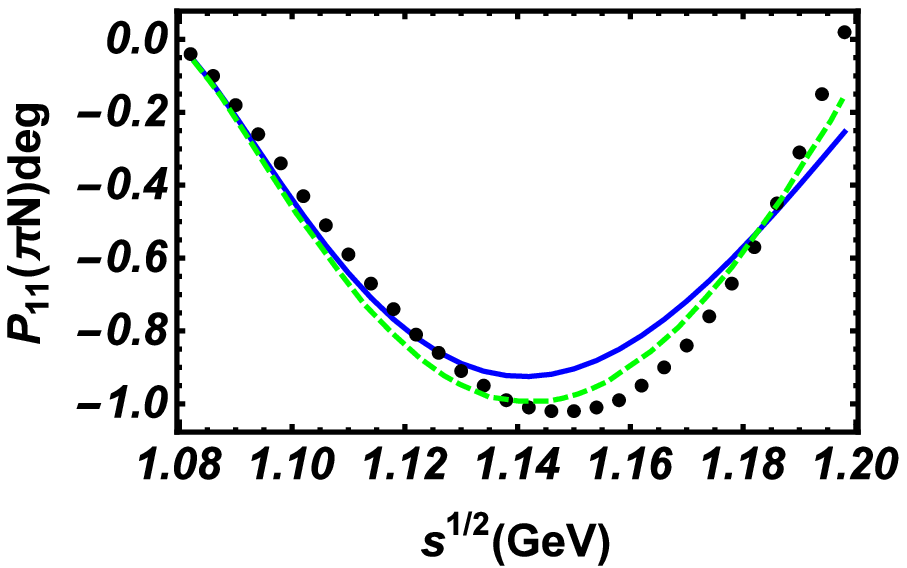}}& {\includegraphics[width=0.32\textwidth]{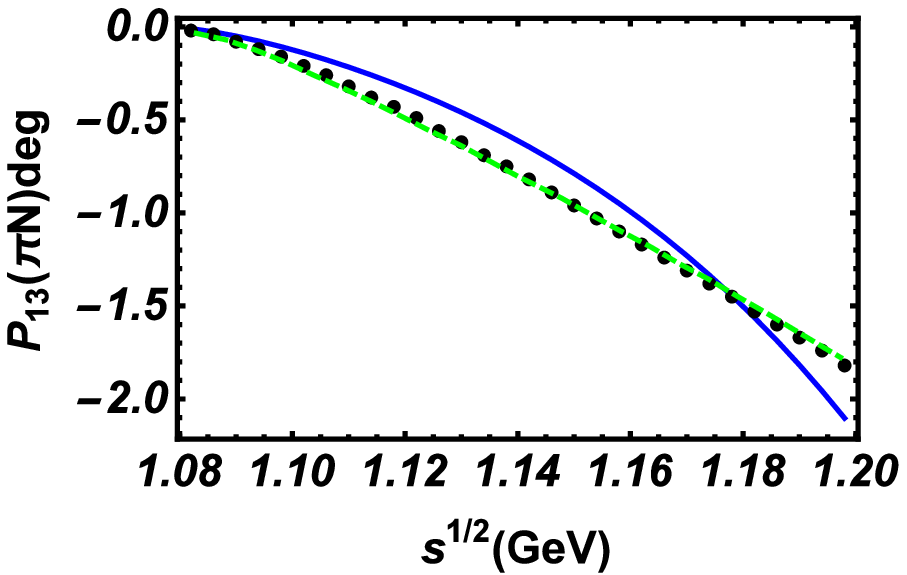}} \\
{\includegraphics[width=0.32\textwidth]{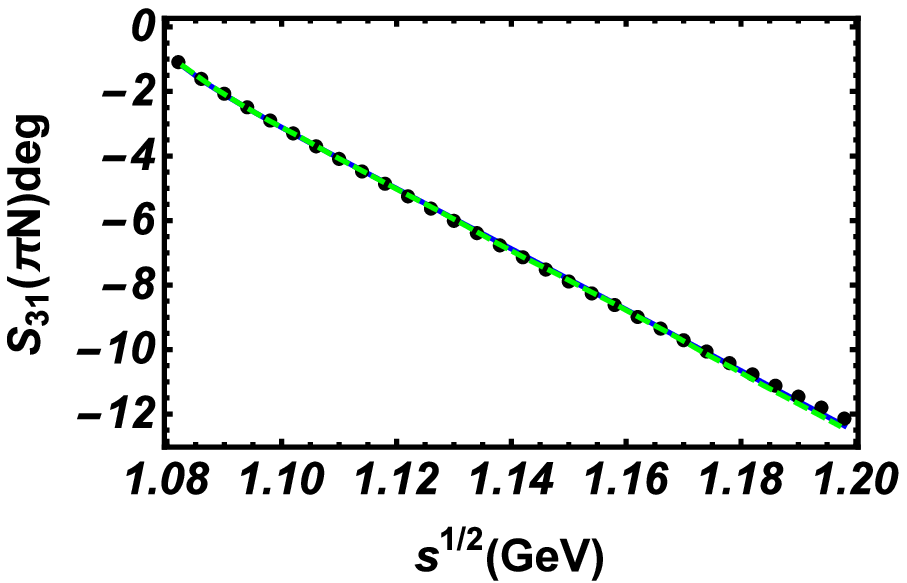}}&{\includegraphics[width=0.32\textwidth]{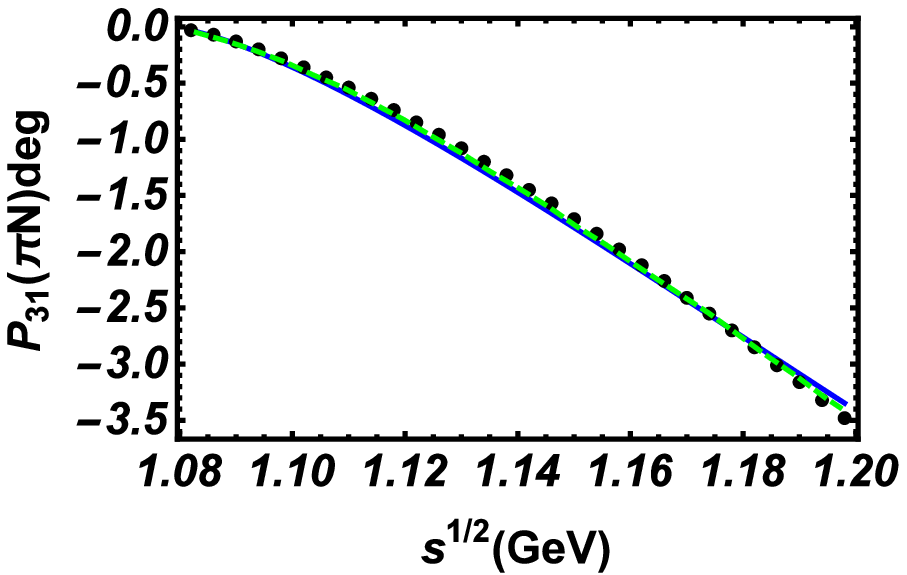}}& {\includegraphics[width=0.32\textwidth]{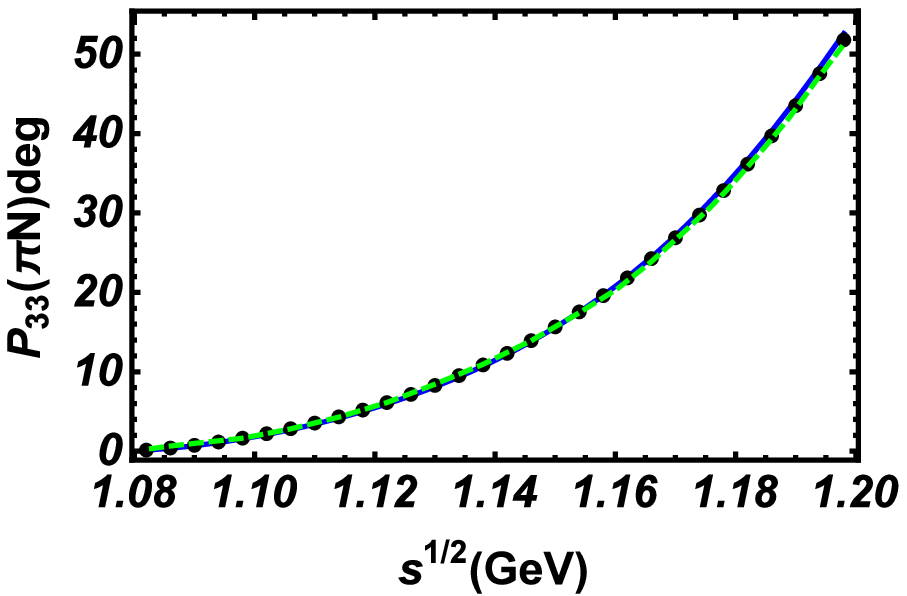}} \\
\end{tabular}
\caption{$\pi N$ phase shifts  with the lowest order decuplet contribution included. The blue lines are our results in SU(3) while the green dashed lines are the results in SU(2) from Ref.~\cite{Chen:2012nx}. The black dots denote the experimental data.}
\label{De:pin}
\end{figure}

\begin{figure}
\centering
\begin{tabular}{ccc}
{\includegraphics[width=0.32\textwidth]{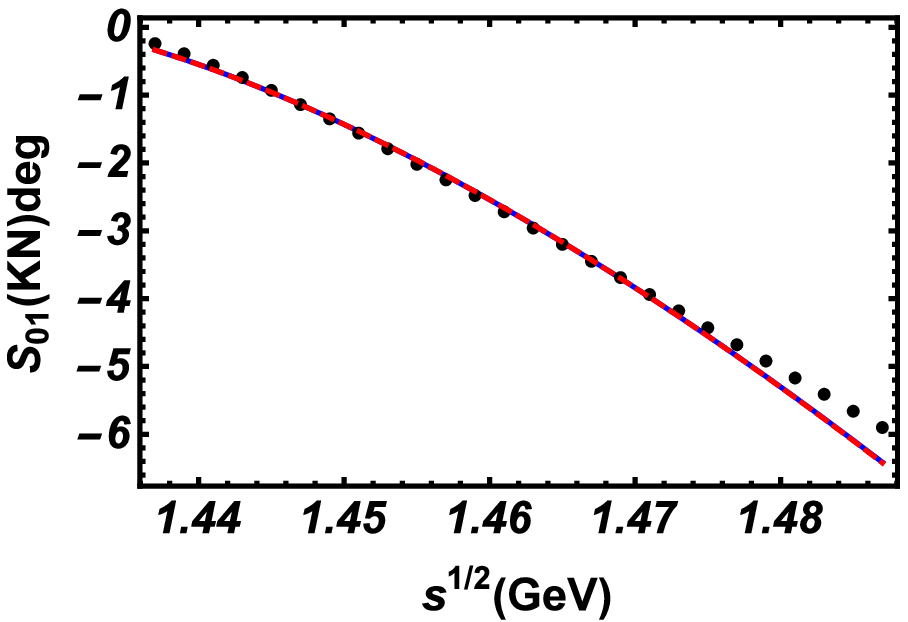}}&{\includegraphics[width=0.32\textwidth]{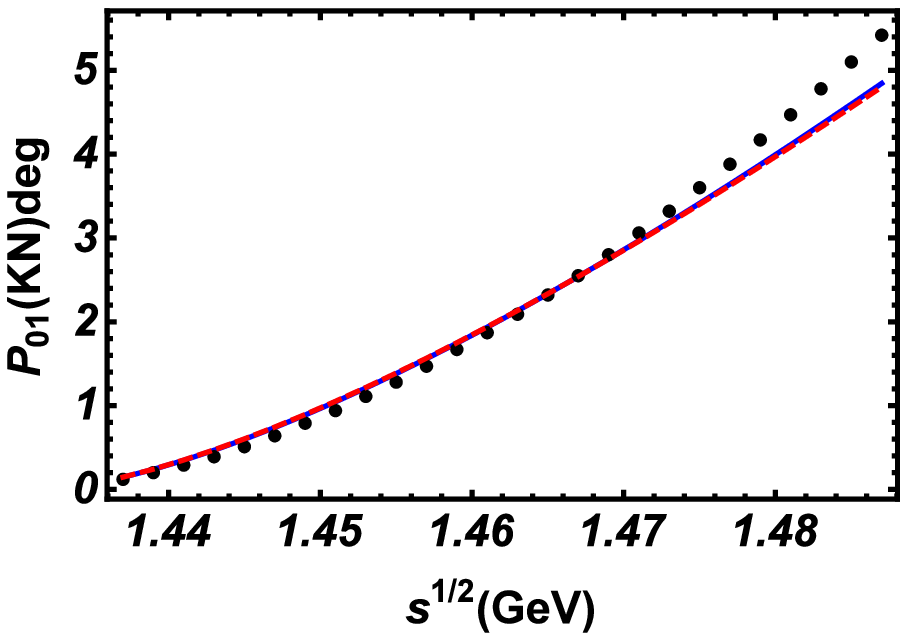}}& {\includegraphics[width=0.32\textwidth]{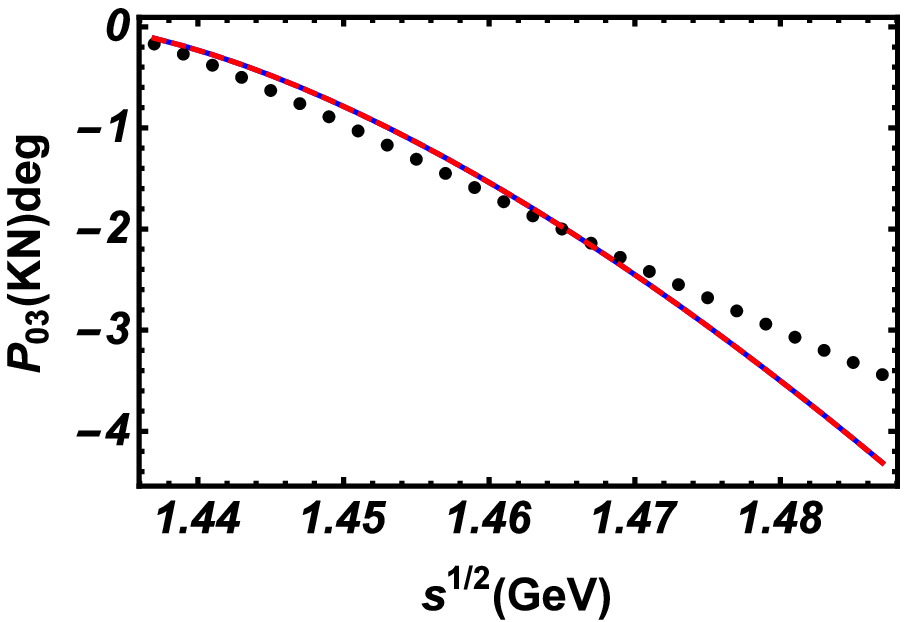}} \\
{\includegraphics[width=0.32\textwidth]{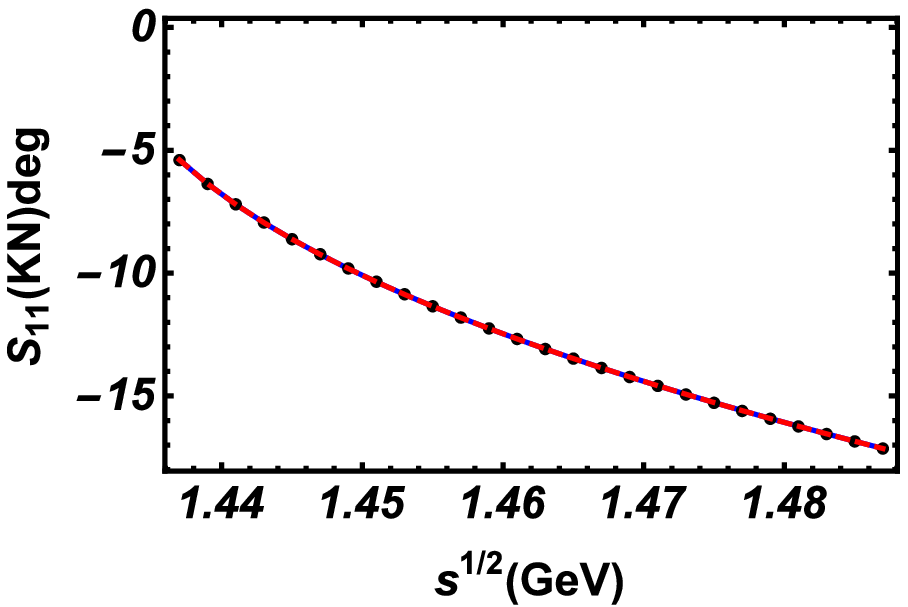}}&{\includegraphics[width=0.32\textwidth]{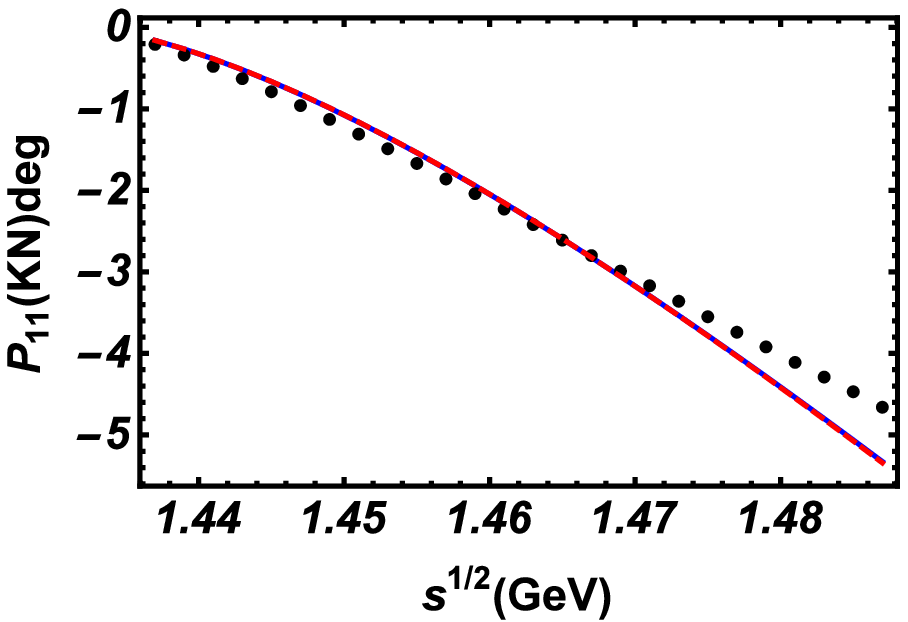}}& {\includegraphics[width=0.32\textwidth]{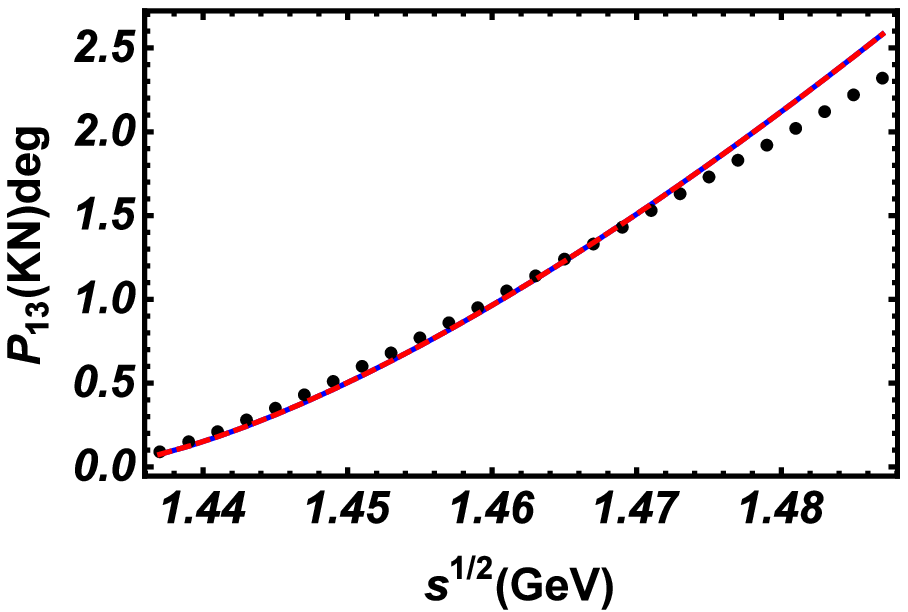}} \\
\end{tabular}
\caption{$K N$ phase shifts with the lowest order decuplet contribution  included. The blue solid lines are our results in SU(3) while the red dashed lines are the results without the decuplet contributions. The black dots denote the experimental data. The blue solid lines and the red dashed lines overlap each other.}
\label{De:kn}
\end{figure}

Taking the lowest order contribution into account, we re-plot the phase shifts with the new set of LECs. The results are shown in Figs.~\ref{De:pin},~\ref{De:kn}. Clearly, the description is compatible with that in the SU(2) case.  Compared with the results without the decuplet contributions, the description is obviously improved up to higher energies, especially for the $P_{11}$ channel, as mentioned in the main text. Meanwhile, for  the
$KN$ channel, taking the decuplet contribution into account or not does not seem to make any appreciable difference.

\begin{table}
\centering
\caption{LECs in the $\pi N$ channel including the lowest order decuplet contribution with $h_A=2.9$.}\label{fitDepiN}
\begin{tabular}{cccccccccc}
\hline\hline
$\alpha_1$  &  $\alpha_2$  &  $\alpha_3$  &  $\alpha_4$ & $\alpha_5$ & $\alpha_6$ & $\alpha_7$ & $\alpha_8$  & $\mathcal{C}$&$\chi^2/d.o.f.$\\
\hline
$-3.737(36)$  &  $0.659(7)$  &  $0.893(4)$  &  $-0.745(21)$ & $0.159(5)$ & $1.066(20)$ & $0.351(10)$ & $0.133(34)$ &$\frac{h_A}{2\sqrt{2}}$& 0.918 \\
\hline
\end{tabular}
\end{table}

\begin{table}
\centering
\caption{LECs in the $I=0$ $KN$ channel including the lowest order decuplet contribution with $h_A=2.9$ at $\mathcal{O}(p^3)^*$.}\label{fitDeKN0}
\begin{tabular}{cccccc}
\hline\hline
$\beta_1$  &  $\beta_2$  &  $\beta_3$  &  $\beta_4$ &  $\mathcal{C}$  & $\chi^2/d.o.f.$ \\
\hline
$-0.831(1)$  &  $0.1535(2)$  &  $0.608(2)$  &  $-0.055(1)$ & $\frac{h_A}{2\sqrt{2}}$ & 1.02\\
\hline
\end{tabular}
\end{table}

\begin{table}
\centering
\caption{LECs in the $I=1$ $KN$ channel including the lowest order decuplet contribution with $h_A=2.9$ at $\mathcal{O}(p^3)^*$.}\label{fitDeKN1}
\begin{tabular}{cccccc}
\hline\hline
$\gamma_1$  &  $\gamma_2$  &  $\gamma_3$  &  $\gamma_4$  &  $\mathcal{C}$  & $\chi^2/d.o.f.$  \\
\hline
$-0.398(22)$  &  $0.420(7)$  &  $0.615(6)$  &  $-0.103(42)$ & $\frac{h_A}{2\sqrt{2}}$ & 0.491 \\
\hline
\end{tabular}
\end{table}

One interesting point is that the consideration of the decuplet contributions changes the values of LECs,  similar to the case  of baryon masses~\cite{Ren:2013dzt}.
In the $\delta$-expansion, the leading order decuplet contribution is counted as of $\mathcal{O}(p^{3/2})$. Up  to $\mathcal{O}(p^3)$, one is supposed to include the NLO contributions, which counts
of as $\mathcal{O}(p^{5/2})$. However, as shown in the SU(2) case~\cite{Alarcon:2012kn}, the inclusion of the NLO decuplet contribution will only introduce redundant parameters which could be absorbed into $\mathcal{C}$ and $b_0,b_D,b_F,b_{1,\ldots,8}$.

\end{document}